\documentclass[twocolumn]{aastex63}
\definecolor{trackchange}{cmyk}{0,1,1,0}
\definecolor{explain}{cmyk}{0.93,0.67,0,0.2}

\usepackage{float}
\usepackage{bookmark}
\bookmarksetup{numbered, open,}
\usepackage{enumitem}
\setlist[enumerate]{itemsep=0mm}
\usepackage{hyperref}
\usepackage{subfigure}
\usepackage{xspace}
\usepackage{natbib}
\usepackage{xcolor}  

\usepackage{CJK}

\graphicspath{{figures/}{./}}

\newcommand{\hyMHlimit}{$-$19.39~mag}
\newcommand{\hyMUVM}{$-18.17^{+0.11}_{-0.27}$~mag}
\newcommand{\hy}{ASASSN-15hy\xspace}
\newcommand{\superc}{03fg-like}
\newcommand{\hyTmax}{2457151.6 $\pm$ 0.4}
\newcommand{\hyBmaxapperent}{15.72 $\pm$ 0.01~mag}
\newcommand{\hyBmaxBeforehost}{$-$19.14 $\pm$ 0.11~mag}
\newcommand{\hyBmaxAfterhost}{$-19.14^{+0.11}_{-0.16}$~mag}
\newcommand{\hyBmVatBmax}{0.18 $\pm$ 0.01~mag}
\newcommand{\hyDM}{0.72 $\pm$ 0.04~mag}

\newcommand{\DmB}{$\Delta{\rm{m}_{15}}(B)$}

\newcommand{\eg}{e.g.,\ }

\newcommand{\Msun}{\ensuremath{M_{\odot}}}
\newcommand{\Zsun}{\ensuremath{Z_{\odot}}}

\newcommand{\kms}{km~s$^{-1}$}

\newcommand{\OI}{O~{\sc i}}

\newcommand{\OIII}{O~{\sc iii}}
\newcommand{\CII}{C~{\sc ii}}
\newcommand{\CI}{C~{\sc i}}

\newcommand{\MgII}{Mg~{\sc ii}}

\newcommand{\SiII}{Si~{\sc ii}}

\newcommand{\CaII}{Ca~{\sc ii}}

\newcommand{\TiII}{Ti~{\sc ii}}

\newcommand{\FeII}{Fe~{\sc ii}}
\newcommand{\FeIII}{Fe~{\sc iii}}
\newcommand{\CoII}{Co~{\sc ii}}

\newcommand{\NiII}{Ni~{\sc ii}}

\newcommand{\Nifs}{$^{56}$Ni}

\newcommand{\ved}{$v_{\rm{edge}}$}

\received{March 29, 2021}
\revised{}
\accepted{July 16,2021}
\submitjournal{ApJ}
\shorttitle{ASASSN-15hy}
\shortauthors{Lu J. et al.}

\begin{document}
\begin{CJK*}{UTF8}{gbsn}

\title{ASASSN-15hy: An  Underluminous, red 03fg-like Type Ia Supernova}

\correspondingauthor{Jing Lu}
\email{jl16x@my.fsu.edu}

\author[0000-0002-3900-1452]{J.~Lu (陆晶)}
\affil{Department of Physics, Florida State University, 77 Chieftan Way, Tallahassee, FL 32306, USA}

\author[0000-0002-5221-7557]{C.~Ashall}
\affil{Institute for Astronomy, University of Hawaii, 2680 Woodlawn Drive, Honolulu, HI 96822,USA}

\author[0000-0003-1039-2928]{E.~Y.~Hsiao}
\affil{Department of Physics, Florida State University, 77 Chieftan Way, Tallahassee, FL 32306, USA}

\author[0000-0002-4338-6586]{P.~Hoeflich}
\affil{Department of Physics, Florida State University, 77 Chieftan Way, Tallahassee, FL 32306, USA}

\author[0000-0002-1296-6887]{L.~Galbany}
\affil{Institute of Space Sciences (ICE, CSIC), Campus UAB, Carrer de Can Magrans, s/n, E-08193 Barcelona, Spain}

\author[0000-0001-5393-1608]{E.~Baron}
\affil{Homer L.~Dodge Department of Physics and Astronomy, University of Oklahoma, Rm 100 440 W. Brooks, Norman, OK 73019-2061, USA}
\affil{Hamburger Sternwarte, Gojenbergsweg 112, D-21029 Hamburg, Germany}

\author[0000-0003-2734-0796]{M.~M.~Phillips}
\affil{Las Campanas Observatory, Carnegie Observatories, Casilla 601, La Serena, Chile}

\author[0000-0001-6293-9062]{C.~Contreras}
\affil{Las Campanas Observatory, Carnegie Observatories, Casilla 601, La Serena, Chile}

\author[0000-0003-4625-6629]{C.~R.~Burns}
\affil{Observatories of the Carnegie Institution for Science, 813 Santa Barbara Street, Pasadena, CA 91101, USA}

\author[0000-0002-8102-181X]{N.~B.~Suntzeff}
\affil{George P. and Cynthia Woods Mitchell Institute for Fundamental
 Physics and Astronomy, Department of Physics and Astronomy, Texas
 A\&M University, College Station, TX 77843, USA}
             
\author[0000-0002-5571-1833]{M.~D.~Stritzinger}
\affil{Department of Physics and Astronomy, Aarhus University, Ny Munkegade 120, DK-8000 Aarhus C, Denmark.}
             
\author[0000-0001-9051-1338]{J.~Anais}
\affil{Las Campanas Observatory, Carnegie Observatories, Casilla 601, La Serena, Chile}

\author[0000-0003-0227-3451]{J.~P.~Anderson}
\affil{European Southern Observatory, Alonso de C\'ordova 3107, Casilla 19, Santiago, Chile}

\author[0000-0001-6272-5507]{P.~J.~Brown}
\affil{George P. and Cynthia Woods Mitchell Institute for Fundamental
 Physics and Astronomy, Department of Physics and Astronomy, Texas
 A\&M University, College Station, TX 77843, USA}
 
\author[0000-0001-9952-0652]{L.~Busta}
\affil{Las Campanas Observatory, Carnegie Observatories, Casilla 601, La Serena, Chile}
 
\author{S.~Castell\'on}
\affil{Las Campanas Observatory, Carnegie Observatories, Casilla 601, La Serena, Chile}
 
\author[0000-0002-2806-5821]{S.~Davis}
\affil{Department of Physics and astronomy, University of California, 1 Shields Avenue, Davis, CA 95616-5270, USA}

\author[0000-0002-0805-1908]{T.~Diamond}
\affil{Department of Physics, Florida State University, 77 Chieftan Way, Tallahassee, FL 32306, USA}

\author[0000-0002-7061-6519]{E.~Falco}
\affil{Harvard-Smithsonian Center for Astrophysics, 60 Garden Street, Cambridge, MA 02138, USA}

\author{C.~Gonzalez}
\affil{Las Campanas Observatory, Carnegie Observatories, Casilla 601, La Serena, Chile}

\author[0000-0001-7981-8320]{M.~Hamuy}
\affil{Vice President and Head of Mission of AURA-O in Chile, Avda Presidente Riesco 5335 Suite 507, Santiago, Chile}
\affil{Hagler Institute for Advanced Studies, Texas A\&M University, College Station, TX 77843, USA}

\author[0000-0002-3415-322X]{S.~Holmbo}
\affil{Department of Physics and Astronomy, Aarhus University, Ny Munkegade 120, DK-8000 Aarhus C, Denmark.}

\author[0000-0001-9206-3460]{T.~W.-S.~Holoien}
\affil{Observatories of the Carnegie Institution for Science, 813 Santa Barbara Street, Pasadena, CA 91101, USA}

\author[0000-0002-6650-694X]{K.~Krisciunas}
\affil{George P. and Cynthia Woods Mitchell Institute for Fundamental Physics and Astronomy, Department of Physics and Astronomy, Texas A\&M University, College Station, TX 77843, USA}
             
\author[0000-0002-1966-3942]{R.~P.~Kirshner}
\affil{Harvard-Smithsonian Center for Astrophysics, 60 Garden Street, Cambridge, MA 02138, USA}
\affil{Gordon and Betty Moore Foundation, 1661 Page Mill Road, Palo Alto, CA 94304, USA}

\author[0000-0001-8367-7591]{S.~Kumar}
\affil{Department of Physics, Florida State University, 77 Chieftan Way, Tallahassee, FL 32306, USA}

\author[0000-0002-1132-1366]{H.~Kuncarayakti}
\affil{Tuorla Observatory, Department of Physics and Astronomy, FI-20014 University of Turku, Finland}
\affil{Finnish Centre for Astronomy with ESO (FINCA), FI-20014 University of Turku, Finland}

\author{G.~H.~Marion}
\affil{University of Texas at Austin, 1 University Station C1400, Austin, TX 78712-0259, USA}

\author[0000-0003-2535-3091]{N.~Morrell}
\affil{Las Campanas Observatory, Carnegie Observatories, Casilla 601, La Serena, Chile}

\author[0000-0003-0554-7083]{S.~E.~Persson}
\affil{Observatories of the Carnegie Institution for Science, 813 Santa Barbara St, Pasadena, CA 91101, USA}

\author[0000-0001-6806-0673]{A.~L.~Piro}
\affil{Observatories of the Carnegie Institution for Science, 813 Santa Barbara St, Pasadena, CA 91101, USA}

\author[0000-0003-0943-0026]{J.~L.~Prieto }
\affil{Nucleo de Astronom\'ia de la Facultad de Ingenier\'ia y Ciencias, Universidad Diego Portales, Av. Ej\'ercito 441, Santiago, Chile}
\affil{Millennium Institute of Astrophysics, Santiago, Chile}

\author[0000-0003-4102-380X]{D.~J.~Sand}
\affil{Steward Observatory, University of Arizona, 933 North Cherry Avenue, Rm.~N204, Tucson, AZ 85721-0065, USA}

\author[0000-0002-9301-5302]{M.~Shahbandeh}
\affil{Department of Physics, Florida State University, 77 Chieftan Way, Tallahassee, FL 32306, USA}

\author[0000-0003-4631-1149]{B.~J.~Shappee}
\affil{Institute for Astronomy, University of Hawaii, 2680 Woodlawn Drive, Honolulu, HI 96822,USA}

\author[0000-0002-2387-6801]{F.~Taddia}
\affil{Department of Physics and Astronomy, Aarhus University, Ny Munkegade 120, DK-8000 Aarhus C, Denmark.}

\begin{abstract}
We present photometric and spectroscopic observations of the \superc\ type Ia supernova (SN~Ia) \hy\ from the ultraviolet (UV) to the near-infrared (NIR). \hy\ shares many of the hallmark characteristics of \superc\ SNe~Ia, previously referred to as ``super-Chandrasekhar'' SNe~Ia. It is bright in the UV and NIR, lacks a clear $i$-band secondary maximum, shows a strong and persistent \CII\ feature, and has a low \SiII\ $\lambda$6355 velocity. However, some of its properties are also extreme among the subgroup. \hy\ is underluminous ($M_{B,\rm{peak}}=$\hyBmaxAfterhost), red ($(B-V)_{Bmax}=0.18^{+0.01}_{-0.03}$~mag), yet slowly declining (\DmB=\hyDM). It has the most delayed onset of the $i$-band maximum of any \superc\ SN. \hy\ lacks the prominent $H$-band break emission feature that is typically present during the first month past maximum in normal SNe~Ia. Such events may be a potential problem for high-redshift SN~Ia cosmology. \hy\ may be explained in the context of an explosion of a degenerate core inside a nondegenerate envelope. The explosion impacting the nondegenerate envelope with a large mass provides additional luminosity and low ejecta velocities. An initial deflagration burning phase is critical in reproducing the low \Nifs\ mass and luminosity, while the large core mass is essential in providing the large diffusion time scales required to produce the broad light curves. The model consists of a rapidly rotating 1.47~\Msun\ degenerate core and a 0.8~\Msun\ nondegenerate envelope. This ``deflagration core-degenerate'' scenario may result from the merger between a white dwarf and the degenerate core of an asymptotic giant branch star.
\end{abstract}

\keywords{supernovae: general -- supernovae: Type Ia supernova  --  supernovae: individual (\hy)}


\section{Introduction}
Type Ia supernovae (SNe~Ia) are one of the most precise extragalactic distance indicators in the cosmos. 
Their high intrinsic luminosity and apparent uniformity enable them to be used in cosmological studies, such as mapping the expansion history of the universe \citep[e.g.,][]{Riess1998,Perlmutter1999}, measuring the Hubble constant, and determining the equation of state of dark energy \citep[e.g.,][]{Riess2007,Sullivan2011,Suzuki2012}. 

SNe~Ia are thought to come from the thermonuclear explosions of at least one carbon-oxygen white dwarf (C/O WD) in a binary system \citep{Hoyle1960}.
The explosive products are mainly dictated by nuclear physics, and, to first order, it is the effectiveness of this burning that determines the observed photometric and spectroscopic properties \citep[\eg][]{Hoeflich96, Mazzali07}.  
These luminous events are powered by the decay of radioactive $^{56}$Ni.
The amount of $^{56}$Ni produced determines the peak luminosity \citep{Arnett82}, while the interplay between $^{56}$Ni, luminosity, and opacity determines the light-curve shape \citep{Mazzali2001}. For normal SNe~Ia, brighter objects produce more $^{56}$Ni and also have broader light curves \citep{Phillips93,Nugent1995,Phillips99}.

However, the ejecta mass and the exact details of the explosions are still debated, and the progenitor systems and explosion mechanisms of SNe~Ia are still not fully understood (e.g., \citealt{Maoz14}). 
Some of the possible progenitor scenarios are single degenerate, double degenerate, triple/quadruple systems, and core degenerate. 
The single-degenerate scenario consists of a C/O WD and a nondegenerate companion star which is either a red-giant, main sequence or He star (\eg \citealt{Whelan73,Nomoto84}), whereas the double degenerate scenario consists of two WDs that merge on dynamical or secular time scales \citep{Iben84,Webbink84}. 
Another scenario consists of collisions of two WDs \citep[\eg][]{Rosswog2009}. Based on secular Lidov-Kozai mechanism, \cite{Thompson2011} suggested that ``something akin to a collision" may occur in a triple system. Going beyond the secular approximations, \cite{Katz2012},\cite{Kushnir2013},\cite{Pejcha2013} found that direct collision between two WDs may result in a thermonuclear explosion in a triple/quadruple system.
Finally, the core-degenerate (CD) scenario describes the merger of a WD and the degenerate core of an asymptotic giant branch (AGB) star within its nondegenerate envelope \citep{Livio03,Kashi11,Ilkov12}.
In each scenario, the mass of the exploding WD may be near, below, or above the Chandrasekhar mass limit ($M_{\text{Ch}}$) depending on the exact configuration of the system or the type of the companion star. Furthermore, the flame front may propagate as a deflagration, detonation or may transition between these two propagation speeds. 

As observations of SNe have accumulated, many peculiar subtypes of SNe~Ia with unique and extreme observational properties have emerged. It is currently unclear which combinations of progenitor systems and explosion mechanisms lead to the observed SN~Ia diversity. One of the most rare peculiar subtypes is the ``super-Chandrasekhar'' or \superc\ SNe.\footnote{In this work we follow the nomenclature of naming a subtype after the first SN of its kind discovered (i.e., \superc), as it is not clear if the ejecta mass of these objects in fact exceeds $M_{\text{Ch}}$.} To date there have only been a handful of well-studied SNe~Ia identified as members of this subtype, 
including: 
SN~2003fg \citep{Howell06},
SN~2006gz \citep{Hicken07,Maeda09}, 
SN~2007if \citep{Scalzo10,Yuan10,Childress11}, 
SN~2009dc \citep{Yamanaka09,Tanaka10,Silverman11,Taubenberger11,Hachinger12},
SN~2012dn \citep{Chakradhari14,Parrent16,Nagao17,Taubenberger19}, 
ASASSN-15pz \citep{Chen19}, and LSQ14fmg \citep{Hsiao2020}.
Furthermore, \citet{Ashall2021} presents data of 5 more 03fg-like SNe.

Generally, \superc\ SNe have high peak luminosities (M$_{B} \leq -19.3$~mag), slow decline rates (\DmB\ $\leq 1.1$~mag), and long rise times (typically $>$ 21~d, \eg \citealt{Silverman11,Taubenberger11}). They also have a primary $i$-band maximum that peaks after that of the $B$ band (\eg \citealt{Ashall20}).
These objects have also been shown to be bright in the ultraviolet (UV) in comparison to normal SNe~Ia \citep{Silverman11,Taubenberger11,Brown14,Chen19}.
However, not all members of this subtype are overluminous. 
For example, SN~2012dn has a relatively fainter peak luminosity (M$_{B} = -19.3$~mag, \citealt{Taubenberger19}) comparing to those SNe~Ia that have similar decline rate, yet it still shares the other similarities of the  \superc\ group mentioned above.

\superc\ SNe exhibit low expansion velocity gradients and have weaker \CaII\ features than normal SNe~Ia at early times (\eg \citealt{Scalzo10,Taubenberger11,Chakradhari14}).
Furthermore, the vast majority of these objects show very strong \CII\ features that persist past maximum light.
Carbon lines, if detected in a normal SN~Ia, are usually only observed in spectra within a few days of the explosion. 
Another peculiar spectroscopic feature of \superc\ SNe is the absence of the $H$-band break near 1.5~$\mu$m within the first month past maximum light \citep{Taubenberger11,Hsiao19}. In normal SN~Ia, the strength and velocity of the $H$-band break is directly linked to the amount and distribution of \Nifs\ in the ejecta \citep{Wheeler98,Hsiao13,Ashall19a,Ashall19b}.

The origins of these \superc\ events are still under debate. Most recently, a class of interaction scenarios has emerged among the possible theoretical interpretations.
Early suggestions of single-degenerate, rapid rotating super-$M_{\text{Ch}}$ WD \citep{Howell06,Hachisu2012} have been challenged since the model simulations could not reproduce the slow expansion velocities on top of the high luminosity regardless of the explosion mechanism, such as pure detonation, pure deflagration and delayed detonation \citep[e.g.,][]{Pfannes2010_A75,Hachinger12,Fink2018}. 
The ``interaction'' \citep[e.g.,][]{Yuan10,Hachinger12,Taubenberger2013,Noebauer2016,Taubenberger19} or ``envelope'' models \citep[e.g.,][]{kmh93,Hoeflich96,Scalzo10,Scalzo14,Hsiao2020}, in which the interaction between the exploding WD and surrounding circumstellar medium (CSM) provides an additional energy source and slows down the expansion, present a promising route to explain these peculiar events.
Recent works have shown photometric evidence of CSM such as the NIR excess in SN~2012dn \citep{Yamanaka2016,Nagao17} and possible wind interaction in LSQ14fmg \citep{Hsiao2020}.
However, the nature of the ``envelope'' is still under debate
\citep[e.g., pre-explosion wind material, the product of double degenerate merger, or the envelope of an AGB star in the core degenerate scenario;][]{Kashi11,Ilkov12,Aznar15} and there is currently no spectral evidence of interaction observed so far.

Extensive follow-up observations of \superc\ SNe are rare. 
The advent of wide-field and untargeted searches created an opportunity for advancement in the studies of \superc\ SNe, as they are generally located in low-mass, low-luminosity environments \citep{Howell06,Childress11,Taubenberger11,Chakradhari14,Hsiao2020}, and therefore may have been missed by previous surveys. 
The \textit{Carnegie Supernova Project}-II (CSP-II; 2011--2015) obtained high-precision, rapid-cadence observations of 
over 100 SNe~Ia \citep{Phillips19,Hsiao19}, mainly from untargeted searches.  
One of the objects observed by the CSP-II was the \superc\ SN~Ia ASASSN-15hy located in a nearby low-luminosity host galaxy.

We present photometric and spectroscopic follow-up observations of ASASSN-15hy, a new \superc\ object sharing the vast majority of the general properties as previous members but also exhibiting extremeness, such as being the faintest, reddest and having the most delayed $i$-band peak time within the subgroup.
The observations and data reductions are summarized in Section~\ref{sec:obs}. 
The host properties and extinction are presented in Section~\ref{sec:host}.
In Sections~\ref{sec:photometry} and \ref{sec:spec}, the photometric and spectroscopic properties of \hy\ are analyzed, respectively.
Model simulation and interpretation are discussed in Sections~\ref{sec:model} and \ref{sec:discussion}.
Finally, we conclude in Section~\ref{sec:conclusion}.


\section{Observations} \label{sec:obs}

\begin{figure}[t!]
\centering
\includegraphics[width=0.98\columnwidth]{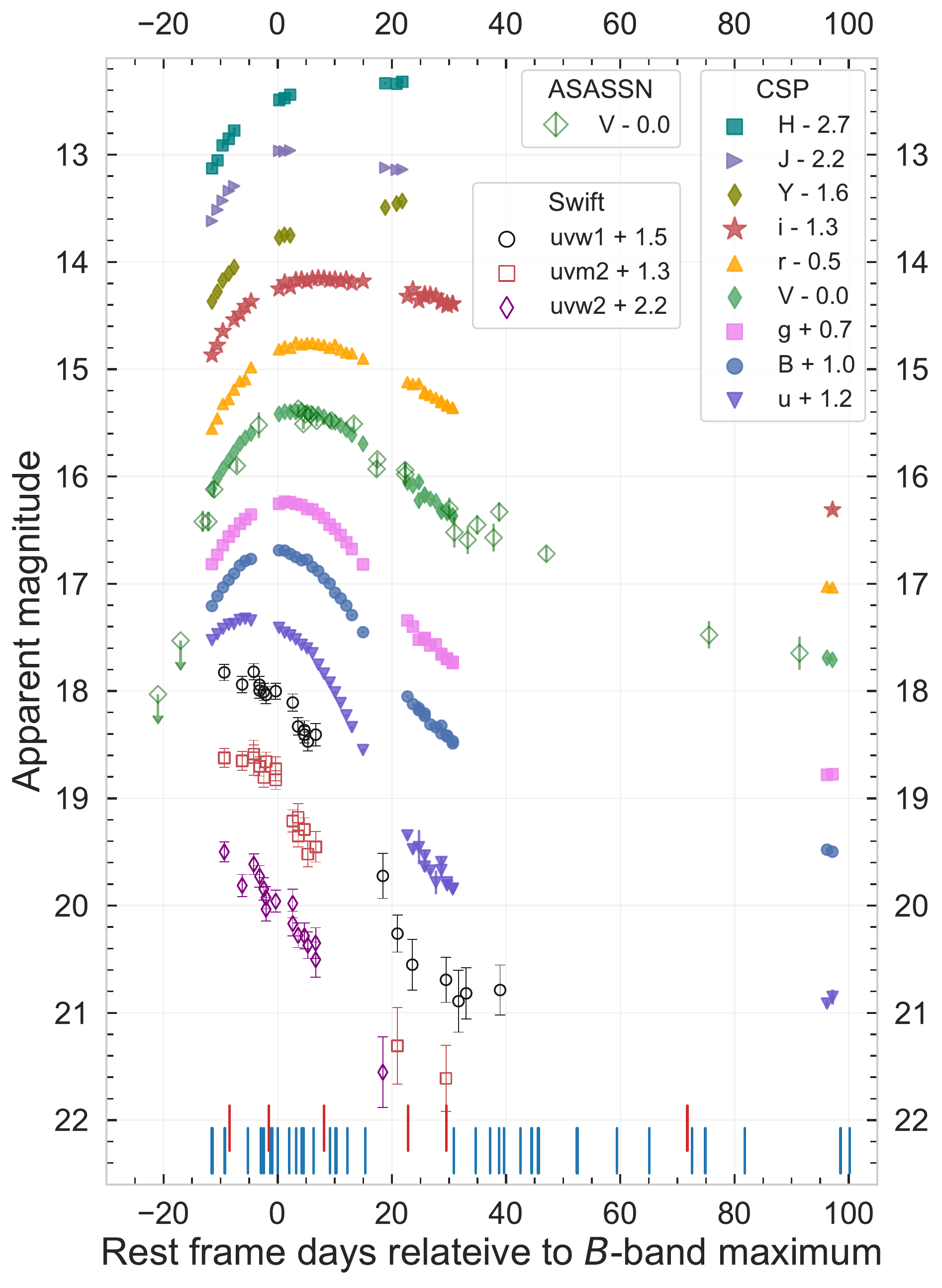}
\caption{The observed UV, optical, and NIR light curves of ASASSN-15hy. The solid markers represent CSP-II photometry. 
$Swift$ UVOT UV and ASAS-SN $V$-band photometry (including detection limits) are plotted with open markers.
The blue vertical lines on the bottom of the figure indicate the epochs of optical spectral observations, whereas the red ones indicate those of NIR spectral observations. 
}
\label{fig:multiband}
\end{figure}

\begin{table}[b!]
\renewcommand{\thetable}{\arabic{table}}
\centering
\setlength{\tabcolsep}{17pt}
\caption{Properties of ASASSN-15hy.\tablenotemark{a}} \label{table:prop}
\begin{tabular}{ll}
\hline
\hline
\textbf{ASASSN-15hy} &  							\\
A.k.a.            	&	PS15aou					\\
$\alpha$ (J2000)	&	$20^{h}10^{m}02^{s}.35$		\\
$\delta$ (J2000)  	&	$-00^{\circ}44^{\prime}21^{\prime\prime}.31$		\\
$JD_{\text{max}}(B)$		&	\hyTmax 	\\
$JD_{\text{max}}(V)$		&	2457155.6 $\pm$ 0.7 	\\
$JD_{\text{max}}(r)$		&	2457157.1 $\pm$ 0.5 	\\
$JD_{\text{max}}(i)$		&	2457158.9 $\pm$ 1.1 	\\
$\Delta m_{15}(B)$	&	0.72 $\pm$ 0.04~mag			\\
$\Delta m_{15}(V)$	&	0.44 $\pm$ 0.03~mag		\\
$\Delta m_{15}(r)$	&	0.28 $\pm$ 0.02~mag			\\
$\Delta m_{15}(i)$	&	0.12 $\pm$ 0.02~mag			\\
$m_{B,\text{max}}$\tablenotemark{b}			&	15.72 $\pm$ 0.01~mag			\\
$m_{V,\text{max}}$\tablenotemark{b}			&	15.38 $\pm$ 0.01~mag			\\
$m_{r,\text{max}}$\tablenotemark{b}			&	15.30 $\pm$ 0.01~mag			\\
$m_{i,\text{max}}$\tablenotemark{b}			&	15.50 $\pm$ 0.01~mag			\\
$M_{B,\text{max}}$\tablenotemark{c}			&	$-19.14^{+0.11}_{-0.16}$~mag			\\
$M_{V,\text{max}}$\tablenotemark{c}			&	$-19.36^{+0.11}_{-0.14}$~mag			\\
$M_{r,\text{max}}$\tablenotemark{c}			&	$-19.38^{+0.11}_{-0.13}$~mag			\\
$M_{i,\text{max}}$\tablenotemark{c}			&	$-19.10^{+0.11}_{-0.13}$~mag			\\
$s_{BV}$        	&	1.24~$\pm$~0.18			\\
$E(B-V)_{\text{MW}}$		&	0.13~mag	   \\				
\hline
\textbf{Host}      & 	 						\\ 
Heliocentric redshift	&	0.0185	$\pm$ 0.0003			\\
CMB redshift        	&	0.0176	$\pm$ 0.0003			\\
Distance modulus\tablenotemark{d}     	&	34.33 $\pm$ 0.11~mag			\\
$E(B-V)_{\text{host}}$		&	$0.00^{+0.03}$~mag			\\
Stellar mass M$_{\star}$       &   (0.77 $\pm$ 0.16) $\times$ 10$^9$ M$_\odot$   \\
log$_{10}$(sSFR)    &   $-$10.21 $\pm$ 0.31 yr$^{-1}$               \\
\hline
\hline
\end{tabular}
\begin{flushleft}
\tablenotetext{a}{All photometric parameters are in the CSP-II natural system.}
\tablenotetext{b}{Peak apparent magnitude without any corrections.}
\tablenotetext{c}{Peak absolute magnitude with $K$-correction, Milky Way and host-galaxy extinction correction.}
\tablenotetext{d}{Obtained from the CMB redshift, including the uncertainty due to peculiar velocity dispersion (assuming 300 \kms). The cosmological parameters assumed for this work are $H_0=73$~\kms\ Mpc $^{-1}$, $\Omega_m=0.27$ and $\Omega_\Lambda=0.73$. }
\end{flushleft}
\end{table}

\hy\ (aka PS15aou) was discovered by the All Sky Automated Survey for SuperNovae (ASAS-SN, \citealt{Shappee14,Kochanek2017}) in an image taken with the double ``Cassius'' telescope on 2015 April 25.21 UT at $V$ $\sim$16.4 mag and was confirmed two days later in another image taken with the Las Cumbres Observatory 1-m robotic telescope in Sutherland, South Africa \citep{Holoien15}.
It was spectroscopically classified by the Public ESO Spectroscopic Survey for Transient Objects \citep[PESSTO;][]{Smartt15} as a young peculiar SN~Ia with similarities to SN~2003fg and SN~2006gz on 2015 April 27.36 UT \citep{Frohmaier15}. 
On the same day as the classification, at $\sim-$11.5~d  relative to the rest-frame $B$-band maximum, the CSP-II started an optical and near-infrared (NIR) follow-up campaign using the 1-m Swope, 2.5-m du Pont, and the 6.5-m Baade telescopes, located at the Las Campanas Observatory (LCO) in Chile.  
In this paper, we present complementary multiwavelength photometric and spectroscopic observations of \hy\ from the CSP-II, ASAS-SN, $Swift$, and Gemini.
\hy\ is only second \superc\ object with a NIR spectroscopic time series and has a multiband light-curve data set from UV to NIR. Note that the photometric observation gap between +31~d and +96~d are due to the lack of available observation time of the CSP-II during the Chilean winter.
The photometry and observation logs are listed in Appendix~\ref{appendix:supplement}.

\subsection{Photometry} \label{subsec:obs-photometric}

The ASAS-SN observations of \hy\ consist of $V$-band photometry, including 3$\sigma$ nondetection limits.
The ASAS-SN images were reduced with an automated pipeline that uses the ISIS package \citep{Alard1998,Alard2000} for image subtraction, the IRAF\footnote{IRAF was distributed by the National Optical Astronomy Observatory, which was managed by the Association of Universities for Research in Astronomy (AURA) under a cooperative agreement with the National Science Foundation.} $apphot$ package for performing aperture photometry, and the AAVSO Photometric All-sky Survey (APASS, \citealt{Henden15}) for calibration. 

CSP-II optical photometric observations in the $uBgVri$ bands of \hy\ were obtained with the e2v 4112 $\times$ 4096 pixel CCD imager attached to the  Swope 1-m telescope at LCO.  
The NIR $YJH$ images were taken using RetroCam on the LCO 2.5-m du Pont telescope. 
RetroCam employs a single chip Rockwell 1024~$\times$~1024 HAWAII-1 HgCdTe detector. 

The CSP-II photometric observations were reduced following the procedure described in \citet{Phillips19}.
Both optical and NIR images of \hy\ were background subtracted, with optical and NIR host-galaxy template images obtained at the Swope and Baade telescopes respectively, more than +300~d past maximum light.
Point-spread-function photometry was used to compute the SN magnitude with respect to local-sequence stars, which were calibrated with standard star fields.
An $i$-band finder chart, with labeled local-sequence stars, is shown in Fig.~\ref{fig:snimage}.

Finally, UV $uvw2$, $uvm2$, $uvw1$, and optical $ubv$ photometry was obtained (PI:~Holoien) with the Ultraviolet Optical Telescope (UVOT) onboard the $Swift$ spacecraft \citep{Gehrels04,Roming05}.
The photometric reduction follows the same basic manner outlined by \citet{Brown09,Brown14}.

Fig.~\ref{fig:multiband} presents the observed light curves of \hy. 
All photometry mentioned above is tabulated in Appendix~\ref{appendix:supplement}. 
The basic observational parameters of \hy\ are summarized in Table~\ref{table:prop}. 
As much as possible, photometry in the $BVri$ bands are placed via $S$-correction in the most recent CSP-II natural system \citep{Phillips19}, referred to as the ``CSP natural system'' in the subsequent text.

\subsection{Spectroscopy}
Spectroscopic observations of \hy\ were obtained using multiple telescopes and instruments.
There are 34 optical spectra (including 22 previously published) from $-$11.6~d to +155.2~d and 6 NIR spectra from $-$8.5~d to +71.7~d relative to rest-frame $B$-band maximum, respectively.
A journal of the spectroscopic observations and the details of the published spectrum are given in Appendix~\ref{appendix:supplement}.

\begin{figure}[htb]
\centering
\includegraphics[width=0.99\columnwidth]{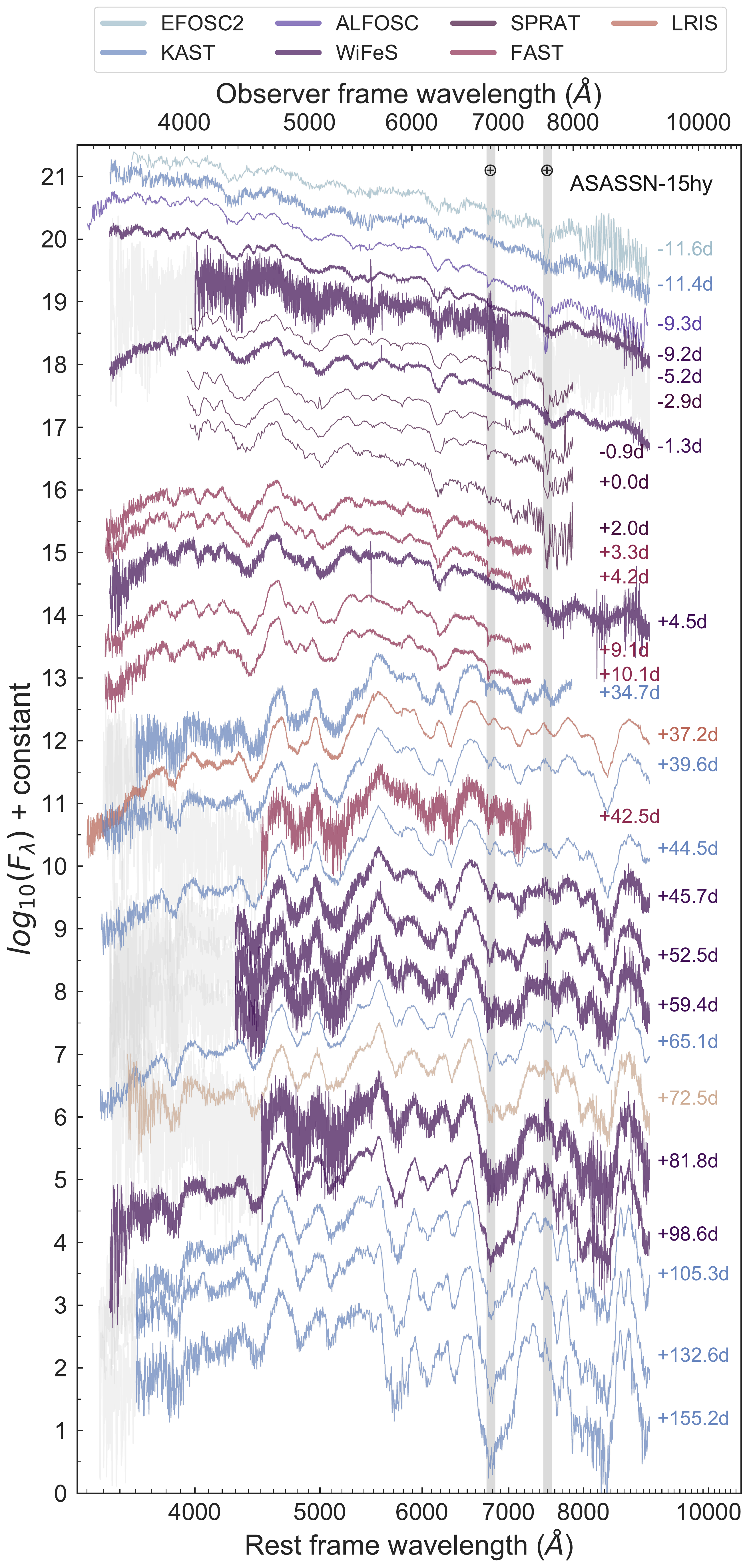}
\caption{Optical spectroscopic time series of \hy\ of a selected sample. 
Regions of strong telluric absorptions are marked with vertical gray bands.
The colors represent various telescopes and instruments. Note that some edges of the spectra are grayed out for presentation purposes due to the high level of noise. }
\label{fig:spec_present_opt}
\end{figure}

\begin{figure}[t]
\centering
\includegraphics[width=0.9\columnwidth]{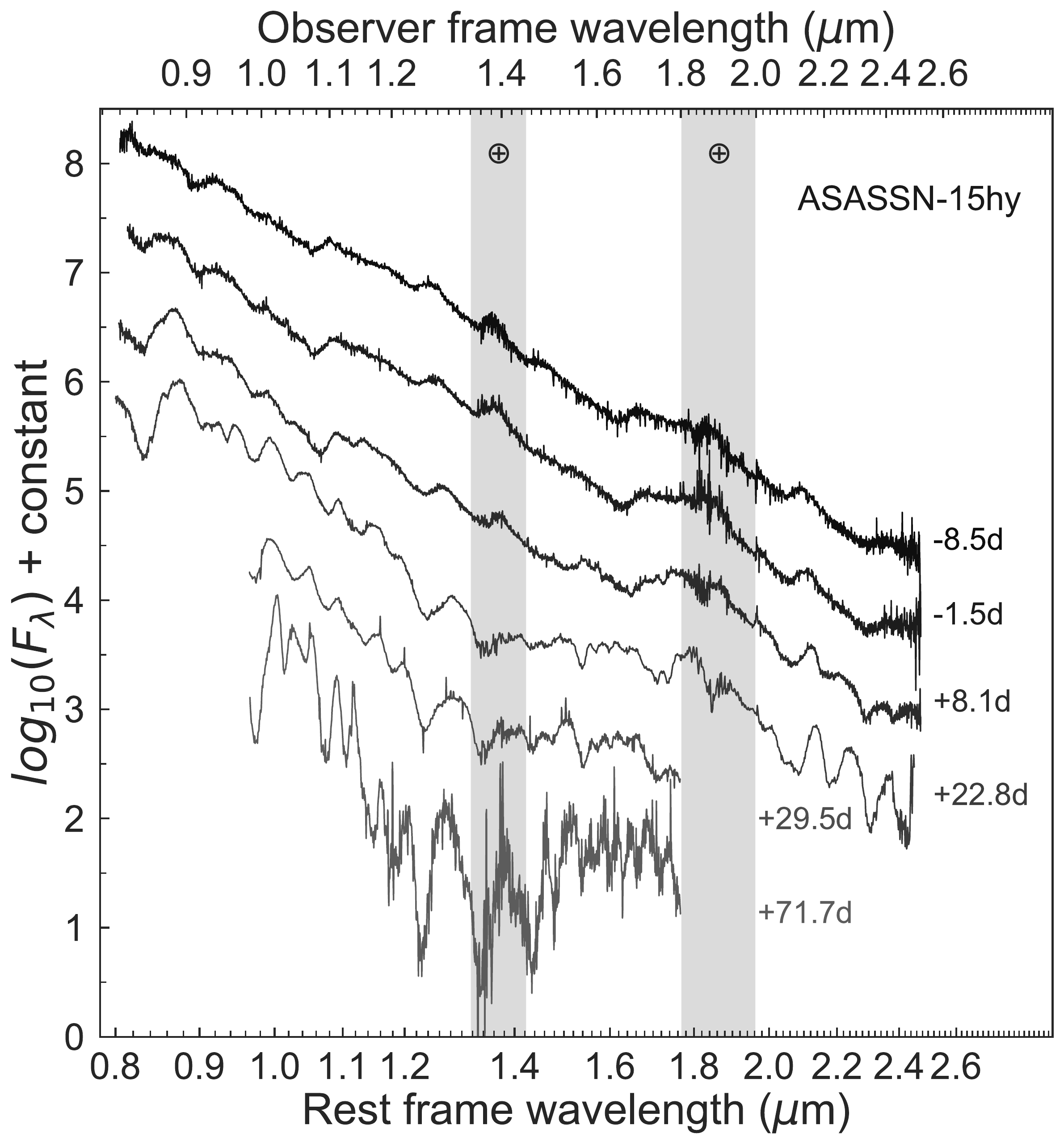}
\caption{NIR spectroscopic time series of \hy.
The strongest telluric absorptions are marked with vertical gray bands.}
\label{fig:spec_present_NIR}
\end{figure}

This work contribute twelve optical spectroscopic observations of \hy. 
Five of them were obtained with the FAST spectrograph \citep{Fabricant1998} on the 1.5-m Tillinghast telescope at the F. L. Whipple Observatory (FLWO 1.5~m) and were reduced using a combination of custom IDL and standard IRAF procedures \citep{Matheson2005}.
Four of the optical spectra were taken with the SPRAT spectrograph  on the 2-m Liverpool Telescope (LT, \citealt{Steele2004}) and reduced using the LT pipeline \citep{Piascik2014}.
Two spectra obtained with ALFOSC on the Nordic Optical 2.5-m Telescope (NOT) and one spectrum taken with WFCCD mounted on the 2.5-m du Pont telescope were all reduced in the standard manner using \textsc{IRAF} scripts.
The optical spectroscopic observations of \hy\ are shown in Fig.~\ref{fig:spec_present_opt}.

The NIR spectral observations were obtained using GNIRS \citep{Elias98} on the 8.2-m Gemini North Telescope (GN), FLAMINGOS-2 (F2, \citealt{Eikenberry08}) on the 8.2-m Gemini South Telescope (GS), and FIRE \citep{Simcoe13} on the 6.5-m Baade Telescope.
The observing techniques and reduction procedures of these three instruments are described in \citet{Hsiao19}. 
The NIR spectra of \hy\ are presented in Fig.~\ref{fig:spec_present_NIR}.

\subsection{Host Observations}
Integral-field spectroscopy of the host galaxy of \hy\ was obtained when the SN had faded, on 2017 July 19 UT,  with the Multi-Unit Spectroscopic Explorer (MUSE, \citealt{2010SPIE.7735E..08B}), mounted on the Unit 4 telescope at the ESO Very Large Telescope (VLT UT4) of the Cerro Paranal Observatory.
These host observations were obtained as part of the All-weather MUSE Supernova Integral-field Nearby Galaxies (AMUSING, \citealt{2016MNRAS.455.4087G}) survey, an ongoing project studying the environments of SNe by means of the analysis of a large number of nearby SN host galaxies.

MUSE provides a wide field-of-view of approximately 1$^{\prime}\times$1$^{\prime}$ and a square spatial pixel size of 0.2\arcsec$\times$0.2\arcsec.
This limits the spatial resolution of the data to the atmospheric seeing during the observations, which was 1.6\arcsec\ while observing the host galaxy of ASASSN-15hy.
The spectroscopy covers a wavelength range from 4750 to 9300~\AA, with a mean spectral resolution $\lambda$/$\Delta\lambda$ $\sim$3000 and a spectral sampling of 1.25~\AA.

\section{Host Properties} \label{sec:host}
The host-galaxy properties of \superc\ SNe can be extreme, and as a group, distinct from those of normal SNe~Ia. Detailed studies of these galaxies are therefore likely pertinent to understanding the origin of \superc\ SNe.
With the integral-field spectroscopy taken by MUSE, we conducted a similar analysis to that presented by \citet{Hsiao2020} in order to obtain several measurements of galaxy properties, both at the location of ASASSN-15hy and for the galaxy as a whole.

\begin{figure*}[t!]
\centering
\includegraphics[width=0.8\textwidth]{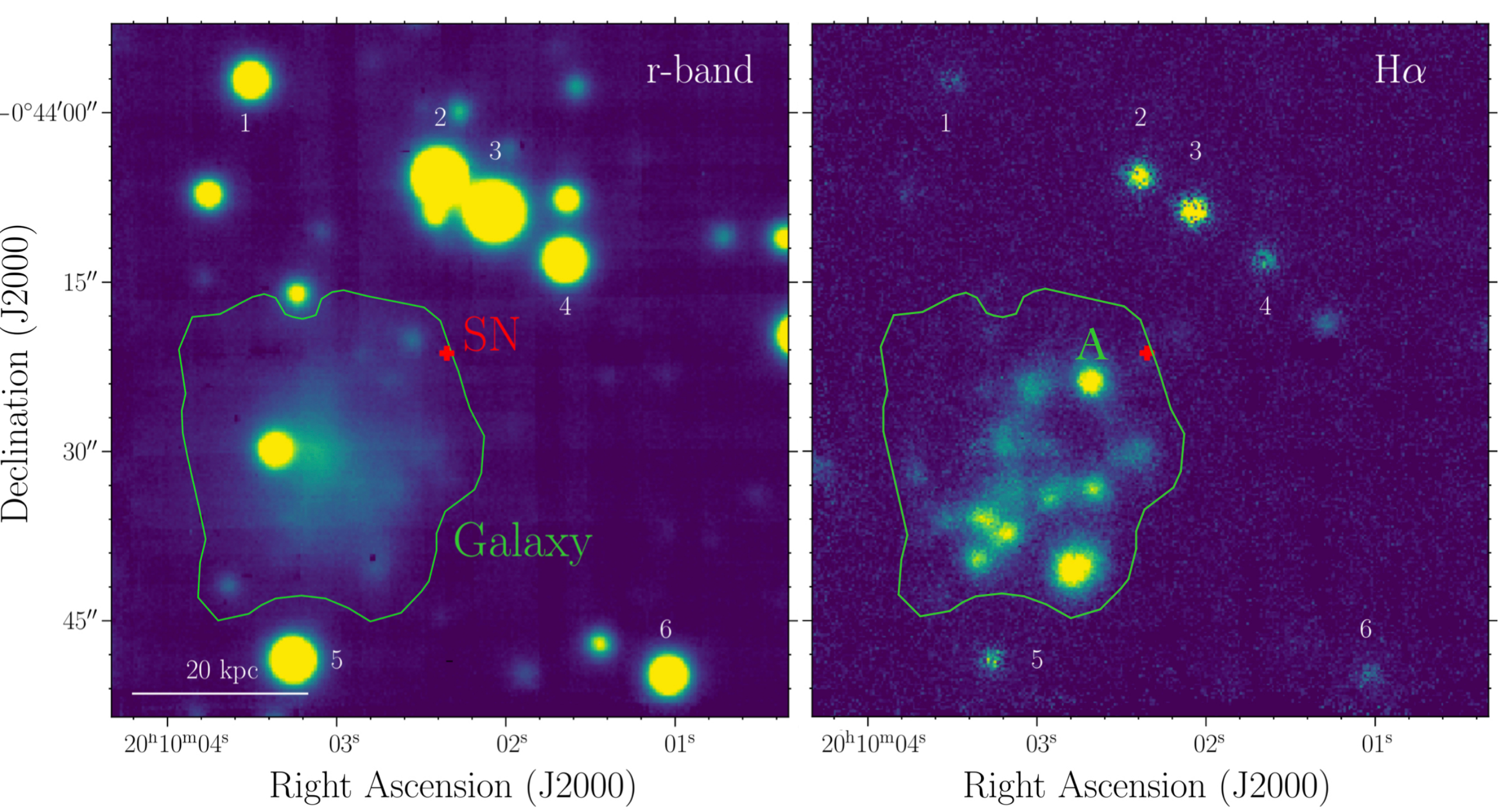}
\caption{Host-galaxy images of ASASSN-15hy extracted from the MUSE datacube. 
The left panel corresponds to the synthetic CSP $r$-band image, and the right panel is the extinction-corrected and stellar population subtracted gas-phase emission line map of H$\alpha$. 
The SN position is marked with a red cross in both panels, and the host galaxy is defined by a green contour (the bright point source inside the contour in the left panel is a foreground star). The numbers from 1 to 6 mark the field stars that show residuals in the H$\alpha$ map. Letter A in the right panel marks the position of the host structure where analysis is performed (see Appendix~\ref{appendix:host}).
\label{fig:host_white}}
\end{figure*}

\subsection{Host Redshift} \label{subsec:host_z}
In most Swope images of \hy, the SN appears hostless, which is remarkable given that the estimated redshift based on the SN spectrum is $z=0.025$, which is relatively nearby \citep{Frohmaier15}.
With the deeper MUSE data, a faint and extended source is apparent in the synthetic $r$-band image (left panel of Fig.~\ref{fig:host_white}), produced by convolving the transmission function of the CSP $r$-band filter with the MUSE datacube.
The extended source observed to the south east (SE) of the SN is presumed to be the host of \hy.
We manually defined a contour (marked with green in Fig.~\ref{fig:host_white}) that includes the extended source but excludes a foreground star.
Note that another bright point source within the contour in the synthetic $r$-band image is a foreground star and is excluded in the analysis.
The extracted spectrum within this contour revealed the typical emission lines of a star-forming galaxy.

A heliocentric redshift for the host galaxy of $z_{\text{helio}}$ = 0.0185 $\pm$ 0.0003 was derived by measuring the positions of the strongest emission lines, such as H$\alpha$, [\OIII]~5007~\AA, and H$\beta$.
Correcting to the cosmic microwave background (CMB) rest frame yielded $z_{\text{CMB}}$ = 0.0176 $\pm$ 0.0003, which translates to a distance modulus ($\mu$) of 34.33 $\pm$ 0.11~mag assuming $H_0=73$~\kms\ Mpc $^{-1}$, $\Omega_m=0.27$ and $\Omega_\Lambda=0.73$ for the cosmological parameters.
Note that the quoted uncertainty in $\mu$ includes an assumed peculiar velocity dispersion of 300 \kms.

\subsection{Host Environment} \label{subsec:host_env}
\label{subsec:host}
To study the host properties of \hy, the MUSE data were analyzed using methods similar to those of \cite{2014A&A...572A..38G,2016A&A...591A..48G} and \cite{Hsiao2020}. 
The right panel of Fig.~\ref{fig:host_white} shows the extinction-corrected H$\alpha$ emission map of the host and reveals structures across the galaxy that are not seen in the synthetic $r$-band image shown in the left panel of Fig.~\ref{fig:host_white}.
Spectra were extracted at several locations, and simple stellar population analysis was performed in order to obtain several measurements characterizing the host-galaxy properties.
These measurements are listed in Table~\ref{table:host_table}.

The host of \hy\ has a low stellar mass and extremely low metallicity indicated by the subsolar oxygen abundance. It contains a remarkably young stellar population component based on the high H$\alpha$ equivalent width.
Furthermore, this low-mass galaxy is very efficient in producing new stars, as shown by the high specific star formation rate (sSFR).
The details of the host environment analysis and the results are presented in Appendix~\ref{appendix:host}.

In the context of \superc\ SNe, the properties of the host environment of \hy\ are largely consistent with other group members. 
The MUSE observation of the LSQ14fmg host shows similar properties, such as low metallicity, low stellar mass, high sSFR, and a relatively young stellar population component \citep{Hsiao2020}.
The host of SN~2007if was also shown to be a low-stellar-mass and metal-poor galaxy \citep{Scalzo10,Childress11}.
The host environment of \superc\ objects indicates that they are likely to originate from low-metallicity and young progenitors \citep{Childress11,Khan2011,Silverman11,Taubenberger11}. However, the current small sample and the lack of a complete and consistent theoretical scenario hinder our ability to disentangle these effects and identify the dominant environmental driver for this class of objects.

\begin{table*}[bt!]
\tabletypesize{\small}
\centering
\caption{Properties of ASASSN-15hy and the host environment from the MUSE data.\label{table:host_table}}
\hspace{-1.8cm}\begin{tabular}{ccccccc}
\hline \hline
Location        & H$\alpha$EW        & \multicolumn{2}{c}{12 + $log_{10}(O/H)$}   	        & Stellar mass			                        & SFR                              & log$_{10}$(sSFR)	 \\ 
	            & (\AA)              & O3N2 (dex)               & D16 (dex)          	    & (log$_{10}$(M$_{\star}$/M$_\odot$)) 	        & (M$_\odot$ yr$^{-1}$)            & (yr$^{-1}$)		 \\ \hline
SN              & 30.48$\pm$0.89     & 8.43~$\pm$~0.11 & 8.04~$\pm$~0.35  & 6.77~$\pm$~0.10 				        & 0.20$\pm$0.09$\times$10$^{-3}$   & $-$10.47~$\pm$~0.61	\\
Structure A		& 75.30$\pm$0.48     & 8.21$\pm$0.14            & 7.84$\pm$0.15      	    & $\dots$ 				                        & 1.21$\pm$0.28$\times$10$^{-3}$   & $\dots$			 \\
Global          & 33.25$\pm$0.28     & 8.34$\pm$0.14            & 7.87$\pm$0.16     		& 8.89~$\pm$~0.09		                        & 0.048$\pm$0.007			   	   & $-$10.21~$\pm$~0.31 \\ \hline
\end{tabular}
\end{table*}

\subsection{Host-galaxy Extinction} \label{subsec:extinction}
Obtaining an accurate value of the host-galaxy extinction of \superc\ SNe is difficult. As described by \citet{Scalzo12}, the Lira relation \citep{Lira1996,Phillips99} does not hold for these objects. 
Nevertheless, \citet{Chen19} derived a color excess for SN~2009dc by utilizing the Lira method and assuming that ASASSN-15pz has zero host-galaxy reddening.
This strategy relies on the assumption that the difference in the late-time $B-V$ color is not intrinsic to the SN~Ia.
In this work, the relation between $E(B-V)$ and the Na~{\sc i}~D  equivalent width (EW) from \citet{Poznanski12} and \citet{Phillips2013} was adopted to estimate the host reddening and extinction of \hy.

Each optical spectrum of \hy\ was inspected for a narrow Na~{\sc i}~D absorption at the redshift of the host galaxy, and no absorption features were found. 
A detection limit of EW $\le$~0.1~\AA\ was estimated using the WiFeS spectrum taken on 2015 May 13, selected for its relatively high resolution and signal-to-noise ratio (S/N).
The detail of the methodology is described in Appendix~\ref{appendix:NaID}.
The Na~{\sc i}~D absorption from the Milky Way (MW) was measured to have an EW of 0.82 $\pm$ 0.25~\AA\ in the same spectrum, corresponding to a MW color excess of $E(B-V)_{\text{MW}}=0.13 \pm 0.09$~mag based on the relation presented in \citet{Poznanski12}.
This is consistent with the value obtained from the Galactic reddening map of \citet{Schlafly11}.

From the nondetection of the host-galaxy Na~{\sc i}~D  lines, we assume zero extinction throughout the paper. 
The extinction estimate from the Balmer decrement of the MUSE observation at the location of the SN is also consistent with zero.  
An extinction uncertainty $\delta A_V$ =  $0.08$~mag is estimated based on the MW Na~{\sc i}~D and extinction measurements compiled by \citet{Phillips2013} and references therein, using the Na~{\sc i}~D EW detection limit of 0.1~\AA.
This translates to an uncertainty of 0.03~mag for the color excess.
Hence, throughout this work, a host-galaxy color excess of $E(B-V)_{\text{host}}=0.00^{+0.03}$~mag is adopted for \hy.

\hy\ is dim in the optical and red in $B-V$ near maximum in comparison to the \superc\ group (see Section~\ref{sec:photometry}) and begs the question whether these properties are intrinsic to the SN or due to host-galaxy reddening.
Comparisons of the spectral energy distributions (SEDs) of \hy\ with those of well-observed \superc\ SNe~Ia show that these properties are likely to be intrinsic to the SN.
To match the optical flux of \hy\ to those of ASASSN-15pz or SN~2009dc by invoking host-galaxy extinction requires an enormous intrinsic UV flux from \hy\ that is $2-5$ times higher than the UV bright SN~2009dc and ASASSN-15pz.
The nondetection of Na~{\sc i}~D reaffirms that the SED differences are dominated by differences in the intrinsic properties.
Details of the SED comparisons are presented in Appendix~\ref{appendix:SED}.

\section{Photometric Properties} \label{sec:photometry}
The observed light curves of \hy\ are presented in Fig.~\ref{fig:multiband}.
In this section, the photometric properties of \hy\ are analyzed by comparing light-curve parameters, such as the $B$-band decline rate\footnote{$\Delta{\rm{m}_{15}}(B)$ is the change in magnitude between maximum light and +15~d  past maximum in a certain band \citep{Phillips93}.} $\Delta{\rm{m}_{15}}(B)$ and the color stretch\footnote{$s_{BV}$ is the difference in time between $B$-band maximum and the reddest point in the $B-V$ color curve, normalized by 30~days \citep{Burns14}.} $s_{BV}$, as well as the shapes of the light and color curves relative to other SNe~Ia.
The basic observational parameters of \hy\ are summarized in Table~\ref{table:prop}.

Six of seven published \superc\ objects so far: SN~2006gz \citep{Hicken07}, SN~2007if \citep{Scalzo10,Friedman15,Krisciunas17}, SN~2009dc \citep{Taubenberger11,Hicken2012,Friedman15,Krisciunas17}, SN~2012dn \citep{Chakradhari14,Yamanaka2016,Taubenberger19}, ASASSN-15pz \citep{Chen19}, and LSQ14fmg \citep{Hsiao2020}, are included for comparison in this section. 
Note that SN~2003fg \citep{Howell06} is not included in the comparison due to the lack of $B$- and $V$-band light curves.
A well observed normal SN~Ia, 2007af \citep{Krisciunas17}, is included for comparison in this section.
All photometry is $K$-corrected (and S-corrected if the filter function is available) following the manner described in Appendix~\ref{appendix:kcorr}.

These SNe were selected for comparison to \hy\ as they allow for the full diversity of properties to be explored, and other subclasses (such as 2002cx-like, 1991bg-like) are not included here due to highly distinct differences with \superc\ objects (see Fig. 1 of \citealt{Taubenberger17}).

\begin{figure}[t!]
\centering
\hspace*{-0.2cm}\includegraphics[width=0.49\textwidth]{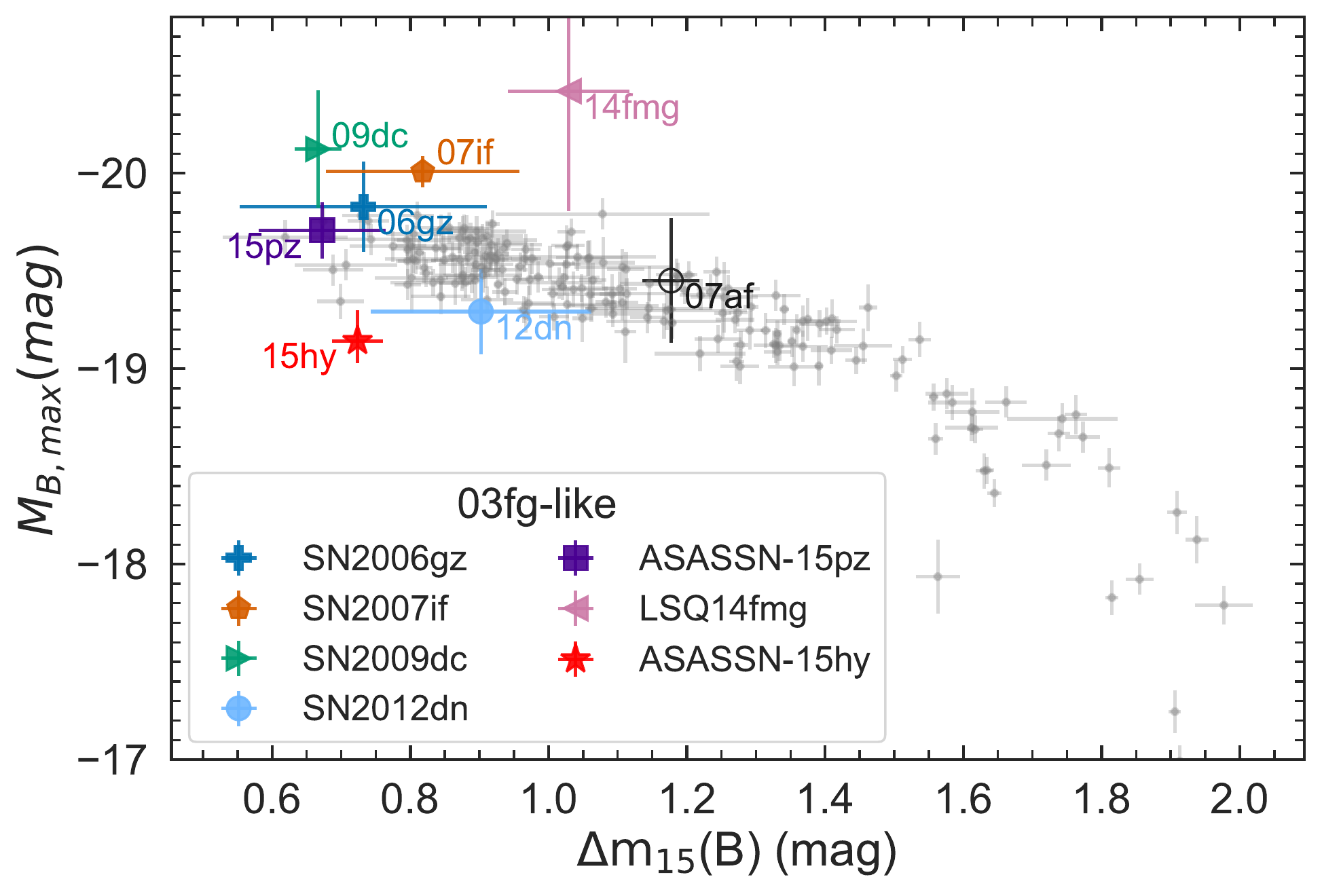}
\caption{The luminosity width relation (Phillips relation) for normal SNe~Ia from the CSP-I \citep{Krisciunas17} is demonstrated by gray markers.
Published \superc\ objects (see references in text) are plotted in colored markers.
All absolute peak magnitudes are corrected for MW extinction and host-galaxy extinction.
}
\label{fig:LWR}
\end{figure}

\subsection{Luminosity Width Relation\label{subsec:LWR}}
SNe~Ia follow a luminosity width relation (LWR), where SNe~Ia with broader light curves are intrinsically brighter, and SNe with narrower light curves are dimmer \citep{Pskovskii1977,Phillips93,Phillips99}.
This relation serves as a cornerstone of modern SN~Ia cosmology.
To determine where \hy\ and other \superc\ SNe are located on the luminosity-width diagram, the $K$-corrected $B$-band light curves were used to directly measure the time of maximum, decline rate $\Delta{\rm{m}_{15}}$(B), and the apparent peak magnitude $m_{B,\rm{max}}$.
These measurements were obtained through a Gaussian process interpolation included in the \textit{SuperNovae in object-oriented Python} (SNooPy; \citealt{Burns11,Burns14}) package. 
Finally, MW and host-galaxy extinction corrections and distance modulus were applied to the $m_{B,\rm{max}}$ to obtain the absolute peak magnitude $M_{B,\rm{max}}$, hereafter $M_{B}$ for short.

\hy\ reached $B$-band maximum at $JD_{\rm{max}}(B)$ = \hyTmax, has a \DmB\ of \hyDM, and a $m_{B,\rm{max}}$ of \hyBmaxapperent.
As suggested in Section~\ref{subsec:extinction}, there is no evidence of \hy\ suffering from substantial host-galaxy extinction.
Hence, only a MW color excess $E(B-V)_{\text{MW}}=0.13$~mag was used to correct the photometry of \hy, which translates to a $B$-band extinction of A$_{B}$ = 0.53~mag based on the extinction law from \citet{Cardelli1989} assuming R$_{V}$ = 3.1. 
Converting to absolute magnitude, using the A$_{B}$ value above, and a $\mu$ of 34.33 $\pm$ 0.11~mag from the host-galaxy redshift in the CMB frame gives $M_{B}=$~\hyBmaxBeforehost. 
This corresponds to a $M_{B}$ of \hyBmaxAfterhost\ if including the uncertainty of the host extinction.

\begin{table}[t!]
\tabletypesize{\small}
\addtolength{\tabcolsep}{-0.8pt}
\caption{Distance modulus and color excess values adopted for comparison sample of \superc\ SNe~Ia. \label{table:SuperC_DM_EBV_stats}}
\resizebox{\columnwidth}{!}{%
\centering
\hspace*{-1.5cm}\begin{tabular}{ccccc}
\hline\hline
SN          & $z_{\text{CMB}}$   & $\mu$\tablenotemark{a} & $E(B-V)$\tablenotemark{c} & $E(B-V)$\tablenotemark{d}\\ 
            &           & (mag)                  & (MW, mag)                          & (host, mag)\\ \hline
2006gz   & 0.023     & 34.95$\pm$0.09       & 0.05          & 0.18$\pm$0.05   \\
2007if   & 0.073     & 37.51$\pm$0.03       & 0.07          & 0.0               \\
2009dc   & 0.022     & 34.79$\pm$0.09       & 0.06          & 0.10$\pm$0.07    \\
2012dn   & 0.009     & 33.28$\pm$0.21\tablenotemark{b} & 0.05          & 0.04$\pm$0.01\\
ASASSN-15pz & 0.014     & 33.89$\pm$0.14       & 0.01          & 0.0\\
LSQ14fmg    & 0.065     & 37.24$\pm$0.03       & 0.05          & 0.13$\pm$0.15\\ 
ASASSN-15hy & 0.0176    & 34.33$\pm$0.11       & 0.13          & $0.00 \pm 0.03$ \\ \hline
\end{tabular}%
}
\tablenotetext{a}{Obtained from the CMB redshift with the same cosmological parameters assumed for \hy\ (see Section~\ref{subsec:host_z}
).}
\vspace*{-0.1cm}
\tablenotetext{b}{Corrected for infall towards the Virgo cluster and the Great Attractor (recession velocity = $3{,}306$~\kms, \citealt{Mould2000}).}
\vspace*{-0.1cm}
\tablenotetext{c}{Obtained from \citet{Schlafly11}.}
\vspace*{-0.1cm}
\tablenotetext{d}{Adopted from \cite{Hicken07}, \cite{Scalzo10}, \cite{Taubenberger11}, \cite{Taubenberger19}, \cite{Chen19},  \cite{Hsiao2020} and this work respectively.}
\end{table}

Figure~\ref{fig:LWR} demonstrates where \hy\ and other \superc\ events are located on the luminosity-width diagram. 
\hy\ is intrinsically the least luminous \superc\ SN and is located below the LWR for normal SNe~Ia; whereas SN~2006gz, SN~2012dn, and ASASSN-15pz blend into the normal population; and SN~2007if, SN~2009dc, and LSQ14fmg are all overluminous compared to normal SNe~Ia.
The corresponding adopted $\mu$ and reddening values of the \superc\ SNe are tabulated in Table~\ref{table:SuperC_DM_EBV_stats}.
A $\mu$ of 32.09 $\pm$ 0.32~mag and $E(B-V)_{\text{host}} = 0.19~\pm~0.01$~mag (obtained from $SNooPy$) are adopted for the normal SN Ia 2007af. 

\hy\ is almost one magnitude dimmer than SN~2009dc despite the fact that they have very similar \DmB\ values, NIR luminosities (see Section~\ref{subsec:LC}), and spectral properties (see Section~\ref{sec:spec}).
If \hy\ followed the LWR, for its absolute $M_{B}$ magnitude of $\sim -$19.14~mag, it would be expected to have a \DmB\ of $\sim1.4$~mag, demonstrating the uniqueness of this event. 
Cosmology light-curve fitters are liable to give incorrect results for SNe similar to ASASSN-15hy because of the red intrinsic color at maximum (see Section~\ref{subsec:color_curve}).

\begin{figure*}[ht!]
\centering
\subfigure[Time axis relative to peak time in each band.]{\includegraphics[width=0.49\textwidth]{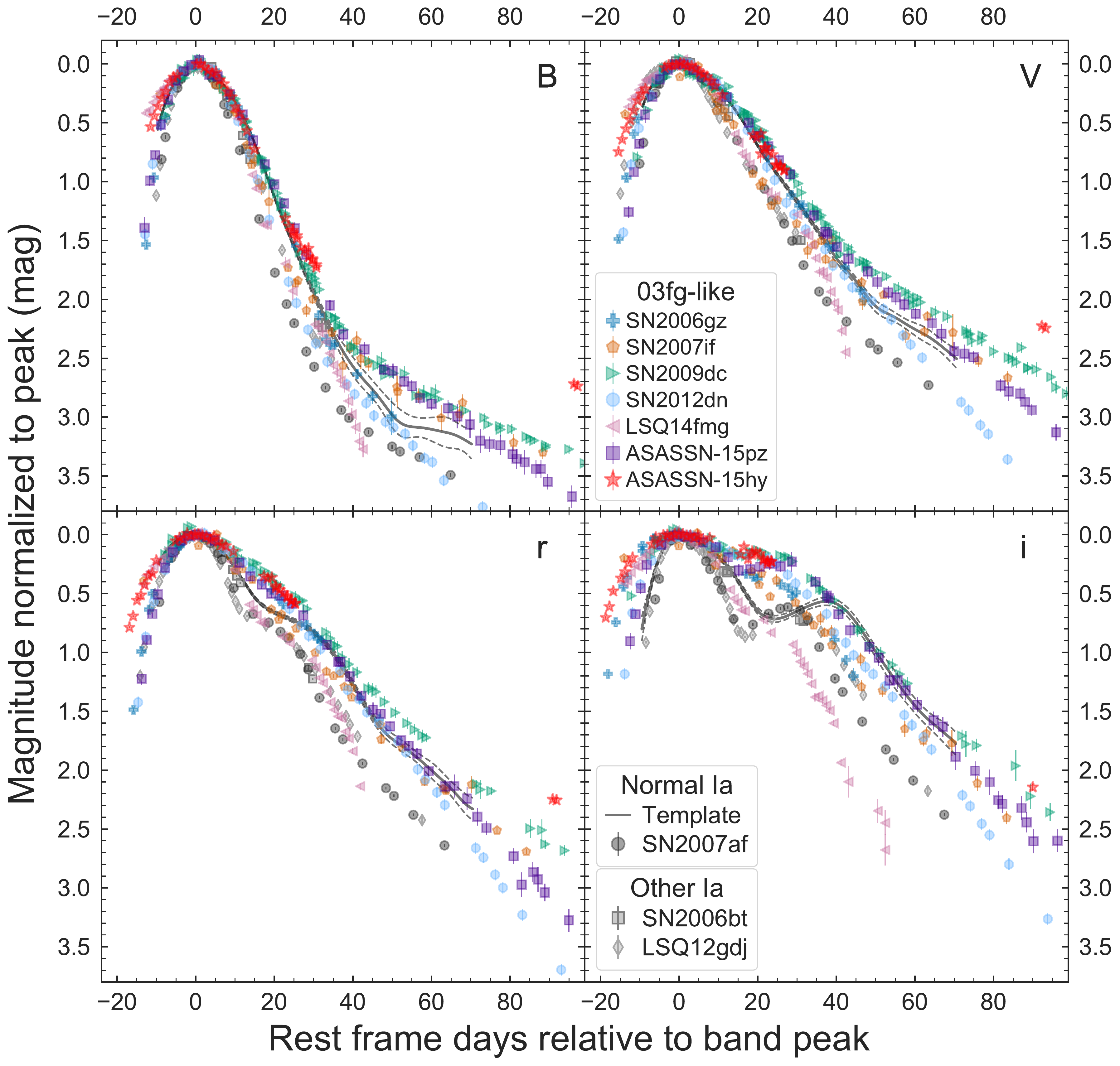}}
\hfill
\subfigure[Time axis relative to peak time in $B$ band.]{\includegraphics[width=0.49\textwidth]{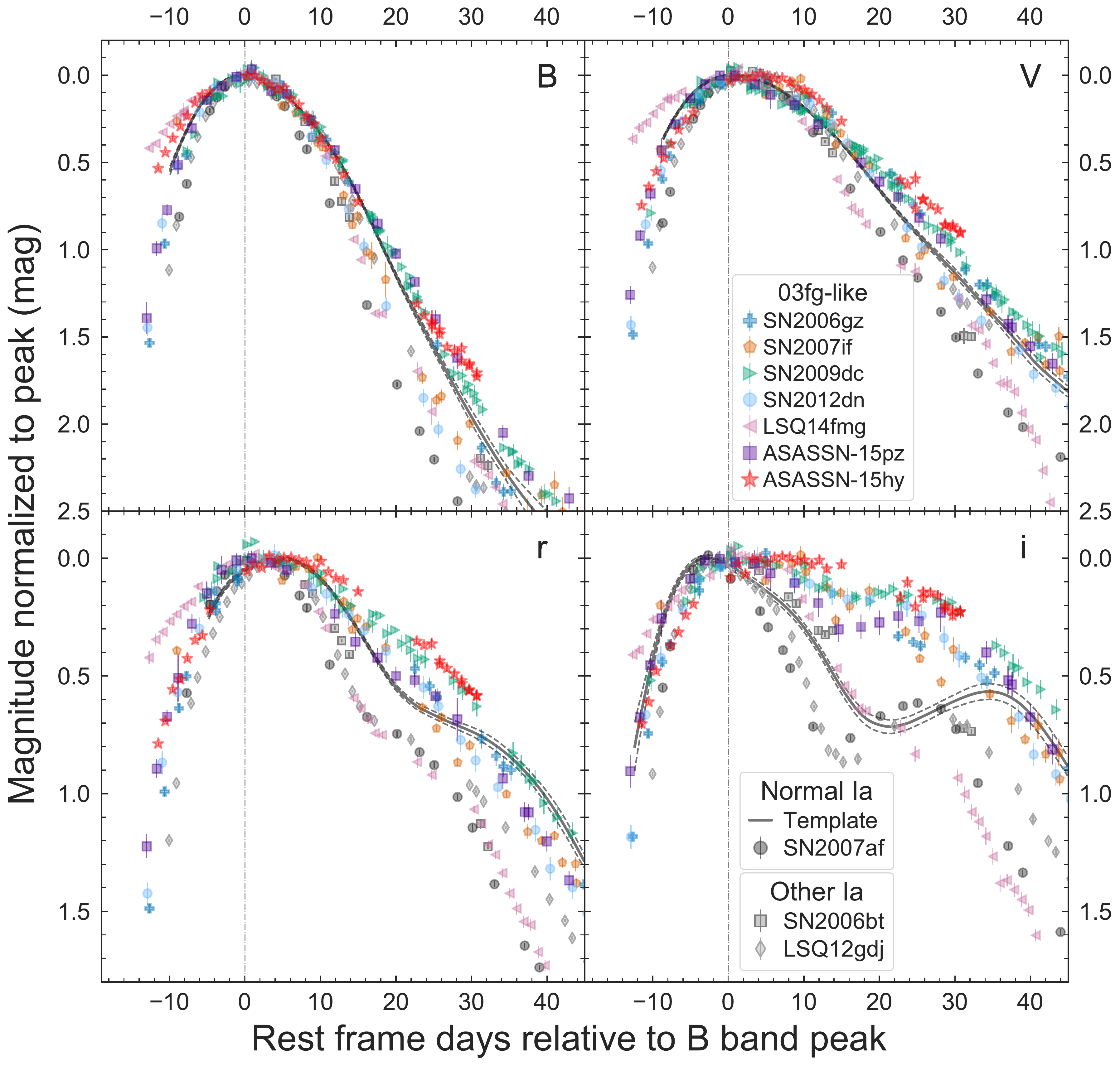}}
\caption{
Comparison of $K$-corrected $BVri$ light curves. 
The magnitudes are normalized to the peak magnitude in each band, and the rest-frame time axes are presented relative to the peak time of each respective band in (a) and to the peak time in the $B$ band in (b). 
The black lines are the Ia template light-curves fits \hy\ using SNooPy, treating it as if it were a normal SN~Ia.
}
\label{fig:BVri_comp}
\end{figure*}

\subsection{Light-curve Evolution} \label{subsec:LC}
Figure~\ref{fig:BVri_comp} compares the optical rest-frame $K$-corrected $BVri$ light curves of \hy\ and other SNe in the comparison sample described earlier.
A 91T-like SN~Ia, LSQ12gdj \citep{Scalzo14} is added to the sample for comparison.
Another peculiar object, SN~2006bt \citep{Foley10,Stritzinger2011,Krisciunas17}, which is underluminous but slowly declining, is also included in the comparisons since it has a similar $i$-band light-curve shape to \superc\ SNe~Ia.
The $B$- and $V$-band light curves of all SNe in the figure have similar shapes but differ in time scale.
\hy\ evolves more slowly on the rise compared to the SNooPy SN~Ia template, but behaves similarly to some other \superc\ SNe. 
The rise time of \hy\ (22.5 $\pm$ 4.6~d), determined by fitting the flux converted from the CSP-II and ASAS-SN $V$-band pre-maximum magnitudes with a second order polynomial, is similar to most of the \superc\ objects where values can be determined. 
In this respect, \hy\ is particularly similar to SN~2007if \citep{Scalzo10} and SN~2009dc \citep{Silverman11,Taubenberger11}, which have rise times of 24.2 $\pm$ 0.4~d and 23 $\pm$ 2~d, and decline rates \DmB\ of 0.71 $\pm$ 0.03~mag and 0.71 $\pm$ 0.06~mag respectively. 
From maximum light to $\sim$+30~d, \hy\ has similar behavior to the comparison SNe.
However, after 30~d past maximum light, \hy\ has among the slowest declining light curves in all $BVri$ bands.
In the $B$ band, \hy\ declines by $\sim$1.0~mag from +30 to +90~d, whereas SN~2009dc declines by $\sim$1.4~mag, for example. 

The $r$- and $i$-band light curves of \hy\ are similar to other \superc\ objects but differ significantly from other SNe~Ia.
In the $r$ band, normal and 91T-like SNe show a shoulder feature at $\sim$+15~d after peak, while \hy\ and other \superc\ objects generally do not show this feature. 
In the $i$ band, \hy\ has a slower rise compared to normal and 91T-like SNe~Ia and lacks a pronounced secondary maximum. 
Note that SN~2006bt also lacks a clear secondary maximum in the $i$ band, however, it differs in NIR light-curve properties and spectra compared to \superc\ objects (see below).
\hy\ has an $i$-band primary maximum that occurs after the $B$-band maximum and has a difference in peak times ($t^{\rm{peak}}_{i-B}$) of 7.2 $\pm$ 1.1~d. 
Although this is a ubiquitous property of \superc\ objects \citep{Ashall20}, \hy\ has the largest value of $t^{\rm{peak}}_{i-B}$ among all \superc\ objects included in \citet{Ashall20}, which have a mean of 3.0 $\pm$ 2.1~d.

In fact, the overall trend of $BVri$ peak times and decline rates indicates the extreme nature of \hy. 
\hy\ declines progressively more slowly and peaks progressively later in time, from blue to redder bands.
This trend is significantly different from the normal SN~Ia 2007af, which does not show the same progressive change especially from $r$ to $i$ band.
On the other hand, the \superc\ objects all show a relatively smooth transition, similar to \hy.
However, \hy\ is on the extreme end in its light-curve properties, including the slowest decline rate and the largest difference in peak times in both $r$ and $i$ bands.

\begin{figure}[!t]
\centering
\includegraphics[width=0.49\textwidth]{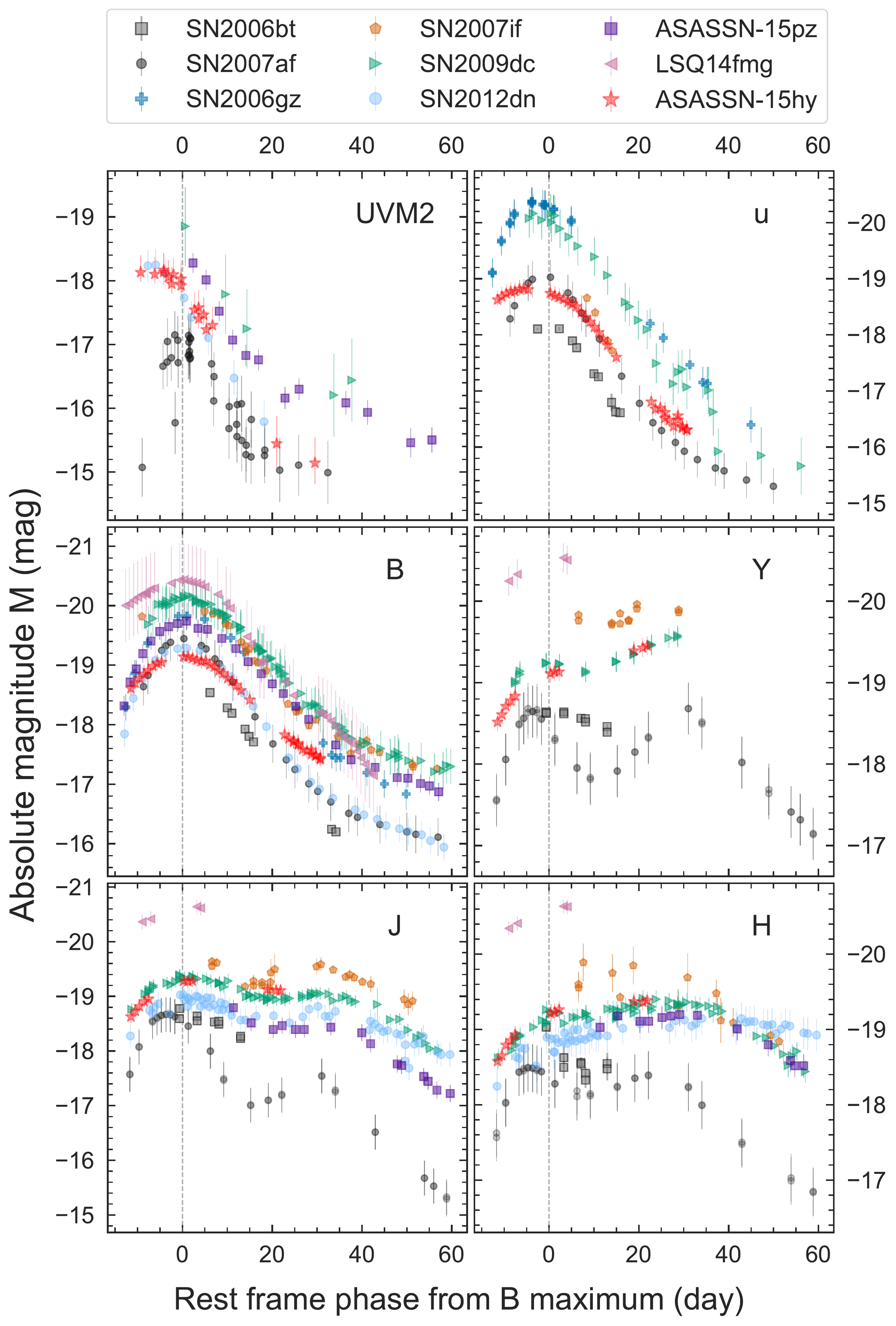}
\caption{Comparison of the absolute magnitudes in UV $uvm2$, optical $uB$ and NIR $YJH$. 
The absolute magnitudes are computed from $M = m -\mu - A_{\text{MW}} - A_{\text{host}}$.
Except for the $B$ band, light curves in this figure are not $K$-corrected due to the lack of spectroscopic coverage. } 
\label{fig:uv_B_NIR_compare}
\end{figure}

To compare the intrinsic brightnesses of \hy\ and others, Fig.~\ref{fig:uv_B_NIR_compare} presents the absolute magnitudes, corrected for MW and host extinction, of the comparison SNe~Ia in the $uvm2$ and $uBYJH$ bands. 
The larger error bars are due to the high uncertainties in the host reddenings for some of the comparison SNe~Ia.
UVOT photometry of the comparison SNe was obtained from the $Swift$ Optical Ultraviolet Supernova Archive (SOUSA; \citealt{Brown14a}) via the Swift Supernova website.\footnote{https://pbrown801.github.io/SOUSA/}

\hy\ is UV bright and has a peak absolute magnitude of \hyMUVM\ in the $uvm2$ band, which is more than 1~mag brighter comparing to the normal SN~Ia 2007af around the same time, consistent with other \superc\ objects \citep{Taubenberger11,Brown14,Chen19}.
The $uvm2$ light curve of \hy\ is most comparable to SN~2012dn, and they both peak earlier than the normal SN~Ia 2007af.
{The early bright UV feature of \superc\ SNe may be explained by the shock interaction \citep[e.g.,][]{Fryer2010,Blinnikov2010,Taubenberger11,Hachinger12,Scalzo12}.}

During the photospheric phase,  \hy\ is one of the faintest \superc\ SNe in the $u$ and $B$ bands at $B$-band maximum. 
It is $\sim$1.4 and $\sim$1.0~mag fainter than SN~2009dc in the $u$ and $B$ band, respectively.
In fact, \hy\ is fainter than the normal SN~Ia 2007af, which has a faster decline rate.
Interestingly, \hy\ has a similar  brightness to other \superc\ SNe in the NIR bands.

As shown in Fig.~\ref{fig:uv_B_NIR_compare}, the NIR light curves of \hy\ are broader, brighter than normal SNe~Ia and do not have two distinct maxima, similar to other \superc\ objects.
In the $Y$ and $H$ bands, \hy\ continued to rise in brightness over the duration of our NIR observations and does not appear to have reached peak by the final observations at $\sim$+22~d past $B$-band maximum, which have magnitudes of $-19.45$ and \hyMHlimit\ respectively.
The \superc\ events SN~2009dc, SN~2012dn and ASASSN-15pz all show weak secondary maxima around +30~d in the $J$ band, similar to the timing of the NIR secondary maximum for a normal SN~Ia.  Unfortunately, our $J$-band light curve does not extend to these epochs. 

SN~2006bt shares some similar properties in the optical with \superc\ objects, but shows a different evolution in the NIR; for example, SN~2006bt showed a decrease in flux in the $Y$ band after the $B$-band peak.  Clearly, NIR light curves can be used as a powerful diagnostic to distinguish \superc\ SNe from other subclasses of SNe~Ia.

\subsection{Color Curves} \label{subsec:color_curve}
\begin{figure}
\centering
\includegraphics[width=0.49\textwidth]{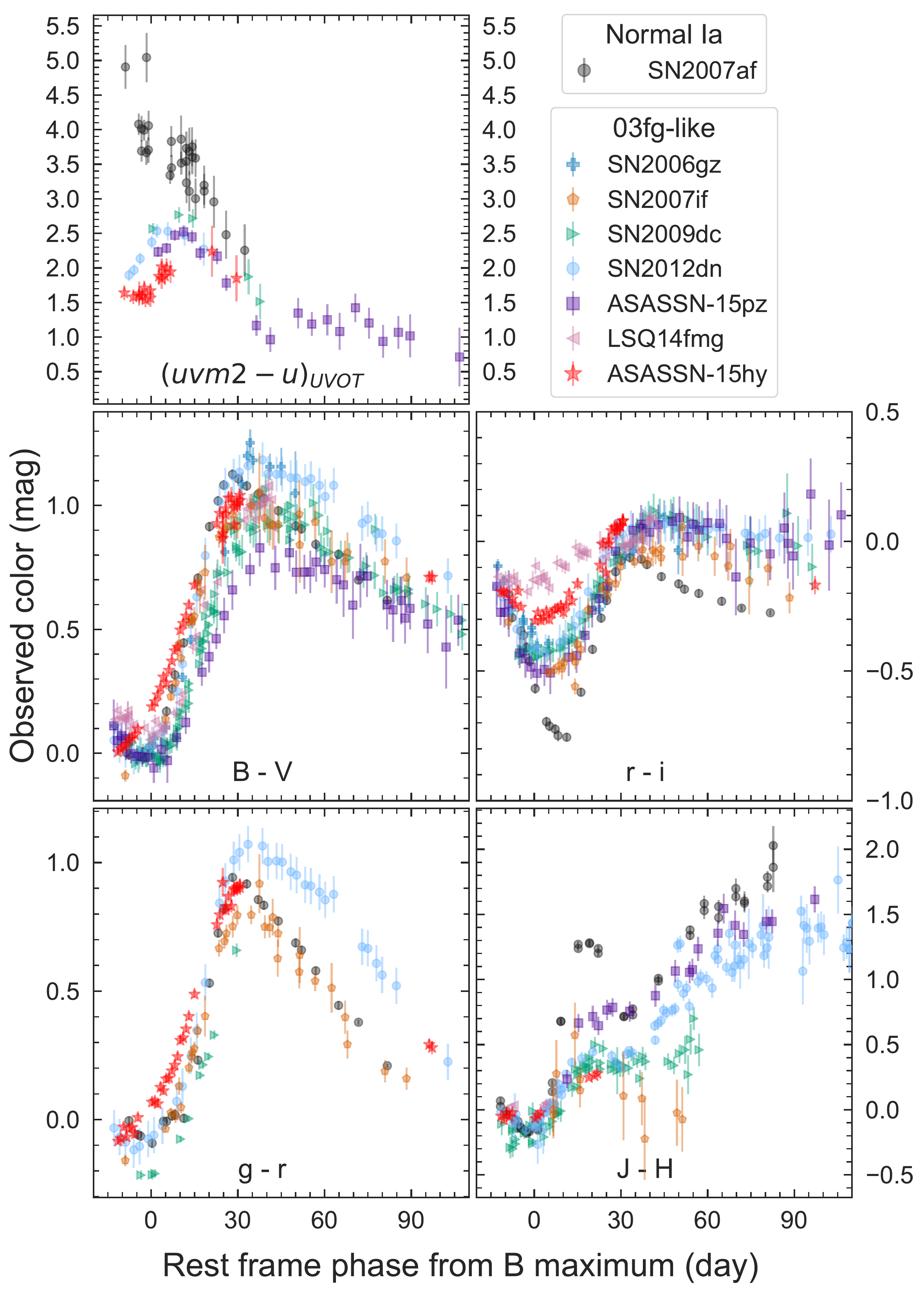}
\caption{Comparison of color curves.
$B-V$, $r-i$, and $g-r$ color curves were obtained from $K$-corrected $BgVri$ photometry.
The color curves in the UV and NIR bands are not $K$-corrected due to the lack of spectroscopic coverage.
All data presented are MW extinction corrected.
}
\label{fig:colors}
\end{figure}

The observed $uvm2-u$, $B-V$, $r-i$, $g-r$, and $J-H$ color curves of \hy\ are presented in Fig.~\ref{fig:colors}, along with those of other \superc\ SNe and the normal SN~Ia 2007af.
Here, host extinction corrections, which are themselves uncertain, are not applied to all comparison SNe.

In $B-V$ and $g-r$, \hy\ shows the reddest observed colors of all \superc\ SNe at maximum light and has values of $0.18 \pm 0.01$~mag and $0.07 \pm 0.01$~mag respectively, compared to an range of $-$0.04 to 0.09~mag and $-$0.21 to $-$0.07~mag for other \superc\ SNe.
This color difference would be even more pronounced after correcting for host-galaxy reddening, since \hy\ has negligible host extinction.  
These colors of \hy\ evolve monotonically redward from the first observation at $-11.5$~d relative to the $B$-band maximum until $\sim$+30~d. 
While most of other \superc\ SNe show local color minima around 0~d similar to the normal SN~Ia 2007af, except for SN~2007if (it is unclear due to the lack of observations).
If \hy\ has such local minimum, the timing of the bluest point would be much earlier than most of the other \superc\ SNe. 
Although any possible host-galaxy extinction may shift the $B-V$ color curve vertically, the shape of the color curve would not change significantly. 
Thus, \hy\ is unique in that it does not have an initial phase where it evolves bluer in $B-V$ and $g-r$, as well as the reddest color among \superc\ SNe at $B$-band maximum.
Our observations may not be early enough to capture the bluest point of $B-V$ and $g-r$; nevertheless, this turn over point is shown to be possibly the earliest among \superc\ SNe.

The color stretch parameter, $s_{BV}$, can also be used to characterize a SN~Ia and is determined by locating the timing of the reddest point in the $B-V$ color curve \citep{Burns14}. 
\hy\ has an $s_{BV}$ of 1.24~$\pm$~0.18 (the large error due to the lack of photometric coverage at the reddest point). 
\hy, along with all other \superc\ SNe, have $s_{BV}$ values larger than 1.0, matching the slowest decliners in the normal population. 
The large $s_{BV}$ together with a late $t^{\rm{peak}}_{i-B}$ are uniform features of these \superc\ SNe, enabling classification between peculiar subgroups of SNe~Ia \citep{Ashall20}.

The $r-i$ color curve of \hy\ is also unique compared to normal SNe~Ia and the majority of \superc\ events. As \citet{Hsiao2020} have shown, the $r-i$ color curve is one of the more obvious ways to distinguish this subclass from normal objects.  
Compared with normal SNe~Ia, these peculiar \superc\ objects evolve more slowly in time, as well as having shallower peaks and troughs in their $r-i$ evolution.
These differences are most likely caused by the unusually broad $i$-band light curve in \superc\ objects that also display very weak or missing $i$-band secondary maxima. 
Within the \superc\ group, there is a large scatter in $r-i$ colors at the epoch of $B$ maximum.
\hy\ is separated from the majority of \superc\ members and is at least $\sim$0.1~mag redder than other \superc\ objects around the time of $r-i$ minimum, but is not as extreme as LSQ14fmg.
The unusually flat $r-i$ color of LSQ14fmg is suggested to be the outcome of  interaction of the SN ejecta with a superwind of an AGB star \citep{Hsiao2020}, which could potentially be the case for \hy\ as well.

The top panel of Fig.~\ref{fig:colors} shows the $Swift$ UVOT $uvm2-u$ color curves. \hy\ evolves to the red up to $\sim$+20~d and then turns blueward.
This is similar to the other members of the \superc\ group (\eg\ SN~2009dc and 2012dn), and \hy\ appears to be the bluest at early times. 
The blue UV$-$optical color ($uvm2-u$ = $1.72\pm0.16$~mag at $B$-band maximum) may indicate a lack of line-blanketing from Fe lines comparing to other members, possibly caused by differences in metallicity (e.g., \citealt{Lentz2000}) or caused by differences in the outer slope of the density profile of the ejecta \citep{Walker2012}.
Conversely, the $uvm2-u$ color of the normal SN~Ia 2007af starts quite red and evolves monotonically blueward over time. 
\superc\ SNe are known to be UV-blue as well as UV-bright at early times \citep{Milne13,Brown14,Chen19}, and this unique early UV evolution separates them from the normal population.

The $J-H$ color curve of \hy\ is similar to that of SN~2009dc and other \superc\ SNe but evolves differently when compared to the normal SN~Ia 2007af. This is not a surprise given the difference between the NIR light curves of \superc\ and normal SNe~Ia. For example, normal SNe~Ia become rapidly redder in $J-H$ until $\sim$+15~d past $B$-band maximum, then drop to a local blue minimum at $\sim$+30~d. They then evolve to the red again as the observations show. The $J-H$ red bump $\sim$+15~d is thought to be associated with the evolution of the $H$-band break and the unveiling of iron group elements in the ejecta \citep{Wheeler98}.
Unlike normal SNe~Ia, \hy\ becomes redder more slowly over time and only increases $\sim$0.3~mag from 0 to +20~d, whereas SN~2007af increases $\sim$1.5~mag during the same time period.
SN~2009dc and SN~2012dn also show a similar slow reddening at early times and then flatten off in color between +15 to +40~d, possibly due to the lack of a $H$-band break in the NIR spectra.  It is noteworthy that SN~2007if evolves bluer in $J-H$ after +15~d until the last observed epoch, although the uncertainties are large.

\begin{figure*}[ht!]
\centering
\subfigure[Pre-maximum comparison to various subgroups.]{\includegraphics[width=0.49\textwidth]{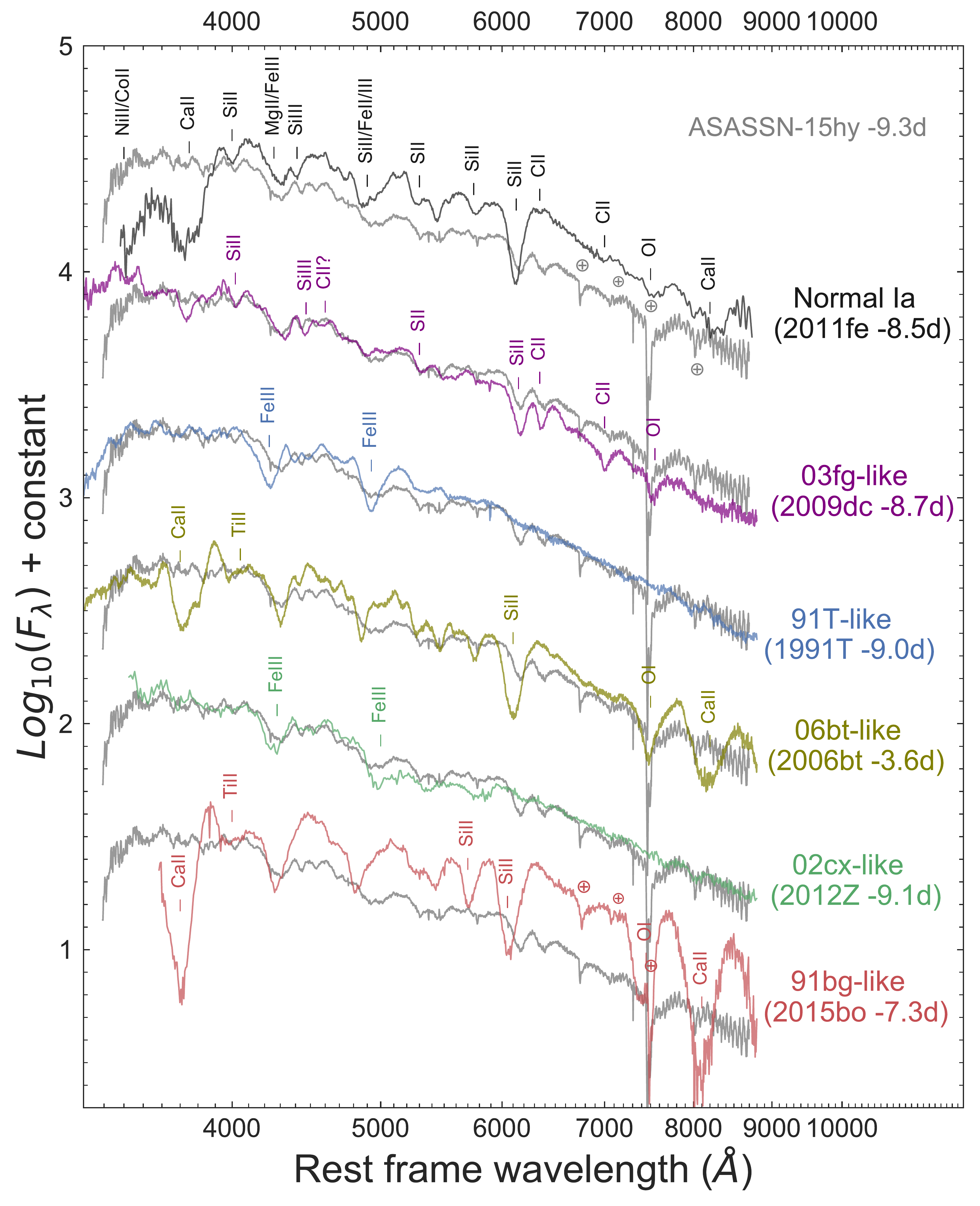}}
\hfill
\subfigure[Post-maximum comparison within the \superc\ subgroup.]{\includegraphics[width=0.49\textwidth]{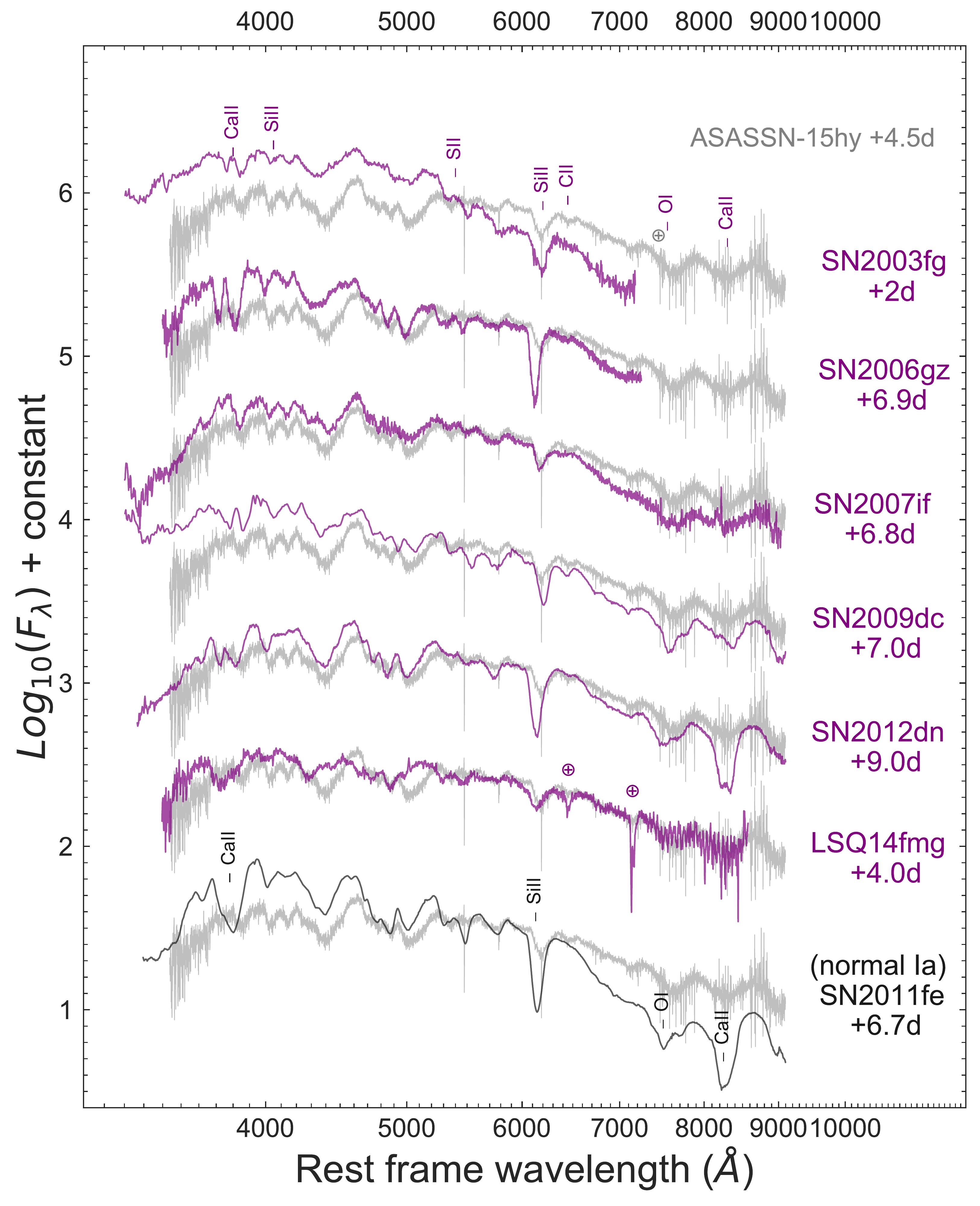}}
\caption{Spectroscopic comparisons of \hy\ and other SNe~Ia. 
Panel (a) compares \hy\ with a normal SN~Ia and other subgroups, including SN~2011fe \citep{Zhang2016}, SN~2009dc \citep{Taubenberger11}, SN~1991T \citep{Filippenko92a}, SN~2006bt \citep{Foley10}, SN~2012Z \citep{Stritzinger2015} and SN~2015bo \citep{Yaron12}, around a week before $B$-band maximum.
The gray spectrum in the background is \hy\ at $-9.3$~d, and the spectra in color are the comparison SNe~Ia around a similar phase.
Panel (b) compares \hy\ with other published \superc\ objects around a week after $B$-band maximum.
The gray spectrum in the background is \hy\ at $+4.5$~d, the purple ones are other \superc\ SNe~Ia around a similar phase, and the normal SN~Ia 2011fe is included for comparison as well.
The telluric features are marked using the color corresponding to the SN. 
}
\label{fig:hy_spec_compare_othergroups}
\end{figure*}

\section{Spectroscopic Properties} \label{sec:spec}
The spectroscopic data set of \hy\ has broad wavelength coverage and high cadence.
In total, there are 42 optical ($-$11.6 to $+$155.2~d) and 6 NIR ($-$8.5 to $+$71.7~d) spectroscopic observations, enabling a detailed look at the spectral features and  evolution of a rare \superc\ object.
Note that all phases in this section are relative to the epoch of $B$-band maximum and corrected for time dilation, unless otherwise stated.

\subsection{Optical Wavelengths}
The first two optical spectra were taken concurrently with the first Swope multifilter observations (Fig.~\ref{fig:spec_present_opt}) and appear blue as confirmed by the $B-V$ color (Fig.~\ref{fig:colors}). The SED then evolves redwards rapidly. The early optical spectra are dominated by P-Cygni profiles of intermediate-mass elements (IME; such as Si, Ca, and S), unburned elements (C and O), and iron group elements (IGEs). Most of these features are also prevalent in the spectra of normal SNe~Ia. However, there is a lack of higher-ionization species such as \FeIII\ in \hy, which are typical of normal, slow-declining SNe~Ia.

\subsubsection{Comparison of Spectral Features }
The left panel of Fig.~\ref{fig:hy_spec_compare_othergroups} presents a pre-maximum optical spectrum of \hy\ with labeled line identifications, as well as comparisons between \hy\ and a variety of subclasses of SNe~Ia. The ions present in \hy\ are similar to those observed in the normal SN~2011fe. 
The notable differences are the overall weaker spectral features in \hy, particularly the \CaII\ features as well as the stronger \CII\ features.
The weak spectral lines of \hy\ resemble those of SN~1991T and SN~2012Z (02cx-like). However, the ionization states of these two objects are much higher than \hy, as the early spectra of SN~1991T and SN~2012Z are dominated by \FeIII\ lines \citep{Filippenko92a,Stritzinger2015}. 
\hy\ does not exhibit the \TiII\ feature around 4000~\AA, which is present in 06bt- and 91bg-like SNe (e.g., \citealt{Filippenko92b,Foley10}). 
Overall, the spectra of \hy\ are remarkably similar to the \superc\ event SN~2009dc \citep{Taubenberger11}, demonstrating that this object belongs in the \superc\ subclass.

The right panel of Fig. \ref{fig:hy_spec_compare_othergroups} presents a comparison between \hy\ and other \superc\ SNe around one week after $B$-band maximum. All of the objects have a similar ionization state, are dominated by IMEs, show weak \CaII\ H\&K features, and most have persistent \CII\ absorptions past maximum light. However, the strength and velocity of the \SiII\ $\lambda$6355 feature vary within this subclass. 

\begin{figure}[t!]
\centering
\hspace*{-0.2cm}\includegraphics[width=0.49\textwidth]{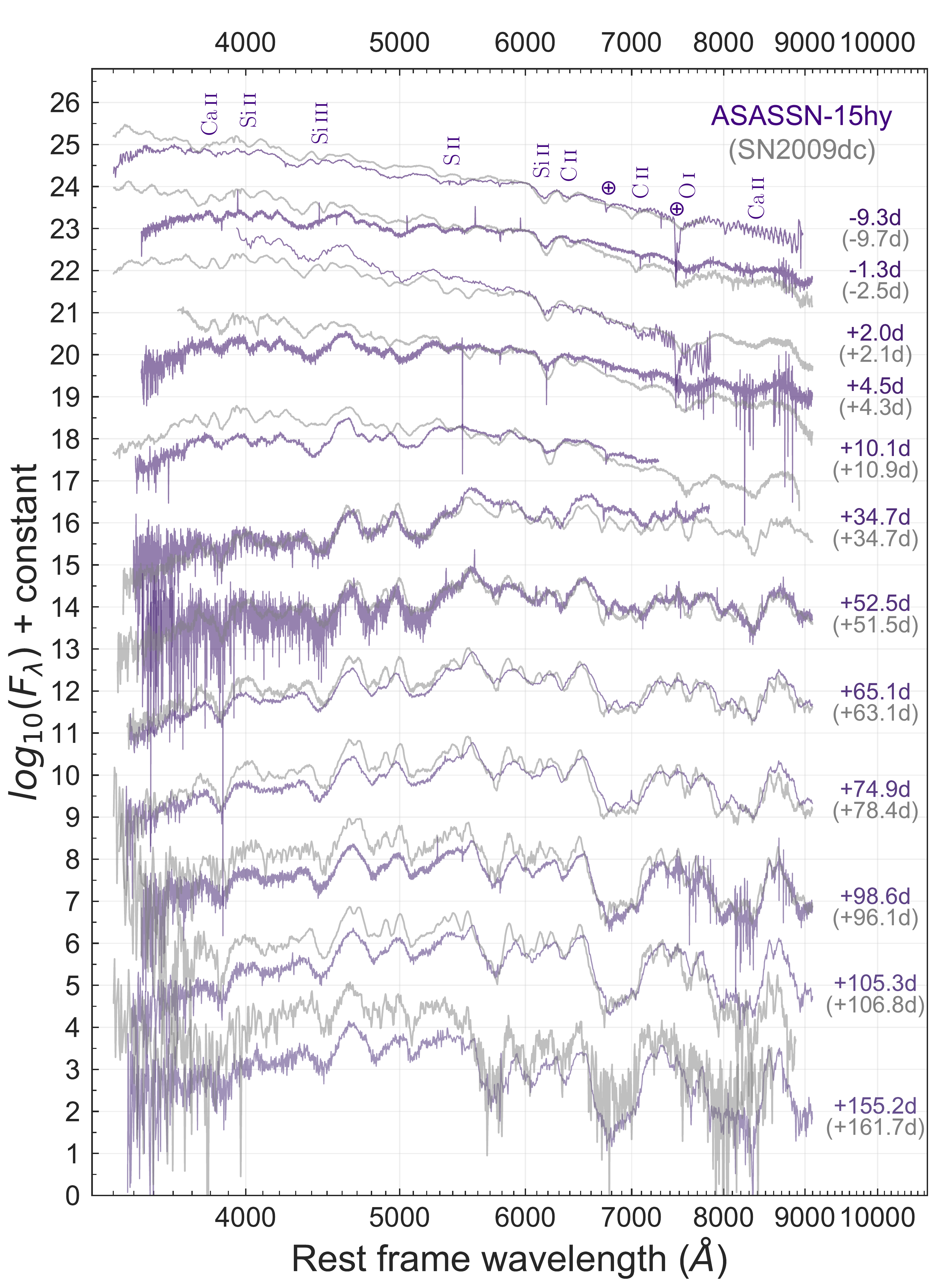}
\caption{Spectroscopic time series comparison of \hy\ and SN~2009dc in the optical.
Spectra of \hy\ are plotted with purple colors and those of SN~2009dc \citep{Taubenberger11} around similar phases are plotted with gray in the background.
Regions of strong telluric adsorptions in the rest frame of \hy\ are marked with the purple symbols at the top. 
}
\label{fig:opt_spec_comp_09dc}
\end{figure}

SN~2009dc and \hy\ are the two \superc\ SNe~Ia that have the most complete optical spectroscopic data sets.
Figure~\ref{fig:opt_spec_comp_09dc} shows the time series comparison between these two \superc\ objects. 
Again, these two objects have very similar evolution in their spectral features, except for the slightly redder color of \hy.
The \CII\ $\lambda$6850 line is still present in both \hy\ and SN~2009dc around one week after maximum, while \CII\ in a normal SN~Ia, if present at all, usually disappears soon after the explosion \citep{Taubenberger11,Folatelli12}.
The strong and persistent \CII\ feature implies substantial unburned material in the ejecta \citep{Howell06,Hicken07}.

\subsubsection{\SiII\ and \CII\ Evolution} \label{subsec:Si_velocity}
To compare the line profiles of \hy\ and other SNe quantitatively, we measured the velocities and pseudo-equivalent widths (pEW) of the \SiII\ $\lambda$6355 and \CII\ $\lambda$6850 features. 
We first divide the flux by the local continuum around the feature, which is constructed by connecting the two local maxima on both sides of the feature. 
The area is integrated to obtain the pEW. 
The position of the absorption minimum is obtained by fitting a Gaussian profile to obtain the blue-shifted velocity. 
To estimate the measurement uncertainties, we bootstrap-resample the flux 500 times repeating the above process. 
The mean and standard deviation of the realizations are adopted as the measured value and corresponding error for each pEW and velocity measurement. 

\begin{figure}[t!]
\centering
\includegraphics[width=0.38\textwidth]{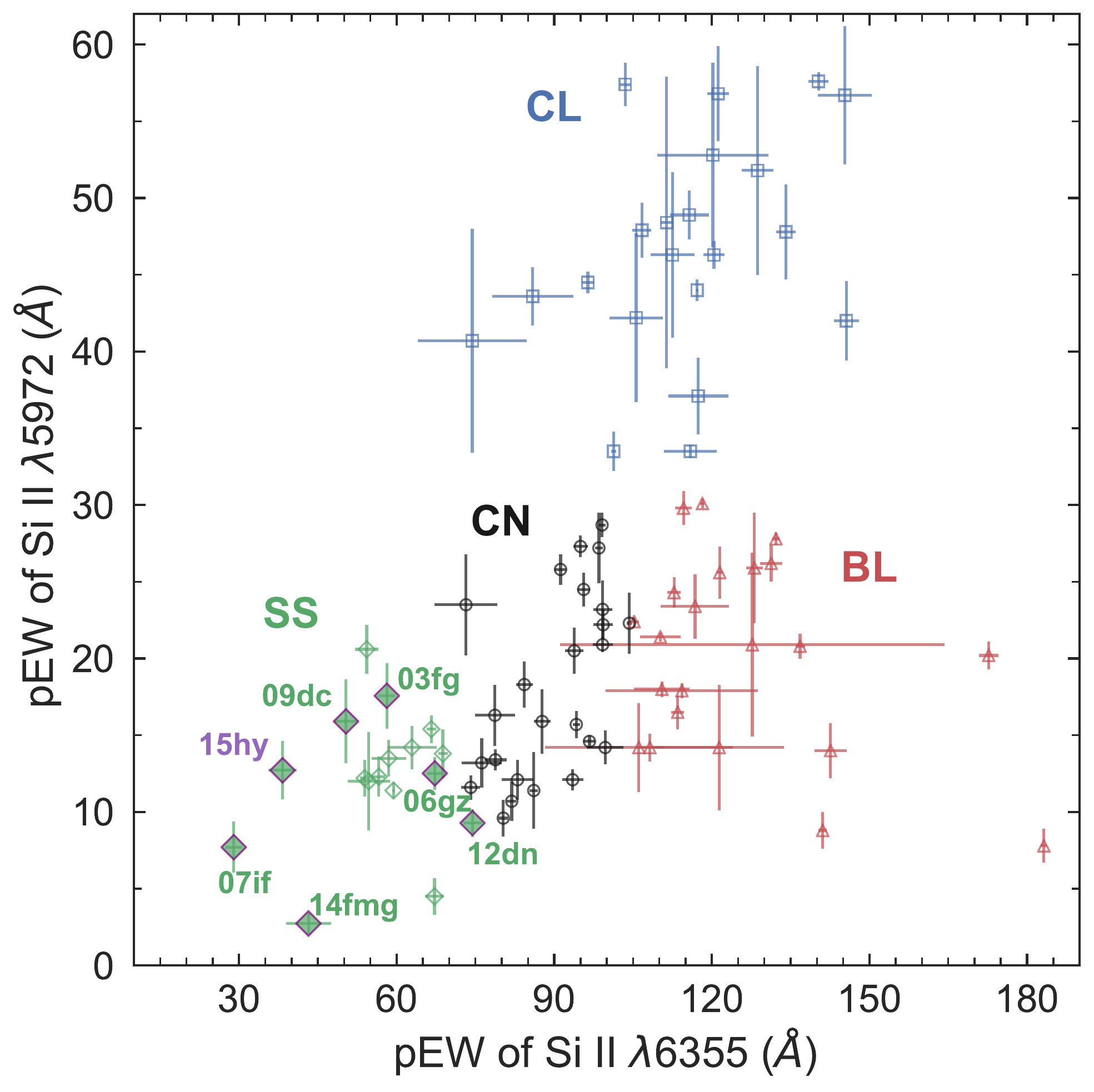}
\caption{
Branch diagram \citep{Branch2006} with the pseudo-equivalent width of \SiII\ $\lambda$6355 and $\lambda$5972,
The open markers are the measurements from \cite{Folatelli2013}, the filled diamonds with purple edges are the measurements from this work of the currently available \superc\ objects.  
}
\label{fig:BD}
\end{figure}

\hy\ has small pEW values for \SiII\ $\lambda$5972 and $\lambda$6355 near maximum light, which places it in the ``shallow silicon" (SS) group on the Branch diagram \citep{Branch2006}. This is a common characteristic of  \superc\ events (e.g., \citealt{Taubenberger11,Hsiao2020}), as shown in Fig.~\ref{fig:BD}. The only \superc\ SN that might sit in a different area of the Branch diagram is SN~2012dn; however, this SN also has a lower luminosity than most of the other \superc\ SNe.
Interestingly, the underluminous \hy\ is sitting closer to the brighter \superc\ SNe, such as SNe~2007if and 2009dc, on the Branch diagram.

\begin{figure}[t!]
\centering
\includegraphics[width=0.38\textwidth]{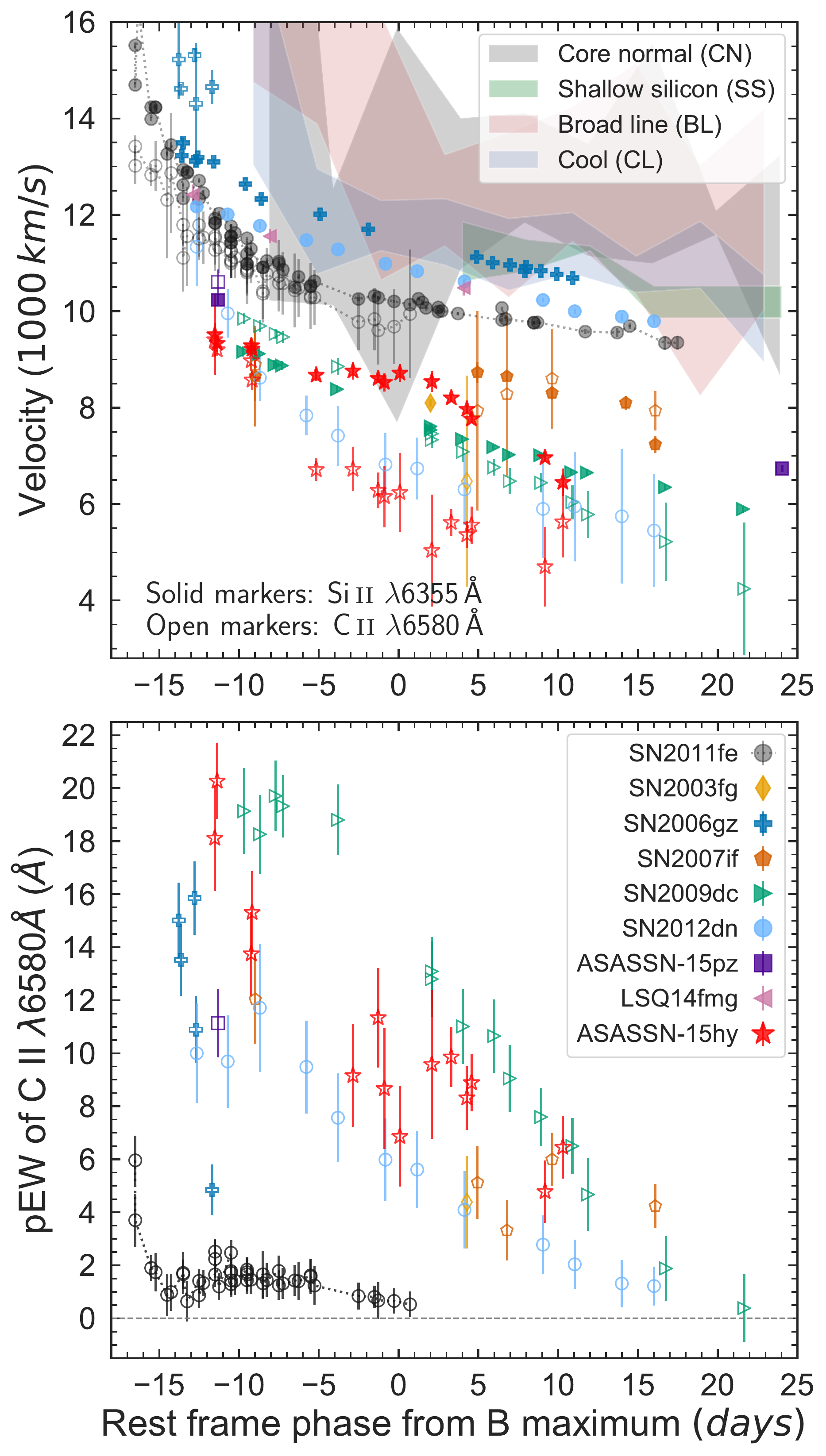}
\caption{
The time evolution of the \SiII\ $\lambda$6355 and \CII\ $\lambda$6580 velocities (top panel) and the \CII\ $\lambda$6580 pseudo-equivalent widths (bottom panel).
\hy\ is shown compared to other \superc\ events and the normal SN~2011fe.
The mean \SiII\ $\lambda$6355 velocity measurements of the four Branch diagram subgroups from \citet{Folatelli2013} are plotted in the top panel for comparison as well. }
\label{fig:Si_v}
\end{figure}

The velocity of \SiII\ $\lambda$6355 for \hy\ ranges from $\sim9{,}300$~\kms\ at -10~d to $\sim8{,}600$~\kms\ at +2~d and is $\sim2{,}000$~\kms\ slower than the normal SN~2011fe.
The lack of a rapid velocity decline at early times appears to be a consistent trait of \superc\ SNe, as shown in the top panel of Fig.~\ref{fig:Si_v}, indicating that the Si layer is confined to a narrow range.
\hy\ shows a velocity plateau at $\sim8{,}600$~\kms\ from $-$5 to +2~d. Similar behavior was also reported in the bright \superc\ object SN~2007if \citep{Scalzo10}.
In general, the velocities of \hy\ and other \superc\ SNe are relatively lower than  the mean velocity of SS SNe~Ia from \citet{Folatelli2013}, however, there is a large spread of velocity among the group.

In \hy, it is not possible to determine the Si velocity evolution after +2~d, because the blending of the low-excitation \FeII~$\lambda$6456, $\lambda$6516 lines with the \SiII~$\lambda$6335 feature causes a sudden drop in in what is measured as the Si velocity. 
Therefore, after +2~d the measurement does not represent the true velocity of the \SiII\ $\lambda$6355 feature. This blending effect is usually seen in  subluminous SNe~Ia at similar early phases \citep{Galbany19} and at later times in normal SNe Ia \citep[\eg][]{Folatelli2013}.  
This velocity drop has been previously mistaken as a distinguishing feature of \superc\ objects \citep{Scalzo12}.
The fact that we see this blending effect so early in \hy\ may be an indication that there is a low ionization state in the line-forming region.

The \CII\ $\lambda$6580 velocity of \hy\ is similar to the  \SiII\ $\lambda$6355 velocity at $-$11~d ($\sim9{,}300$~\kms), but diverges from the \SiII\ velocity shortly after this time period. 
By $B$-band maximum, the \CII\ velocity is $\sim2{,}500$~\kms\ lower than the \SiII velocity.
This is in contrast with the majority of the comparison SNe which show similar velocity gradients for \CII\ $\lambda$6580 and \SiII\ $\lambda$6355.

Previous studies have shown that for normal SNe~Ia the \CII\ and \SiII\ velocity evolution is similar \citep{Parrent2011,Folatelli12}.
However, \hy\ has a pre-maximum \CII\ velocity decline rate of $\sim3{,}000$~\kms\ per 10~days but only $\sim$500~\kms\ per 10~days for \SiII.
This diverging behavior in velocities is also seen in SN~2012dn \citep{Chakradhari14,Parrent16,Taubenberger19}.
This may indicate that there is carbon-rich material below the Si layer, or more likely, it could be a projected velocity and ionization effect.  If the \CII\ is in a confined region just above the photosphere, more absorption is seen from the material not moving directly towards the observer, which results in a projection effect that makes the material appear to be at a lower velocity \citep{Hoeflich1990}.

Despite the differences in velocity of \CII\ $\lambda$6580 within the \superc\ subgroup, \hy\ and most other members of this subgroup have large pEW values for this ion, which continues past maximum light. It persists much longer than in normal SNe~Ia, such as SN~2011fe, as shown in the bottom panel of Fig.~\ref{fig:Si_v}. The \CII\ pEW of \hy\ has a value of $\sim$19~\AA\ at $-$11~d and gradually decreases with time, disappearing around 2 weeks past maximum light. In contrast, the normal SN~Ia 2011fe only has a pEW of $\sim$5~\AA\ at $-$17~d, which rapidly declines to $\sim$2~\AA\ within 5~days, and then gradually disappears while approaching maximum light. This is evidence of substantial unburned material within the ejecta of \hy.

\subsection{Near-infrared Wavelengths\label{subsec:NIR_spec}}
\begin{figure}[ht!]
\centering
\subfigure[NIR spectra of \hy.]{\includegraphics[width=0.49\textwidth]{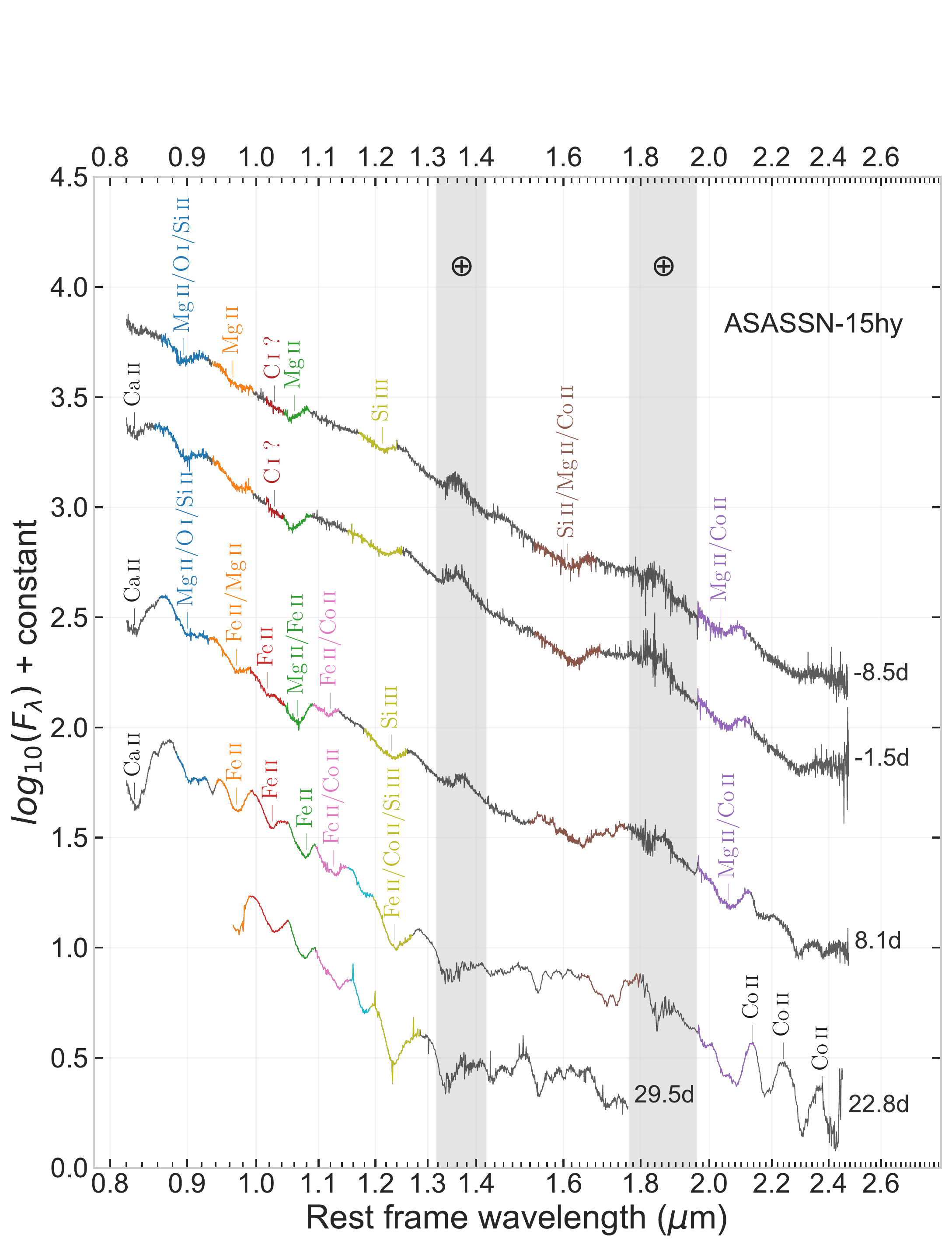}}
\vfill
\subfigure[Velocity measurements of assumed ions.]{\includegraphics[width=0.49\textwidth]{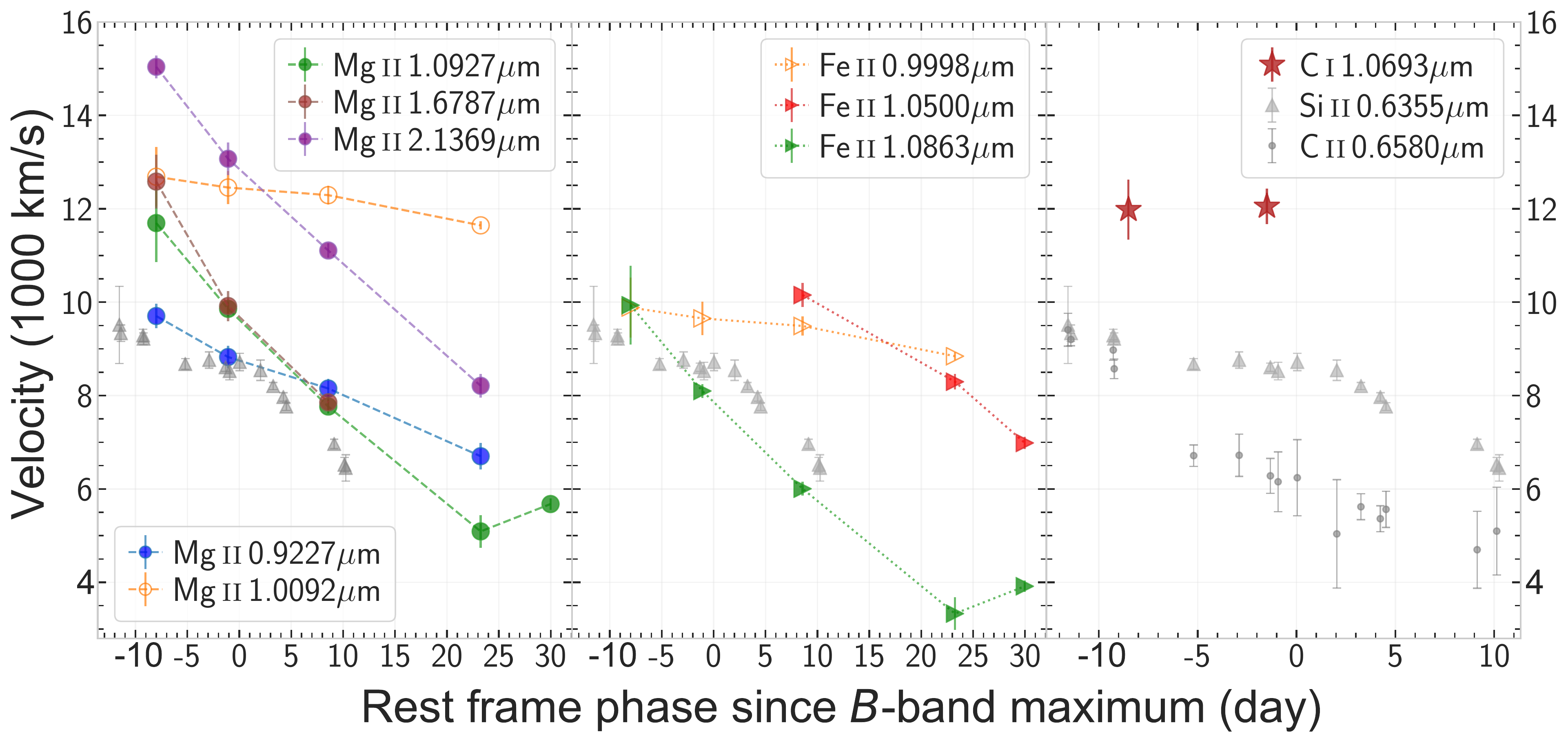}}
\caption{The line identifications of the NIR spectra of \hy. Panel (a) shows the first five NIR spectra of \hy\ and the most likely ion labels for the corresponding features. The vertical gray bands mark the regions of strong telluric adsorptions.
Panel (b) presents the velocity measurements of several assumed ions. The open markers represent features with possible blending.
The velocities of optical \SiII\ and \CII\ are also plotted as reference.
The colors of the features are consistent in the two panels.
}
\label{fig:NIR_lineID}
\end{figure}

\begin{figure*}[t!]
\centering
\includegraphics[width=0.99\textwidth]{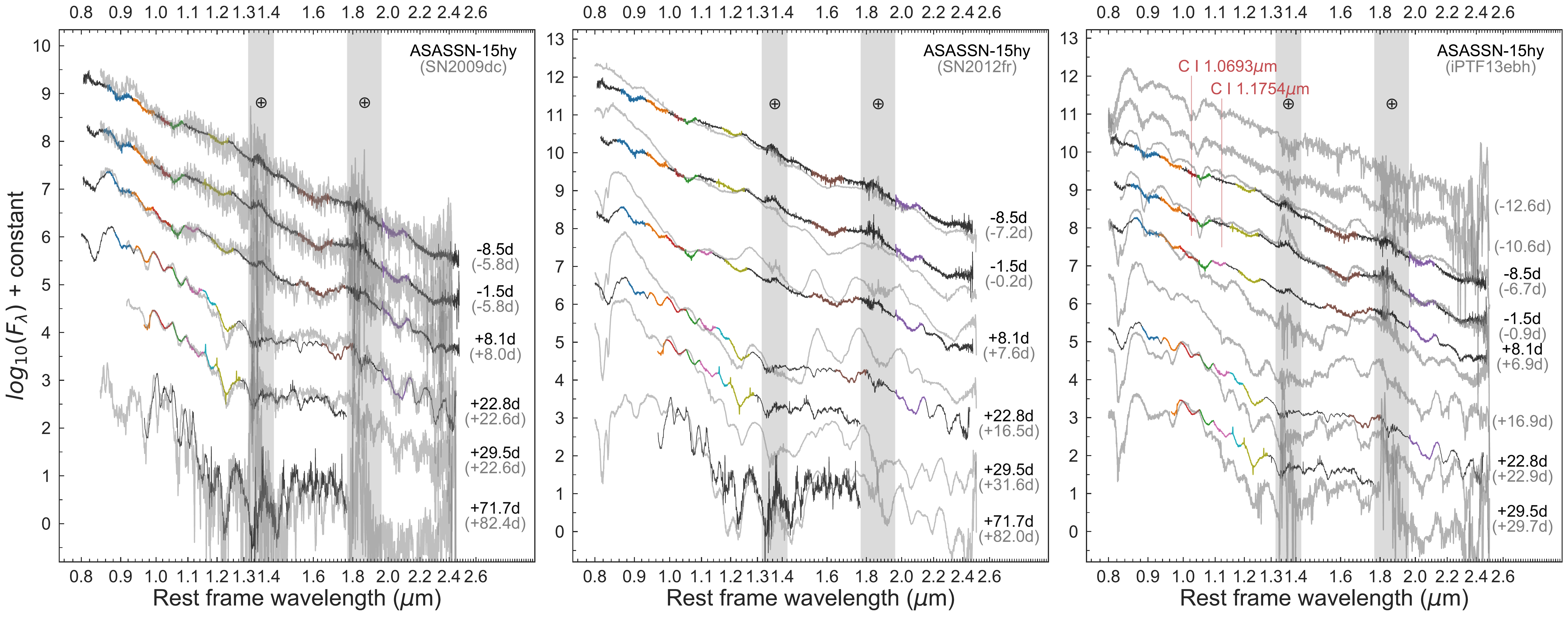}
\caption{NIR spectral comparison of \hy\ with three SNe~Ia: the \superc\ object SN~2009dc \citep{Taubenberger11} (left panel), the normal SN~2012fr \citep{Contreras2018,Hsiao19} (middle panel), and the transitional iPTF13ebh \citep{Hsiao15} (right panel).
The vertical red lines in the right panel mark the positions of prominent \CI\ lines in iPTF13ebh.
The regions of the strongest telluric adsorptions are marked with vertical gray bands.}
\label{fig:NIRspec_comp}
\end{figure*}

\hy\ has 6 NIR spectra covering epochs from $-$8.5 to +71.7~d, making it only the second \superc\ object with NIR spectral observations (the first one being SN~2009dc). All spectra have high S/N allowing for unambiguous identifications of several lines and their evolution. 

\subsubsection{Line Identification and Evolution \label{subsubsec:NIR_lineID}}
A number of spectral lines in \hy\ are identified and labeled in Fig. \ref{fig:NIR_lineID}~(a).
The line list from \citet{Marion09}, as well as line identifications from  \citet{Taubenberger11}, \citet{Hsiao15,Hsiao19} and \citet{Gall2012}, were used as a guide to identify the NIR features in \hy. 
The NIR spectra of \hy\ mainly consist of \CaII, \MgII, \FeII, and \CoII\ lines with a possible contribution from \CI. 

The strongest feature in the NIR is the \CaII\ 0.8538~$\mu$m IR triplet, which is also present in the region overlapping with the optical spectra.
Multiple \MgII\ lines also were identified: 0.9227, 1.0927, 1.6787, and 2.1369~$\mu$m. 
These line identifications were confirmed by examining these features in velocity space, where they all follow a similar velocity trend, shown as solid symbols in the left panel of Fig.~\ref{fig:NIR_lineID}~(b).
The \MgII\ 0.9227~$\mu$m feature may also have contributions from \OI\ and \SiII, but in \hy\ these are minor as the velocity evolution of this feature matches the other \MgII\ lines.  
We attribute the absorption feature seen at $\sim$0.97~$\mu$m (marked orange in Fig.~\ref{fig:NIR_lineID}) to be a blend of \MgII\ 1.0092~$\mu$m and \FeII\ 0.9998~$\mu$m.
Thus, the blending likely affects the velocity evolution of these two ions presented in Fig.~\ref{fig:NIR_lineID}~(b). 
This feature is likely to have a significant contribution from \FeII\ by +8.1~d. The presence of \FeII\ at this time period is consistent with the optical spectroscopy of \hy, where a velocity drop in the \SiII\ $\lambda$6355 feature due to the blending with \FeII\ $\lambda$6456 and $\lambda$6516 was observed around the same time.

The IGE features become stronger with time. 
As shown in the +22.8~d spectra, \FeII\ 0.9998, 1.0500, and 1.0863~$\mu$m are all clearly visible and are also seen in SN~2009dc \citep{Taubenberger11}.
\CoII\ 2.1365, 2.2205, 2.3613, and 2.4596~$\mu$m emission features are also present in the 2.1 $-$ 2.4~$\mu$m region in the same spectrum.

\subsubsection{Possible \CI\ Detection} \label{subsubsec:NIR_CI}
Throughout the NIR spectral observations, there are no strong \CI\ features such as those observed in SN~1999by \citep{Hoeflich02}, iPTF13ebh \citep{Hsiao15} or SN~2015bp \citep{Wyatt2020}, but a weak \CI\ 1.0693~$\mu$m line may be present in  \hy\ at $-$8.5 and $-$1.5~d. 
the right panel of Fig.~\ref{fig:NIRspec_comp} shows a notch at $\sim$1.03~$\mu$m (marked with dark red) next to the absorption of \MgII\ 1.0927~$\mu$m.
This is similar to the position of the \CI\ 1.0693~$\mu$m feature in iPTF13ebh at early times, where the feature is strong and unambiguous.
The weak notch in \hy\ would correspond to a velocity of $\sim12{,}000$~\kms\ assuming that it originated from \CI\ 1.0693~$\mu$m.
Such a velocity at $-1.5$~d is $\sim6{,}000$~\kms\ faster than the optical \CII\ $\lambda$6580 line.
This velocity difference could be due to a large carbon envelope in the ejecta that extends to higher velocities. An ionization change throughout the envelope could cause \CI\ and \CII\ to have distinctly different observed velocities. Furthermore, if the \CII\ layer is small, the velocity could appear lower than the location of the bulk of the \CII\ material due to projected velocity effects (see Section~\ref{subsec:Si_velocity}). 

Unfortunately, there are no other clear \CI\ features, such as \CI\ 1.1754~$\mu$m, in the first two NIR spectra.
The next observation at +8.1~d contains two newly formed features aroun 1.02 and 1.12~$\mu$m, marked red and pink in Fig.~\ref{fig:NIR_lineID}~(a), respectively. 
These may be attributed to \CI\ 1.0693~$\mu$m and 1.1754~$\mu$m. 
However, these lines are more likely produced through absorption from IGEs, due to their similarity in velocity to the \FeII\ lines.
We do not exclude the possibility that these two features are blended with weak \CI.

\subsubsection{Missing H-band Break\label{subsubsec:NIR_missing_Hband_break}}
Interestingly, \hy\ does not show an $H$-band break near 1.5~$\mu$m until +72~d. For a normal SN~Ia, the $H$-band break \citep{Kirshner73} appears around +3~d and peaks in strength around $\sim$+10 to +12~d \citep[\eg][]{Hsiao13}, as demonstrated by SN~2012fr \citep{Contreras2018,Hsiao19} in the middle panel of Fig.~\ref{fig:NIRspec_comp}.
The $H$-band break is produced by a multiplet of allowed \CoII, \NiII, and \FeII\ emission lines that emerge when the photosphere recedes deep into the \Nifs\ rich region \citep{Wheeler98,Hoeflich02}. The strength of the $H$-band break has been found to correlate with SN~Ia light-curve shape \citep{Hsiao13}, and the outer blue velocity (\ved) of the break has been found to directly trace the outer \Nifs\ region in the ejecta \citep{Ashall19a,Ashall19b}. The absence of an $H$-band break suggests that the photosphere has not reached the core of the \Nifs\ region by +30~d. This absence was also noted in SN~2009dc \citep{Taubenberger11}.

Despite the peculiar behavior of the $H$-band break, \CoII\ features are apparent in the 2.1$-$2.4~$\mu$m region at +23~d, which is a similar phase to the onset of these features in all three comparison SNe in Fig.~\ref{fig:NIRspec_comp}. 
It has been proposed that these \CoII\ features form in the same region as the $H$-band break \citep{Wheeler98}. 
We rule out the lack of the $H$-band break being caused by an ionization effect, since 91T-like SNe~Ia show strong $H$-band breaks but have a higher-ionization state than \hy\ (M.M.~Phillips et al. in preparation).
The missing $H$-band break in \hy\ could be evidence of an opacity difference between the $H$ and $K$ bands for this peculiar object (see Appendix~\ref{model:spec_constrain} for more details).

The NIR spectra of \hy\ and SN~2009dc are nearly identical in both color and evolution of the line profiles, as shown in the left panel of Fig.~\ref{fig:NIRspec_comp}. 
\hy\ shows both similarities and discrepancies when compared to a normal SN~Ia.
The early time NIR spectra of \hy\ have smooth continua and show shallow IME features such as \MgII, which is consistent with both SN~2009dc and SN~2012fr.
However, after maximum light the features of \hy\ are different from the normal SN~Ia 2012fr, especially in the 1.2 to 1.6~$\mu$m region. As discussed above, part of this is due to the delayed onset of the $H$-band break and distribution of \Nifs.
After maximum light, SN~2012fr presents strong absorption features near 1.2~$\mu$m, thought to be the cause of the rapid decrease in $J$-band magnitude that is seen in normal SNe~Ia \citep{Branch1983,Elias1985}. However, \hy\ shows a limited decrease of the flux in this region, which is consistent with its broad and slowly-declining $J$-band light curve (see Fig.~\ref{fig:uv_B_NIR_compare}).

 

\section{Model Results}
\label{sec:model}
In this section, we briefly discuss the physics of \hy, provide one possible interpretation of the data, and compare the data to explosion models.
A more in depth description of our modeling assumptions and methodology can be found in Appendix~\ref{appx:models}.

\subsection{{Envelope models}}
In this work, we employ the parameterized framework of spherical envelope
models \citep{Hoeflich96} and make use of the observed spectral evolution
and photometric properties of \hy to constrain model
parameters. This class of models has been previously shown to provide a consistent picture for the \superc\ object LSQ14fmg \citep{Hsiao2020} and the observed slow rise and decline of SN~2009dc \citep{Noebauer2016}.
The envelope models may be consistent with the core degenerate (CD) scenario \citep[e.g.,][]{Kashi11}.
The CD scenario consists of the explosion of a degenerate core within a nondegenerate envelope.  
In this scenario, the ignition process and thermonuclear explosion may be triggered by the merger between the core of an AGB star and a WD: 1) on dynamical time scale which results in a detonation \citep{Kashi11,Aznar15,Taubenberger17}, or 2) on secular time scales which may lead to an early deflagration phase with a transition to a detonation, subsequently called ``Deflagration-CD'' (DCD) \citep{Hoeflich2019}. We stress that in our models the nature of the envelope is unknown. It may be the nondegenerate outer envelope of an AGB star in the CD scenario, but we do not rule out other possibilities.

The main model parameters are: 
    the mass of the nondegenerate envelope $M_{\text{env}}$;   the
    radius of the nondegenerate envelope $R_{\text{env}}$;  the extent
    of the He and C layers in mass and velocity space;  the initial
    metallicity $Z$;  the mass of the hydrostatic, possibly rotating,
    core which is referred to as the core mass, $M_{\text{core}}$;
    the amount of mass burnt during the deflagration phase,
    $M_{\text{defl}}$, which is controlled by the transition density
    at which the 
    deflagration transforms into a  detonation $\rho_{\text{tr}}$; and 
    possible interaction with the nearby environment or wind.

The central density ($\rho_c$) is calculated using the equations of hydrostatic
equilibrium for models with $M_{\text{core}}$ up to $M_{\text{Ch}}$ \citep{hwt98}. 
For models with $M_{\text{core}}$ in excess of
$M_{\text{Ch}}$, we assume fast rotating cores that ignite at a
central density of $7\times10^9$~g~cm$^{-3}$ \citep{YL05a}. These densities are found
at the lower end of the classical delayed-detonation models and may be
attributed to He- or C-accreting WDs within the $M_{\text{Ch}}$ class
of explosions \citep{Telesco15,tiara15,Diamond18}. For these values of
$\rho_c$, approximately $0.08$~\Msun\ of electron rich iron-group elements are
produced.

To first order, specific observational parameters are directly linked
to model parameters. This produces a set of selection criteria that
allows us to determine the best matching model. The main criteria are:
$M_{\text{env}}$, which determines the final ``shell'' velocity indicated by
the Si line region formed in quasi statistical equilibrium (QSE)
\citep{Hoeflich96,Quimby07}; and $R_{\text{env}}$, which determines the width of
the shell. With increasing radius, the shell becomes more confined in
velocity space (Fig. \ref{fig:models}); and $M_{\text{core}}$
determines the diffusion time scales and therefore the rise time to
maximum light \citep{Hoeflich96,Dessart2014,Shen14}. The detonation in
a sub-$M_{\text{Ch}}$ ($M_{\text{core}}\sim$1~\Msun) model produces a rise time that
is too short, and models with $M_{\text{core}} = 1.8~M_\odot$ produce
a rise time that is too 
long by several days. The best agreement was found with a model of
$M_{\text{core}} =1.47~M_\odot$ that puts ASASSN-15hy near, but over $M_{\text{Ch}}$.

\begin{figure*}[ht]
\centering
\includegraphics[width=0.97\textwidth]{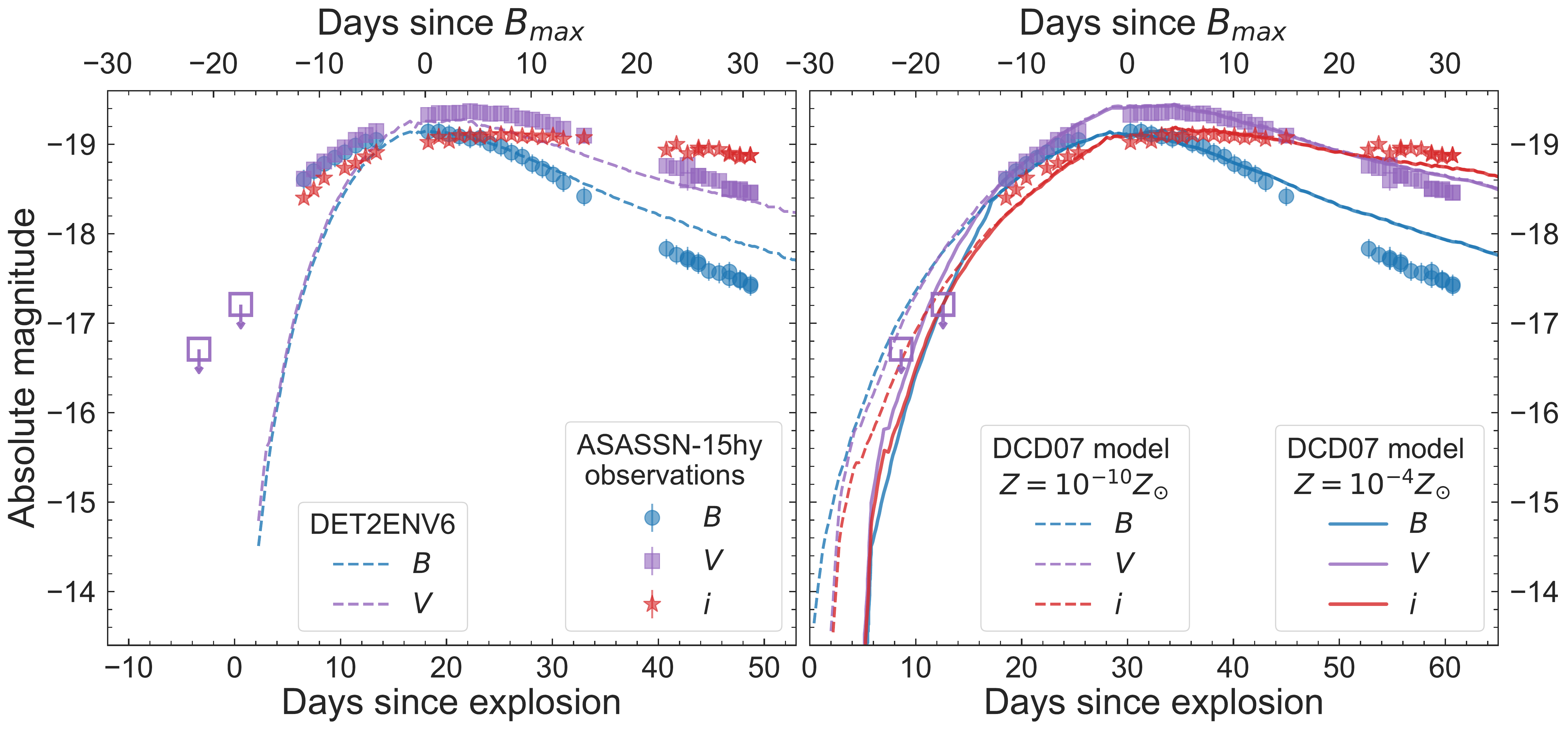}
\caption{Comparison of the absolute $BVi$-band light curves between models and \hy. 
The light curves are shown relative to explosion and $B$-band maximum. 
The observations have been $K$-corrected (except for the nondetection limits in open squares) and corrected for Galactic reddening.
\textit{Left panel}: The low-mass high-metallicity DET2ENV6 model ($M_{\rm{core}}=1.2~M_\odot$ and $Z=Z_\odot$) from \citet{Hoeflich96} is presented as an illustrative case. 
\textit{Right panel}: The best matching DCD07 model with $Z = 10^{-4}~Z_{\odot}$ (solid lines) is shown in comparison to the DCD07 model with $Z = 10^{-10}~Z_{\odot}$ (dashed lines).}
\label{fig:LC}
\end{figure*}

The secondary selection criteria in the case of DCD are:
$\rho_{\text{tr}}$ or $M_{\text{defl}}$, which regulates the $^{56}$Ni
production and therefore the luminosity, if all other parameters are
kept the same, in the same fashion as in the classical delayed detonation models
\citep[\eg][]{Hoeflich02}. 
In general within the CD scenario, the metallicity, $Z$ and the relative
amount of $M_{\text{He}}$ control the color $B-V$, the UV and NIR
flux, and the formation of CO.

\subsection{Best Matching Model}
\hy's relatively low peak luminosity and broad light curves place it into a specific, and somewhat unusual, area of parameter space.
As \hy\ has a low Si velocity at maximum light, which translates into a relative large $M_{\text{env}}$, previous sets of envelope models which detonate \citep{kmh93,Hoeflich96,Quimby07} predict a higher peak luminosity than majority of other \superc\ events and generally slower rise and decline \citep{Noebauer2016}. The secondary model criteria such as $M_{\text{defl}}$ and $Z$ may be the keys to explain the peculiarity of \hy.
The explanation under the DCD envelope model, the underlying physics, and the path to arrive at the
best matching parameters are presented in Appendix~\ref{appx:models}.
Here we summarize the main results.

The best matching model (a slightly modified version of DCD07, see Appendix~\ref{appx-model:LC_cons}) has the following parameters:
$M_{\text{env}}=0.7~M_\odot$, $R_{\text{env}}=10~R_{\text{core}}$,
$M_{\text{He,env}}=0.1~M_\odot$, $M_{\text{core}}=1.47~M_\odot$,
$Z=10^{-4}~Z_{\odot }$, $M_{\text{defl}}=0.42~M_\odot$, and a total
$^{56}$Ni mass $=$ $0.87~M_\odot$.  The large $M_{\text{core}}$ was
required to produce long diffusion time scales and a broad light
curve; a large $M_{\text{env}}$ is required to produce the low
observed ejecta velocities exemplified by \SiII; the low metallicity 
produces the long rise and lack of line blanketing relative to the normal SNe~Ia; and the low
transition density or large $M_{\text{defl}}$ ensures that the
\Nifs\ mass is lower than that which is produced in other
\superc\ objects and thus, the 
luminosity of the SN is low. Finally, a small $R_{\text{env}}$, which was determined by the velocity range of \SiII, produces
an extended density enhancement in the outer layers of the ejecta
rather than a confined density shell, which would be produced by larger
envelope radii. The chemical structure is largely consistent with the observed
velocity and velocity range of the ions (see Appendix~\ref{appx-model:structure} 
and \ref{appx-model:spectral_cons}). Note that up to 40\% of the total
luminosity comes in the form of hard X-rays and UV at early times in our
model, as discussed in Appendix
\ref{appx-model:LC_cons}.

In Fig.~\ref{fig:LC}, the effects of metallicity $Z$ and core mass $M_{\text{core}}$ are illustrated using the following models: the high $Z$ ($Z = \Zsun$) and low $M_{\text{core}}$ ($M_{\text{core}} = 1.2~\Msun$) DET2ENV6 model of \citet{Hoeflich96} in the left panel and our best matching DCD07 with $Z=10^{-4}~Z_{\odot}$ in comparison to DCD07 with $Z=10^{-10}~Z_{\odot}$ in the right panel.
The low-mass DET2ENV6 model has $0.6~\Msun$ of $M_{\text{env}}$, similar to the best matching parameter of \hy, but shows a rise time significantly shorter than the observation and a very asymmetric light curve.
Moreover, its high $Z$ leads to an optically thick C/O shell, and the low $M_{\text{core}}$ leads to a shorter rise time and faster evolution that is inconsistent with \hy. 
The DCD07 models ($M_{\text{core}}=1.47~M_\odot$), as shown in the right panel of Fig.~\ref{fig:LC}, provide overall matching light-curve shapes and luminosities as the observed ones.
Comparing the $Z=10^{-4}~Z_{\odot}$ and $Z=10^{-10}~Z_{\odot}$ DCD07 models shows a more extended dark phase as $Z$ increases, and the $Z=10^{-4}~Z_{\odot}$ model is more consistent with the early ASAS-SN nondetections.

\hy\ requires a rather large envelope mass that
has a significant amount of He left in the outer layers combined
with a low metallicity and a late deflagration to detonation
transition. 
The significant amount of helium and the low progenitor
metallicity below that of the host galaxy points to an old, low-mass
progenitor.  
Note that, despite the considerable amount of He, narrow He lines may not be detected due to the low opacity in the outer layer and gamma-ray trapping, such that the necessary gamma-rays to excite the He are not available \citep[e.g.,][]{Graham1988}.
\hy\ is not a typical \superc\ object 
compared to the current public sample.
An analysis on larger sample is needed to determine whether the model 
parameter space of \hy\ is common for \superc\ SNe or not.

\subsection{DCD Models in Context \label{sec:discussion}}

In the following discussion, we put \hy\ and our model in a larger context. 
The basic characteristics are the explosion of a degenerate core of
$\approx M_{\text{Ch}}$ embedded in a compact high-mass envelope with
a significant fraction of He $\approx 0.1-0.2~M_\odot$ and low
$Z$. The resulting explosion has a red color at maximum light for a
\superc\ object, a low luminosity, and took place in a low-metallicity
galaxy with ongoing star formation.

The best matching model has a low transition density from deflagration
to detonation (DDT) which results in an extend phase of the deflagration burning 
similar to the sublumious classical SNe~Ia models.
As discussed in Appendix~\ref{appx-model:LC_cons} a low metallicity and C/O ratio can be
expected to shift the DDT to low densities \citep{polundenko19}. 
Thus, the very low Z may be the key for understanding why
\hy\ is a  subluminous CD explosion.  

Our analysis suggests that \hy\ can be described by the DCD scenario
in which the explosion starts as a rather long deflagration phase and
turns into a detonation.  An initial deflagration requires a WD
companion merging with the  central
core of an AGB star on secular time scales rather than on dynamical
ones.  This is 
similar to classical delayed detonation $M_{\text{Ch}}$ models, which
have a long deflagration phase and low transition densities.  These
models may explain transitional and  subluminous SNe~Ia
\citep{Hoeflich02,Patat12,Hsiao15,Ashall16b,Ashall18,Galbany19}.

In normal SNe~Ia, a low transition density is rare and produces
intrinsically red optical colors similar to that observed in \hy.
Within the DCD framework,  subluminous objects that behave like \hy\ should
be rare  and the low metallicity suggests
that the progenitor may have been a Pop~II/III star.  Low-mass AGB stars
in metal-poor environments are carbon stars with highly enhanced Ba
and other s-process elements \citep{Aaronson80,Cohen2006,Kirby2015}.
Even though hydrogen is not apparent in the spectra and the
limits for ongoing mass loss are rather low, hydrogen and s-process
elements may be detectable when the ejecta have cleared the
low-density cocoon that has a size of a few tenths to a few light years
and was produced by prior mass-loss episodes \citep{dragulin2016}.  We do not
expect Balmer lines to appear right away, but when the envelope 
collides with the edge of the cocoon in $\sim 1-10$~yr. This
collision may lead to a revival of X-ray emission.

For degenerate cores (or WDs), the
C/O ratio increases with decreasing main sequence mass $M_{\text{MS}}$
as a result of stellar evolution.  At lower mass and hence lower
central temperatures, during the central stellar He burning, $\alpha$
capture on $^{12}$C dominates triple-$\alpha$, resulting in low C/O
ratios (C/O $\sim 0.1-0.2$).  The subsequent He-shell burning
 occurs at higher temperatures. 
Thus, shell burning results in ratios close to the statistical equilibrium
(C/O $\approx 1$).  Because of the size of the convective He burning core
decreases with $M_{\text{MS}}$ and $Z$ \citep{dominguez01}, the
overall C/O ratio also decreases with $M_{\text{MS}}$ and $Z$.  
The CD scenario requires an explosion during the AGB
phase. A common-envelope phase is more common for more massive stars. Stars
of mass $\sim 5-7~\Msun$ have evolutionary times of $\sim 10^8$ years
and may be expected to be related to starburst episodes.  
Although there is a strong correlation between low-metallicity,
starburst galaxies and \superc\ in general (L.~Galbany et al, in
preparation), the progenitor system of \hy\ may have been produced in a previous episode of star formation. 
This may indicate that \superc\ SNe come from a diverse group of progenitors.

\subsection{Alternative Scenarios}
Under the class of interaction or envelope models, several studies on \superc\ objects
have suggested that the origin of the envelope material could come from WD-WD mergers  \citep[e.g.,][]{Scalzo10,Hachinger12,Taubenberger2013,Noebauer2016}. Here we briefly discuss the alternative progenitor systems based on the parameter space described above. However, numerical simulations and comparisons are beyond the scope of this work.

The dynamical merger of two WDs may be an alternative possible scenario for
\hy\ and \superc\ objects \citep[e.g.,][]{Pakmor10,Pakmor12}.  This
scenario may produce a similar overall structure, light curves, and
spectra when compared to CDs of $\sim2~M_\odot$ that undergo
detonation burning.  However, it is not obvious how, in this
framework, one would produce an explosion with similar conditions as
\hy. Our models of \hy\ require an extended nondegenerate envelope
consisting of He/C/O without extensive mixing and high density outer
material, which favors a CD/AGB scenario. Furthermore, a long
deflagration phase is required to produce the low \Nifs\ mass, which
cannot be achieved in a dynamical merger, since these models
detonate. Finally a total ejecta mass exceeding the $M_{\text{Ch}}$ is
required to produce the observed light curves. 
This may disfavor the secular merger scenario \citep{Piersanti2003}, 
which explodes near $M_{\text{Ch}}$, as a viable scenario for \hy.

Though there is no polarimetric observations available for \hy, the lack of polarization in the other two \superc\ explosions, SN~2009dc \citep{Tanaka10} and SN~2007if \citep{Cikota19}, may disfavor explosion mechanisms with strong, large
scale asymmetries, such as dynamical mergers \citep{Bulla16a}. 
However, this does not exclude the possibility that a majority
of \superc\ SNe are the result of dynamical mergers.
\hy\ and LSQ14fmg \citep{Hsiao2020} are consistent with the DCD scenario, however, these two are on the extreme ends among the current \superc\ objects.
A larger sample of \superc\ events and larger scale of model comparison are needed for future analysis.

\section{Conclusion \label{sec:conclusion}}
Photometric and spectroscopic observations of \hy\ have been presented, covering $-$12 to +155~d relative to $B$-band maximum.
\hy\ has many similar characteristics to previously discovered \superc\ SNe~Ia. 
However, \hy\ also shows its own uniqueness and provides additional insight into this peculiar subgroup.

\hy\ has the following properties that are similar to other \superc\ SNe:
\begin{itemize}
    \item Slowly evolving (rise time = 22.5 $\pm$ 4.6~d in $V$-band) and broad light curves (\DmB\ = \hyDM),
    \item No distinct secondary maxima and peaking later than normal SNe~Ia in the $iYJH$ bands,
    \item A relatively blue UV$-$optical color ($uvm2-u$ = $1.72\pm0.16$~mag at $B$-band maximum),
    \item Bright peak luminosity in both UV (M$_{uvm2}=$~\hyMUVM) and NIR ($M_H \leq$~\hyMHlimit),
    \item Weak \CaII\ but strong and persistent \CII\ spectral feature,
    \item A relatively low \SiII\ velocity ($V_{\text{Si}}=8{,}600 \pm 200$~\kms\ at maximum light) and lacks an early time rapid velocity decline,
    \item No prominent $H$-band break feature in the first month past $B$-band maximum.
\end{itemize}

AMUSING observations show that the host of \hy\ is a low-mass, low-metallicity galaxy with a relatively young stellar population. These properties are  generally consistent with other \superc\ events. The host galaxy of \hy\ is barely visible in optical images, but in the H$\alpha$ map, the host is clearly visible.

Despite similarities mentioned above with previously published \superc\ SNe, \hy\ is unique in several ways. 
At optical wavelengths, it is dim compared with classical \superc\ events and even when compared with normal SNe~Ia with similar decline rates.
It has a peak absolute magnitude of M$_B=$~\hyBmaxAfterhost, which is a $\sim1$~mag dimmer than SN~2009dc and located below the LWR of normal SNe~Ia. This makes \hy\ the faintest SN among the current \superc\ SNe~Ia sample. 
\hy\ is also the reddest and has a observed $B-V$ color of \hyBmVatBmax\ at $B$-band maximum, which is $\sim$0.1~mag redder than the average of other \superc\ objects. 
Furthermore, \hy\ has the longest delayed onset of its $i$-band primary peak relative to that in the $B$ band at 7.2 $\pm$ 1.1~d.

We find that \hy\ is consistent with the ``deflagration core-degenerate” scenario, which consists of the secular merger of the degenerate core of an AGB star and a WD inside the nondegenerate envelope of the AGB star.
The best matching model has the parameters: $M_{\text{env}}=0.7~M_\odot$, $R_{\text{env}}=10~R_{\text{core}}$, $M_{\text{He,env}}=0.1~M_\odot$,  $M_{\text{core}}=1.47~M_\odot$, $Z=10^{-4}~Z_\odot$, $M_{\text{defl}}=0.42~M_\odot$, and total $M(^{56}\rm{Ni}) = 0.87~M_\odot$.
A large core mass is required in order to produce a broad light curve.
To produce a slow expansion velocity and a narrow IME shell, a large envelope mass is needed.  
A major portion of this envelope consists of He, which does not add to the line opacity but contributes to the hydrodynamics in the ejecta. 
Critically, a long deflagration phase is required to produce the observed low luminosity and red $B-V$ color. 
Finally, a metallicity of $Z\sim10^{-4}~Z_\odot$ is required to provide a small opacity and a dark phase on the rise, in order to match the early light curve and avoid producing significant line-blanketing to block the UV flux and over-redistribute the flux to the NIR.

The predominant host-galaxy type for \superc\ objects has been low mass, low metallicity and star-forming, including \hy, meaning that these objects may be more common at high redshifts. 
This has direct consequences for high-redshift dark energy experiments.
\superc\ SNe~Ia have similar light-curve shapes in the $B$ and $V$ bands compared to normal SNe~Ia and therefore cannot be distinguished at high redshifts in these rest-frame bands alone. 
Observations of redder bands in the rest frame, such as $i$ band to NIR band, are required to identify \superc\ SNe~Ia through purely photometric means.
Further studies on larger samples of \superc\ SNe~Ia are necessary to fully understand their impact on dark energy experiments.
Overall, more observations are required to test the consistency of the model simulations to understand the physics of the progenitors and the explosions.


\section*{Acknowledgments}
The authors would like to thank the anonymous referee for their useful comments.
We thank the technical, scientific staff, and support astronomers of the Las Campanas Observatory and the ESO Telescopes at the Paranal Observatory.
The CSP-II has been supported by NSF grants AST-1008343, AST-1613426, AST-1613455, AST-1613472, and the Danish Agency for Science and Technology and Innovation through a Sapere Aude Level 2 grant (PI: M.S.)
C.A. is supported by the NSF grant AST \#1908952.
J.L., S.K., M.S., and E.Y.H acknowledge the support provided by the Florida Space Research Program.
P.A.H. acknowledges the support by the National Science Foundation grant AST-1715133. 
L.G. acknowledges financial support from the Spanish Ministry of Science, Innovation and Universities (MICIU) under the 2019 Ram\'on y Cajal program RYC2019-027683 and from the Spanish MICIU project PID2020-115253GA-I00.
E.B. was supported in part by NASA grant 80NSSC20K0538.
M.D.S. is supported by generous grants from Villum FONDEN (13261, 28021) and by a project grant (8021-00170B) from the Independent Research Fund Denmark.
Support for T.W.-S.H. was provided by NASA through the NASA Hubble Fellowship grant HST-HF2-51458.001-A awarded by the Space Telescope Science Institute, which is operated by the Association of Universities for Research in Astronomy, Inc., for NASA, under contract NAS5-26555.
H.K. was funded by the Academy of Finland projects 324504 and 328898.
Support for J.L.P. is provided in part by ANID through the Fondecyt regular grant 1191038 and through the Millennium Science Initiative grant ICN12$\_$009, awarded to The Millennium Institute of Astrophysics, MAS.
Time domain research by D.J.S. is supported by NSF grants AST-1821987, 1813466, \& 1908972, and by the Heising-Simons Foundation under grant \#2020-1864.
This paper includes data gathered with the 6.5 meter Magellan Telescopes located at Las Campanas Observatory, Chile, and is also
based upon observations made with ESO Telescopes at the La Silla or Paranal Observatories under programme ID(s) 191.D-0935 and 099.D-0022(A).
This work has been partially supported by the Spanish grant PGC2018-095317-B-C21 within the European Funds for Regional Development (FEDER).
Finally, this paper is also includes observations obtained at the international Gemini Observatory (GN-2015A-Q-8, GS-2015A-Q-5), a program of NSF’s NOIRLab, which is managed by the Association of Universities for Research in Astronomy (AURA) under a cooperative agreement with the National Science Foundation. on behalf of the Gemini Observatory partnership: the National Science Foundation (United States), National Research Council (Canada), Agencia Nacional de Investigaci\'{o}n y Desarrollo (Chile), Ministerio de Ciencia, Tecnolog\'{i}a e Innovaci\'{o}n (Argentina), Minist\'{e}rio da Ci\^{e}ncia, Tecnologia, Inova\c{c}\~{o}es e Comunica\c{c}\~{o}es (Brazil), and Korea Astronomy and Space Science Institute (Republic of Korea).

$Facilities$: Swope (e2v), du Pont (RetroCam, WFCCD), $Swift$, ASAS-SN, VLT (MUSE), NOT (ALFOSC), 
LT (SPART), FLWO 1.5m (FAST), Baade (FIRE), GN (GNIRS), GS (F2).

$Software$: $SNooPy$ \citep{Burns11}; HYDRA \citep{Hoeflich_HYDRA_2003,Hoeflich_HYDRA_2009}.

\bibliographystyle{aasjournal}
{\footnotesize
\bibliography{ref_15hy,15hy_add}}


\appendix
\restartappendixnumbering 
\renewcommand{\thefigure}{A\arabic{figure}}
\renewcommand{\theHfigure}{A\arabic{figure}}

\section{Photometry and Observational Logs} \label{appendix:supplement}
\begin{figure}[h!]
\centering
\includegraphics[width=0.7\textwidth]{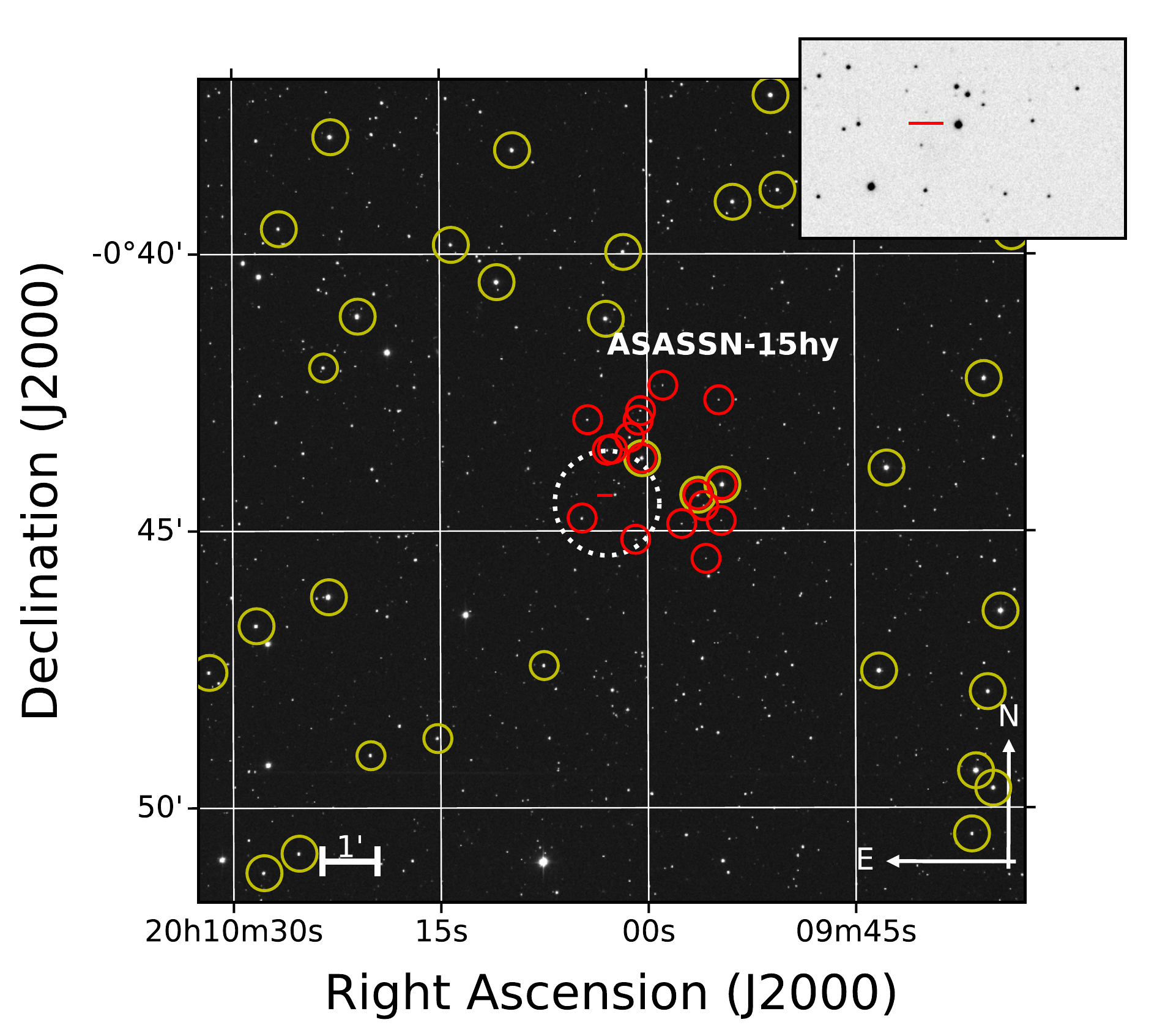}
\caption{The $i$-band finding chart of ASASSN-15hy. 
The optical (in yellow circles) and NIR (in red circles) local 
sequence stars are indicated. The location of the SN is labeled with a red line.
}
\label{fig:snimage}
\end{figure}

Figure~\ref{fig:snimage} presents the $i$-band finding chart of \hy with the positions of the local-sequence stars, used for CSP-II photometric observations at LCO. 
ASAS-SN $V$-band nondetection limits and photometric observations, in the standard system, are tabulated in Table~\ref{table:ASAS_lc}.
The CSP-II optical photometry of ASASSN-15hy is presented in the natural system (see Section 5.1 of \citealt{Krisciunas17}) of the Swope+e2v \citep{Phillips19} in Table~\ref{table:opt_lc}. 
The host subtracted $Swift$ UVOT photometry is shown in Table~\ref{table:swift_lc}.
The CSP-II NIR photometry is presented in the natural system of the du Pont+RetroCam in Table~\ref{table:nir_lc}. 

Among the previously published optical spectra, 
twelve were taken with KAST double spectrograph on the 3-m Shane telescope at the Lick Observatory and the LRIS \citep{Oke95} on the 10-m Keck-I telescope at the W.M. Keck Observatory and were published by \citet{Stahl20}. 
Another previously published series was obtained with the WiFeS instrument \citep{Dopita2007,Dopita2010} on the Australian National University 2.3-m telescope (ANU) as part of the ANU WiFeS SuperNovA Programme (AWSNAP, \citealt{Childress2016}). 
Note that red and blue spectra observed on the same night were merged, yielding nine spectra in total from WiFeS.
One classification spectrum taken with the EFOSC2 on the ESO New Technology Telescope (NTT) by the PESSTO program \citep{Smartt15} is also included and is publicly available on WISeREP \citep{Yaron12}.
A log of the spectroscopic observations of \hy\ is given in Table~\ref{table:specs_log}.
The phases in all tables are expressed in rest-frame phase in days with respect to $B$-band maximum JD = \hyTmax.

\renewcommand{\arraystretch}{1.2} 
\begin{deluxetable}{cccc}[H]
\centering
\tablecaption{ASAS-SN $V$-band photometry.\tablenotemark{a}\label{table:ASAS_lc}}
\tablehead{
\colhead{JD} & \colhead{UT Date} & \colhead{Phase} & \colhead{mag(err)}}
\startdata
2457100.2	& 2015-03-18 &	$-$50.5 & 	$>$17.11         \\
2457108.9	& 2015-03-27 &  $-$42.0 & 	$>$17.90         \\
2457115.9	& 2015-04-03 &  $-$35.1 & 	$>$16.73         \\
2457116.9	& 2015-04-04 &  $-$34.1 & 	$>$16.97         \\
2457117.9	& 2015-04-05 &  $-$33.1 & 	$>$17.19         \\
2457129.8	& 2015-04-17 &  $-$21.4 & 	$>$18.03         \\
2457133.8	& 2015-04-21 &  $-$17.5 & 	$>$17.53         \\
2457137.8	& 2015-04-25 &  $-$13.6 & 	16.42(0.10)      \\
2457138.8	& 2015-04-26 &  $-$12.6 & 	16.42(0.09)      \\
2457139.8	& 2015-04-27 &  $-$11.6 & 	16.12(0.08)      \\
2457143.9	& 2015-05-01 &  $-$7.6  &  15.90(0.06)       \\
2457147.8	& 2015-05-05 &  $-$3.8  &  15.52(0.12)       \\
2457154.8	& 2015-05-12 &  3.1	    &  15.37(0.05)       \\
2457155.8	& 2015-05-13 &  4.1	    &  15.51(0.05)       \\
2457156.1	& 2015-05-13 &  4.4	    &  15.42(0.05)       \\
2457157.1	&   2015-05-14 &    5.4	     & 15.42(0.05)   \\
2457158.1	&	2015-05-15 &	6.4		 & 15.48(0.05)   \\
2457160.8	&	2015-05-18 &	9.0		 & 15.48(0.06)   \\
2457164.7	&	2015-05-22 &	12.8	 & 15.51(0.06)   \\
2457168.9	&	2015-05-26 &	17.0	 & 15.88(0.06)   \\
2457173.9	&	2015-05-31 &	21.9	 & 15.96(0.09)   \\
2457181.8	&	2015-06-08 &	29.6	 & 16.30(0.10)   \\
2457182.7	&	2015-06-09 &	30.5	 & 16.52(0.14)   \\
2457185.0	&	2015-06-11 &	32.8	 & 16.59(0.13)   \\
2457186.8	&	2015-06-13 &	34.5	 & 16.45(0.09)   \\
2457189.7	&	2015-06-16 &	37.4	 & 16.57(0.13)   \\
2457190.7	&	2015-06-17 &	38.4	 & 16.33(0.09)   \\
2457199.1	&	2015-06-25 &	46.6	 & 16.72(0.07)   \\
2457228.1	&	2015-07-24 &	75.1	 & 17.48(0.13)   \\
2457244.2	&	2015-08-09 &	90.9	 & 17.65(0.15)   \\
\enddata
\tablenotetext{a}{Magnitudes in the standard system.}
\end{deluxetable}
\clearpage

\renewcommand{\arraystretch}{1.2}
\begin{deluxetable}{cccccccc}[h!]
\centering
\tablecaption{Swope e2v optical photometry of ASASSN-15hy.\tablenotemark{a}\label{table:opt_lc}}
\tablehead{
\colhead{JD} & \colhead{Phase} & \colhead{$u$} & \colhead{$g$} & \colhead{$B$} & \colhead{$V$} & \colhead{$r$} & \colhead{$i$}}
\startdata
2457139.9 & $-$11.5 & 16.33(0.01) & 16.12(0.01) & 16.21(0.01) & 16.11(0.01) & 16.05(0.01) & 16.17(0.01) \\
2457140.9 &	$-$10.6 & 16.27(0.01) & 16.03(0.01) & 16.11(0.01) & 16.01(0.01) & 15.96(0.01) & 16.08(0.01) \\
2457141.9 &	$-$9.6  & 16.22(0.01) & 15.94(0.01) & 16.03(0.01) & 15.93(0.01) & 15.82(0.01) & 15.95(0.01) \\
2457142.9 &	$-$8.6  & 16.18(0.01) & 15.86(0.01) & 15.96(0.01) & 15.85(0.01) & 15.78(0.02) & \dots \\
2457143.8 &	$-$7.6  & 16.18(0.01) & 15.81(0.01) & 15.90(0.01) & 15.78(0.01) & 15.69(0.01) & 15.84(0.01) \\
2457144.9 &	$-$6.6  & 16.14(0.01) & 15.74(0.01) & 15.83(0.01) & 15.69(0.01) & 15.61(0.01) & 15.79(0.01) \\
2457145.8 &	$-$5.7  & 16.13(0.01) & 15.70(0.01) & 15.79(0.01) & 15.64(0.01) & 15.59(0.01) & 15.71(0.01) \\
2457146.9 &	$-$4.7  & 16.14(0.01) & 15.65(0.01) & 15.77(0.01) & 15.60(0.01) & 15.48(0.01) & 15.67(0.01) \\
2457151.9 &	0.2   & 16.21(0.01) & 15.55(0.01) & 15.69(0.01) & 15.42(0.01) & 15.31(0.01) & 15.55(0.01) \\
2457152.9 &	1.2   & 16.25(0.01) & 15.53(0.01) & 15.69(0.01) & 15.40(0.01) & 15.29(0.01) & 15.49(0.01) \\
2457153.9 &	2.2   & 16.28(0.01) & 15.54(0.01) & 15.72(0.01) & 15.40(0.01) & 15.30(0.01) & 15.53(0.01) \\
2457154.9 &	3.2   & 16.32(0.01) & 15.56(0.01) & 15.75(0.01) & 15.40(0.01) & 15.25(0.01) & 15.47(0.01) \\
2457155.9 &	4.2   & 16.37(0.01) & 15.56(0.01) & 15.78(0.01) & 15.38(0.01) & 15.27(0.01) & 15.46(0.01) \\
2457156.9 &	5.1   & 16.40(0.03) & 15.60(0.01) & 15.78(0.02) & 15.40(0.02) & 15.26(0.01) & 15.48(0.01) \\
2457157.8 &	6.1   & 16.45(0.01) & 15.61(0.01) & 15.84(0.01) & 15.42(0.01) & 15.26(0.01) & 15.46(0.01) \\
2457158.9 &	7.1   & 16.55(0.01) & 15.65(0.01) & 15.88(0.01) & 15.42(0.01) & 15.27(0.01) & 15.45(0.01) \\
2457159.9 &	8.1   & 16.64(0.01) & 15.69(0.01) & 15.95(0.01) & 15.44(0.01) & 15.27(0.01) & 15.45(0.01) \\
2457160.9 &	9.1   & 16.72(0.01) & 15.75(0.01) & 16.00(0.01) & 15.48(0.01) & 15.30(0.01) & 15.47(0.01) \\
2457161.9 &	10.1  & 16.81(0.01) & 15.79(0.01) & 16.08(0.01) & 15.49(0.01) & 15.27(0.01) & 15.46(0.02) \\
2457162.9 &	11.0  & 16.91(0.01) & 15.85(0.01) & 16.14(0.01) & 15.52(0.01) & 15.32(0.01) & 15.48(0.01) \\
2457163.9 &	12.0  & 17.03(0.01) & 15.91(0.01) & 16.20(0.01) & 15.56(0.01) & 15.34(0.01) & 15.46(0.01) \\
2457164.9 &	13.0  & 17.14(0.01) & 15.97(0.01) & 16.29(0.01) & 15.61(0.01) & 15.35(0.01) & 15.50(0.01) \\
2457166.9 &	14.9  & 17.35(0.01) & 16.12(0.01) & 16.45(0.01) & 15.70(0.01) & 15.40(0.01) & 15.48(0.01) \\
2457174.8 &	22.7  & 18.15(0.03) & 16.64(0.01) & 17.05(0.01) & 16.06(0.01) & 15.62(0.01) & 15.62(0.01) \\
2457175.7 &	23.7  & 18.27(0.04) & 16.70(0.01) & 17.12(0.01) & 16.08(0.01) & 15.64(0.01) & 15.56(0.01) \\
2457176.8 &	24.7  & 18.25(0.16)	& 16.82(0.05) &	17.17(0.06)	& 16.14(0.07) &	15.63(0.02)	&	15.66(0.04) \\
2457177.9 &	25.7  & 18.39(0.05)	& 16.81(0.01) &	17.22(0.02)	& 16.18(0.01) &	15.72(0.01)	&	15.61(0.01) \\
2457178.8 &	26.7  & 18.48(0.03)	& 16.88(0.01) &	17.31(0.02)	& 16.21(0.01) &	15.74(0.01)	&	15.61(0.01) \\
2457179.8 &	27.7  & 18.58(0.11)	& 16.86(0.02) &	17.34(0.04)	& 16.23(0.01) &	15.77(0.01)	&	15.62(0.01) \\
2457180.8 &	28.7  & 18.43(0.06)	& 16.96(0.01) &	17.36(0.02)	& 16.33(0.02) &	15.80(0.02)	&	15.67(0.02) \\
2457181.9 &	29.7  & 18.60(0.03)	& 17.00(0.01) &	17.42(0.01)	& 16.34(0.01) &	15.84(0.01)	&	15.70(0.01) \\
2457182.9 &	30.7  & 18.65(0.03)	& 17.04(0.01) &	17.48(0.01)	& 16.37(0.01) &	15.86(0.01)	&	15.69(0.01) \\ 
2457249.6 &	96.2  & 19.71(0.04) & 18.08(0.01) & 18.48(0.01) & 17.69(0.01) & 17.53(0.01) & \dots \\
2457250.6 &	97.2  & 19.65(0.07) & 18.07(0.01) & 18.50(0.02) & 17.71(0.02) & 17.53(0.01) & 17.61(0.02) \\
\enddata
\tablenotetext{a}{Magnitudes in the Swope+e2v natural system.}
\end{deluxetable}

\renewcommand{\arraystretch}{0.91}
\begin{deluxetable}{cccccccc}[h!]
\tabletypesize{\small}
\centering
\small\addtolength{\tabcolsep}{-1pt}
\tablecaption{$Swift$ UVOT photometry of ASASSN-15hy.\tablenotemark{a} \label{table:swift_lc}}
\tablehead{
\colhead{JD} & \colhead{Phase} & \colhead{$uvw2$} & \colhead{$uvm2$} & \colhead{$uvw1$} &  \colhead{$u$} & \colhead{$b$} & \colhead{$v$}}
\startdata
2457142.1	&	$-$9.4	&	17.30(0.09)	&	17.32(0.09)	&	16.33(0.08)	&	15.20(0.06)	&	16.11(0.06)	&	15.98(0.08)	\\
2457145.3	&	$-$6.2	&	17.61(0.10)	&	17.35(0.09)	&	16.44(0.08)	&	15.26(0.06)	&	15.88(0.06)	&	15.67(0.08)	\\
2457147.3	&	$-$4.3	&	\dots	&	17.32(0.16)	&	16.32(0.08)	&	\dots	&	\dots	&	\dots	\\
2457147.4	&	$-$4.2	&	17.41(0.10)	&	17.29(0.09)	&	\dots	&	15.21(0.06)	&	15.76(0.06)	&	15.61(0.07)	\\
2457148.4	&	$-$3.2	&	17.53(0.10)	&	17.40(0.09)	&	16.47(0.08)	&	15.33(0.06)	&	15.88(0.06)	&	15.62(0.08)	\\
2457149.1	&	$-$2.5	&	\dots	&	\dots	&	\dots	&	15.28(0.06)	&	\dots	&	\dots	\\
2457149.2	&	$-$2.4	&	17.64(0.10)	&	17.51(0.09)	&	16.51(0.08)	&	15.32(0.06)	&	15.70(0.06)	&	15.55(0.07)	\\
2457149.5	&	$-$2.1	&	17.73(0.10)	&	\dots	&	16.54(0.08)	&	\dots	&	\dots	&	\dots	\\
2457149.6	&	$-$2.0	&	17.83(0.11)	&	17.36(0.09)	&	\dots	&	15.34(0.06)	&	15.89(0.06)	&	15.55(0.07)	\\
2457151.3	&	$-$0.3	&	17.76(0.10)	&	17.48(0.10)	&	16.50(0.08)	&	15.38(0.06)	&	15.72(0.06)	&	15.57(0.07)	\\
2457154.3	&	2.6	&	17.87(0.12)	&	17.91(0.10)	&	16.61(0.08)	&	15.55(0.07)	&	15.88(0.06)	&	15.45(0.07)	\\
2457155.3	&	3.6	&	18.08(0.11)	&	17.96(0.12)	&	16.83(0.08)	&	15.55(0.07)	&	15.81(0.06)	&	15.51(0.07)	\\
2457156.4	&	4.7	&	18.08(0.12)	&	17.99(0.11)	&	16.89(0.08)	&	15.63(0.07)	&	15.88(0.06)	&	15.45(0.07)	\\
2457157.0	&	5.3	&	18.17(0.12)	&	18.22(0.12)	&	16.97(0.09)	&	15.72(0.06)	&	15.95(0.06)	&	15.42(0.07)	\\
2457158.4	&	6.6	&	18.23(0.15)	&	18.15(0.14)	&	16.91(0.10)	&	15.73(0.08)	&	16.04(0.07)	&	15.43(0.08)	\\
2457170.4	&	18.4	&	19.35(0.33)	&	\dots	&	18.22(0.21)	&	17.02(0.14)	&	17.04(0.10)	&	15.82(0.10)	\\
2457173.0	&	21.0	&	\dots	&	20.01(0.36)	&	18.76(0.17)	&	17.29(0.10)	&	17.03(0.07)	&	16.04(0.10)	\\
2457175.6	&	23.5	&	\dots	&	\dots	&	19.05(0.24)	&	17.38(0.12)	&	17.30(0.09)	&	16.13(0.09)	\\
2457181.6	&	29.4	&	\dots	&	\dots	&	19.19(0.21)	&	17.98(0.13)	&	17.59(0.08)	&	16.39(0.08)	\\
2457181.7	&	29.5	&	\dots	&	20.31(0.31)	&	\dots	&	\dots	&	\dots	&	\dots	\\
2457183.9	&	31.7	&	\dots	&	\dots	&	19.39(0.29)	&	17.83(0.14)	&	17.64(0.10)	&	16.48(0.10)	\\
2457185.2	&	33.0	&	\dots	&	\dots	&	19.32(0.24)	&	17.97(0.14)	&	17.68(0.09)	&	\dots	\\
2457185.3	&	33.1	&	\dots	&	\dots	&	\dots	&	\dots	&	\dots	&	16.40(0.13)	\\
2457187.8	&	35.5	&	\dots	&	\dots	&	\dots	&	18.26(0.16)	&	17.75(0.10)	&	16.51(0.09)	\\
2457191.3	&	38.9	&	\dots	&	\dots	&	19.29(0.23)	&	18.27(0.16)	&	17.72(0.09)	&	16.59(0.09)	\\
2457192.6	&	40.2	&	\dots	&	\dots	&	\dots	&	18.52(0.29)	&	18.11(0.16)	&	16.97(0.17)	\\
2457197.0	&	44.5	&	\dots	&	\dots	&	\dots	&	18.49(0.19)	&	17.77(0.10)	&	16.84(0.11)	\\
\enddata
\tablenotetext{a}{Magnitudes are host subtracted.}
\end{deluxetable}

\onecolumngrid


\begin{deluxetable}{cccccccc}[h!]
\centering
\addtolength{\tabcolsep}{2pt}
\tablecaption{du Pont RetroCam NIR photometry.\tablenotemark{a}\label{table:nir_lc}}
\tablehead{
\colhead{JD} & \colhead{Phase} & \colhead{$Y$} & \colhead{$J$} & \colhead{$H$}}
\startdata
2457139.9 & $-$11.5	 & 15.97(0.01) & 15.82(0.01) & 15.83(0.02) \\
2457140.9 & $-$10.5	 & 15.88(0.01) & 15.71(0.01) & 15.75(0.02) \\
2457141.8 & $-$9.6	 & 15.77(0.01) & 15.63(0.01) & 15.61(0.03) \\
2457142.9 & $-$8.6	 & 15.71(0.01) & 15.54(0.01) & 15.55(0.03) \\
2457143.9 & $-$7.6	 & 15.65(0.01) & 15.49(0.01) & 15.47(0.03) \\
2457151.9 & 0.3	     & 15.37(0.01) & 15.16(0.01) & 15.19(0.02) \\
2457152.8 & 1.2	     & 15.35(0.01) & 15.17(0.01) & 15.17(0.02) \\
2457153.8 & 2.2	     & 15.35(0.01) & 15.16(0.01) & 15.14(0.02) \\
2457170.8 & 18.8	 & 15.09(0.01) & 15.32(0.01) & 15.03(0.01) \\
2457172.8 & 20.8	 & 15.06(0.01) & 15.34(0.01) & 15.04(0.01) \\
2457173.8 & 21.8	 & 15.03(0.01) & 15.34(0.01) & 15.02(0.02) \\
\enddata
\tablenotetext{a}{Magnitudes in the du Pont+RetroCam natural system.}
\end{deluxetable}

\renewcommand{\arraystretch}{0.94} 
\begin{deluxetable}{lrcccc}[h!]
\tabletypesize{\small}
\centering
\addtolength{\tabcolsep}{1.pt}
\caption{Spectroscopic observations of ASASSN-15hy.\label{table:specs_log}} 
\tablehead{
\colhead{JD} & \colhead{UT Date} & \colhead{Phase} & \colhead{Telescope} & \colhead{Instrument} & \colhead{Ref.\tablenotemark{a}}}
\startdata
\multicolumn{6}{c}{\textbf{Optical}} \\
\hline
2457139.9 & 2015-04-27 & $-$11.6 & ESO-NTT     & EFOSC2      & 1    \\
2457140.0 & 2015-04-27 & $-$11.4 & Shane     & KAST          & 2    \\
2457142.2 & 2015-04-29 & $-$9.3  & NOT         & ALFOSC      & 3    \\
2457142.2 & 2015-04-29 & $-$9.2  & ANU    & WiFeS     		 & 4    \\
2457146.3 & 2015-05-03 & $-$5.2  & ANU    & WiFeS     		 & 4    \\
2457148.7 & 2015-05-06 & $-$2.9  & LT   & SPRAT     		 & 3    \\
2457150.3 & 2015-05-07 & $-$1.3  & ANU    & WiFeS      		 & 4    \\
2457150.7 & 2015-05-08 & $-$0.9  & LT   & SPRAT     		 & 3    \\
2457151.7 & 2015-05-09 & 0.0	  & LT   & SPRAT    		 & 3    \\
2457153.7 & 2015-05-11 & 2.0	  & LT   & SPRAT    	     & 3    \\
2457155.0 & 2015-05-12 & 3.3   & FLWO 1.5m & FAST      		 & 3    \\
2457155.9 & 2015-05-13 & 4.2	  & FLWO 1.5m & FAST         & 3    \\
2457156.2 & 2015-05-13 & 4.5	  & ANU    & WiFeS     		 & 4    \\
2457160.9 & 2015-05-18 & 9.1   & FLWO 1.5m & FAST       	 & 3    \\
2457161.9 & 2015-05-19 & 10.1  & FLWO 1.5m & FAST            & 3    \\
2457187.0 & 2015-06-13 & 34.7  & Shane     & KAST            & 2    \\
2457189.5 & 2015-06-16 & 37.2  & Keck-I       & LRIS         & 2    \\
2457191.1 & 2015-06-17 & 38.7  & NOT         & ALFOSC    	 & 3    \\
2457192.0 & 2015-06-18 & 39.6  & Shane     & KAST       	 & 2    \\
2457194.9 & 2015-06-21 & 42.5  & FLWO 1.5m & FAST      	     & 3    \\
2457196.9 & 2015-06-23 & 44.5  & Shane     & KAST      		 & 2    \\
2457198.2 & 2015-06-24 & 45.7  & ANU    & WiFeS     		 & 4    \\
2457205.1 & 2015-07-01 & 52.5  & ANU    & WiFeS     		 & 4    \\
2457212.1 & 2015-07-08 & 59.4  & ANU    & WiFeS     		 & 4    \\
2457217.9 & 2015-07-14 & 65.1  & Shane     & KAST    	     & 2    \\
2457225.5 & 2015-07-22 & 72.5  & du Pont   & WFCCD   	     & 3    \\
2457227.9 & 2015-07-24 & 74.9  & Shane     & KAST    	     & 2    \\
2457234.9 & 2015-07-31 & 81.8  & ANU    & WiFeS     		 & 4    \\
2457252.0 & 2015-08-17 & 98.6  & ANU    & WiFeS     		 & 4    \\
2457258.9 & 2015-08-24 & 105.3 & Shane     & KAST     	     & 2    \\
2457286.7 & 2015-09-21 & 132.6 & Shane     & KAST       	 & 2    \\
2457305.6 & 2015-10-10 & 151.2 & Shane     & KAST      		 & 2    \\
2457306.7 & 2015-10-11 & 152.3 & Shane     & KAST      	 	 & 2    \\
2457309.7 & 2015-10-14 & 155.2 & Shane     & KAST       	 & 2    \\
\hline
\multicolumn{6}{c}{\bf Near-infrared}\\ 
\hline
2457143.0 & 2015-04-30 &  $-$8.5  & GN  & GNIRS             & 3    \\
2457150.1 & 2015-05-07 &  $-$1.5  & GN  & GNIRS             & 3    \\
2457159.9 & 2015-05-17 &   8.1  & GN  & GNIRS               & 3    \\
2457174.9 & 2015-06-01 &  22.8  & Baade & FIRE           & 3    \\
2457181.7 & 2015-06-08 &  29.5  & GS  & F2                  & 3    \\
2457224.7 & 2015-07-21 &  71.7  & GS  & F2                  & 3    \\
\enddata
\tablenotetext{a}{1: \citealt{Frohmaier15}; 2: \citealt{Stahl20}; 3: this work; 4: \citealt{Childress2016}.}
\end{deluxetable} 

\onecolumngrid

\restartappendixnumbering 
\renewcommand{\thefigure}{B\arabic{figure}}
\renewcommand{\theHfigure}{B\arabic{figure}}

\section{$K$- and $S$-corrections} \label{appendix:kcorr}
In this work, $K$-corrections \citep{Oke1968} were performed in the optical $BVri$ bands for all the comparison SNe~Ia presented in Section~\ref{sec:photometry}.
Note that for SN~2007if and SN~2012dn, the SNIFS absolute-flux-calibrated spectra \citep{Scalzo10,Taubenberger19} were used to calculate the magnitudes by integrating the flux directly under the CSP Swope+e2v passbands, hence no $K$- or $S$-corrections were needed. 
The $K$-corrections of SN~2007af, SN~2006bt and LSQ12gdj were computed using SNooPy, which uses the Hsiao templates \citep{Hsiao07} for SNe Ia.

For other peculiar \superc\ objects, such as SN~2006gz, SN~2009dc, and ASASSN-15pz, the spectral time series of SN~2009dc \citep{Silverman11,Taubenberger11}, SN~2012dn \citep{Taubenberger19}, and the Hsiao normal SN~Ia templates \citep{Hsiao07} were used as different options for the assumed SED in order to determine the $K$-corrections and the corresponding errors. 
The following procedures were followed for each SN~Ia:
(1) warp all the spectral sets (SN~2009dc, SN~2012dn, Hsiao templates, and the observed spectra of the SN~Ia in question) to match the observed photometric colors of the SN~Ia;
(2) calculate the $K$-correction using the warped spectra from the above step; 
(3) interpolate the $K$-corrections to the photometric epochs with Gaussian process \citep{GP_book} and with constant extrapolation if the desired epoch is outside the covered range;
(4) choose the SED option that produces the smallest $\chi^2$ in $K$-correction values compared to the ones obtained from the observed spectra of the SN~Ia in question, in order to produce the final $K$-correction;
and (5) at each photometric epoch, adopt the standard deviation of the $K$-correction from all SED options as the final $K$-correction errors. 
Following the above procedures, the optimal SED set for SN~2006gz, SN~2009dc, and ASASSN-15pz were spectral time series of SN~2012dn, SN~2009dc, and SN~2009dc, respectively.

For photometry published in the natural system with known filter response functions, such as that obtained by the Harvard-Smithsonian Center for Astrophysics (CfA) Supernova Group \citep{Hicken07,Hicken2012}, the light curves are also $S$-corrected to the CSP-II Swope+e2v natural system.
For photometry published in the standard systems, the light curves are left as published without removing or applying color terms, such as the Las Cumbres Observatory $BVri$ light curves of ASASSN-15pz from \citet{Chen19} and the multi-instrument S-corrected $BV$ light curves of SN~2009dc from \citet{Taubenberger11}.
When computing the $K$-corrections in step 2), the rest-frame filters were set to be the CSP-II Swope+e2v $BVri$ filters.

\begin{figure}[!h]
\centering
\includegraphics[scale=0.38]{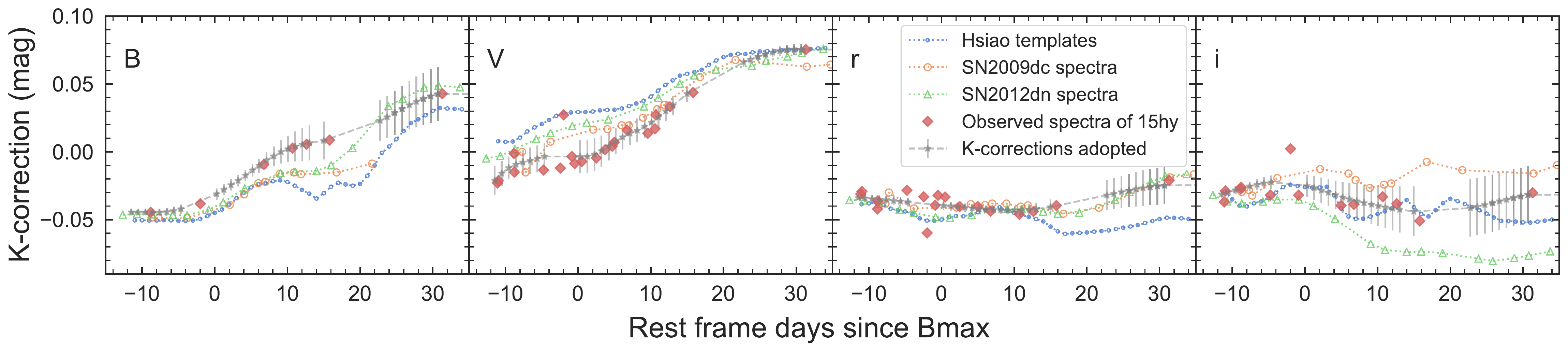}
\caption{The comparison of $K$-correction estimates for ASASSN-15hy, using various spectral time series (SN~2009dc, SN~2012dn, Hsiao normal SN~Ia templates, and observed spectra of \hy) as the assumed SED.
The final adopted $K$-correction values and the corresponding errors are also plotted. 
}
\label{fig:kcorr}
\end{figure}

The optical spectroscopic data set of \hy\ are quite complete in time coverage and rapid in cadence. 
Thus, it forms the base set of SEDs for the $K$-correction calculations for \hy.
Figure~\ref{fig:kcorr} presents the comparison of $K$-corrections computed with various SED options for \hy.
The different sets of $K$-corrections are similar at early times but diverge after +10~d past maximum light, especially in the $i$ band where \superc\ SNe~Ia show peculiar light-curve evolution compared to normal SNe~Ia.
This divergence is captured in the adopted $K$-correction errors.
Among four \superc\ objects, SN~2006gz, SN~2009dc, ASASSN-15pz, ASASSN-15hy, the average adopted $K$-correction error is 0.01 mag. 
This is on the same level as the errors attributed to the diversity of spectroscopic features in normal SNe~Ia \citep{Hsiao07}.

\restartappendixnumbering 
\renewcommand{\thefigure}{C\arabic{figure}}
\renewcommand{\theHfigure}{C\arabic{figure}}
\renewcommand{\thetable}{C\arabic{table}}
\renewcommand{\theHtable}{C\arabic{table}}

\section{Host-Galaxy Analysis}
\subsection{Integral-field Spectroscopy}
\label{appendix:host}
The analysis of the MUSE integral-field spectroscopy of the \hy\ host is detailed here, while the main results were presented in Section~\ref{subsec:host}.
First, the MW extinction-corrected maps of the most prominent emission lines were constructed.
Spectra were then extracted using a 1.6\arcsec\ diameter aperture, similar to the FWHM of the observation, and centered at two positions: one at the brightest nearby host H$\alpha$ structures toward the SE of the SN (marked as Structure A in the right panel of Fig.~\ref{fig:host_white}) and the other at the position of \hy.  
A simple stellar populations analysis was performed on these two spectra and on the integrated global spectrum of the galaxy by summing up the spectra in all spaxels while accounting for spatial covariance in the error budget.
Several galactic parameters were obtained from the three spectra including the ongoing SFR determined from the H$\alpha$ emission line, the ratio of young-to-old populations determined from the H$\alpha$ EW, and the oxygen abundance (12 + $log_{10}(O/H)$) in both the O3N2 \citep{2013A&A...559A.114M} and D16 \citep{2016Ap&SS.361...61D} scales.

The spectrum extracted at the location of \hy\ has a low S/N with a very noisy stellar continuum. 
The Balmer H$\alpha$ and $H\beta$ lines were detected, and they indicate the presence of gas ionized by ongoing star formation at that location.
This location show subsolar oxygen abundance implied by both O3N2 and D16 calibrators and relatively high sSFR.
From the spectrum extracted at the location of Structure A, the oxygen abundance is substantially subsolar.
The oxygen line at $\lambda$5007 is as bright as H$\alpha$, pointing to very low metal content. 
This is supported by the location of the line ratios in the BPT diagram \citep{1981PASP...93....5B}, falling below the \cite{2001ApJ...556..121K} demarcation line at the edge where low-metallicity regions fall, similar to the host of LSQ14fmg \citep{Hsiao2020}.
The H$\alpha$EW of Structure A is relatively high, indicating a significant contribution from populations as young as $\sim$10 Myr (See \citealt{2018A&A...613A..35K} for example).
Regarding global properties, we found a SFR of 0.048~$\pm$~0.007 M$_\odot$ yr$^{-1}$ and a stellar mass of (0.77~$\pm$~0.16) $\times$ 10$^9$~M$_\odot$, corresponding to a log$_{10}$(sSFR) of $-$10.21~$\pm$~0.31~yr$^{-1}$.
The global spectrum also yielded relatively high H$\alpha$EW and low subsolar oxygen abundances in both the O3N2 and D16 calibrators.
As discussed in Section~\ref{subsec:host_env}, such low-mass and low-metallicity  host environment are typical for the \superc\ SNe.  Analysis with a larger sample are underway (L.~Galbany et al, in preparation).

\subsection{Detection Limit of Na~{\sc i}~D}
\label{appendix:NaID}


A detection limit of EW $\le$~0.1~\AA\ was estimated using the WiFeS spectrum taken on 2015 May 13, selected for its relatively high resolution and S/N.
Here, we outline the procedures for estimating the detection limit.
First, an idealized spectrum was created by Gaussian smoothing and oversampling the observed WiFeS spectrum. 
Artificial absorption lines with varying depths and widths, simulating the narrow Na~{\sc i}~D absorptions, were then added to the idealized spectrum. 
The resolution was degraded to match the resolution of WiFeS with the R3000 grating. 
The spectrum was then resampled at the same wavelength grid as the observed WiFeS spectrum. 
Finally, random flux noise was added using the uncertainties measured from the observed spectrum. 

The strength of the simulated Na~{\sc i}~D absorption was decreased until the EW measured from the degraded spectrum was on the same level as the $1\sigma$ EW measurement error determined through 100 realizations of degraded spectra. 
The EW measured using the corresponding idealized spectrum is then taken as the detection limit. 
We estimated for ASASSN-15hy that the Na~{\sc i}~D EW is 0.0 $\pm$ 0.1~\AA. 
Note that the derived detection limit is robust against varying width of the simulated Na~{\sc i}~D absorption, as well as resolution further degraded by seeing, for example, by a factor of 2.

\subsection{Comparisons of Spectral Energy Distribution}
\label{appendix:SED}

\begin{figure}[h!]
\centering
\subfigure{\includegraphics[width=0.9\textwidth]{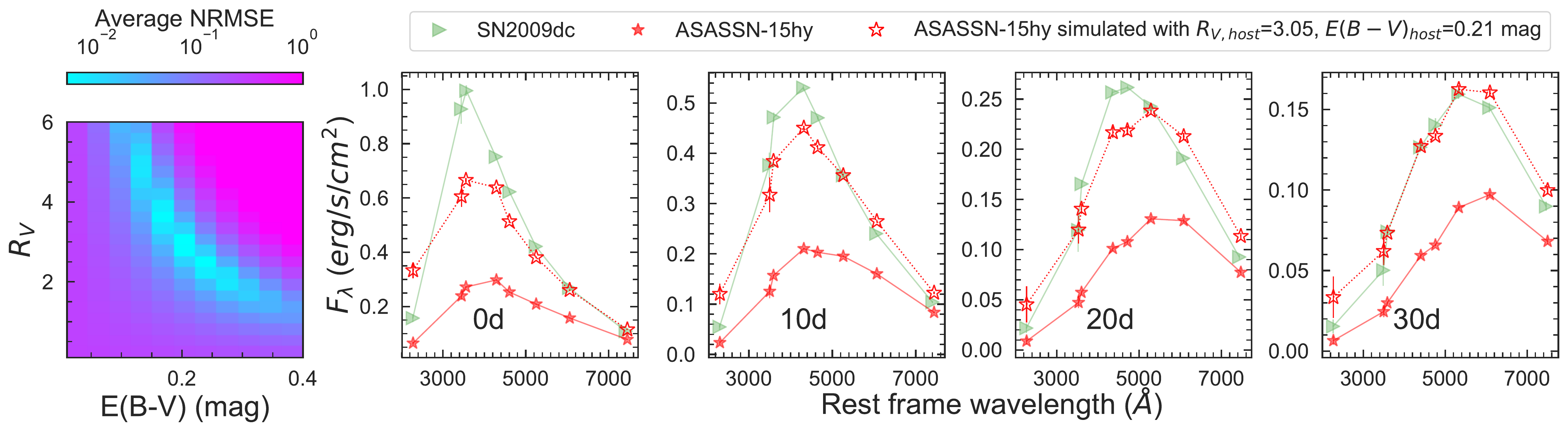}}
\subfigure{\includegraphics[width=0.9\textwidth]{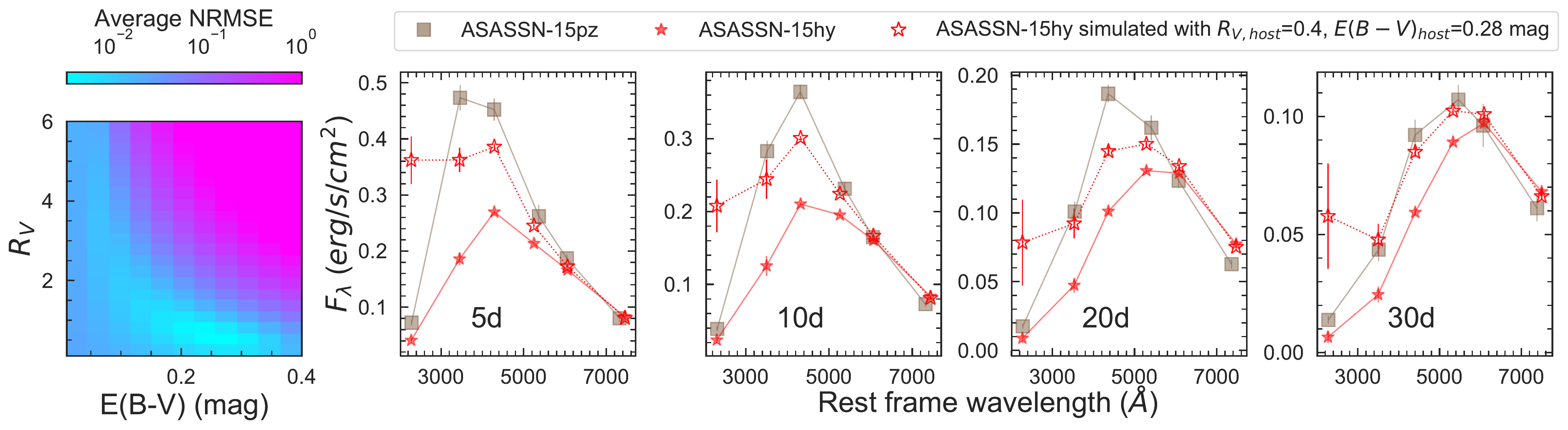}}
\caption{The host-galaxy extinction simulation of \hy\ by matching the SED to SN~2009dc (top row, using $Swift$ $uvm2~u$ and CSP $uBgVri$ bands) and ASASSN-15pz (bottom row, using $Swift$ $uvm2~u$ and Las Cumbres Observatory $BVri$ bands).
The left-hand panels of each row show the averaged NRMSE between the reference SED and simulated SED of \hy\ with corresponding extinction parameters, ratio $R_V$ and color excess $E(B-V)$, ranging from 0.1 to 6.0 and 0.01 to 0.04~mag respectively. 
Note that it would require unusually high UV flux if \hy\ were to have a similar flux as SN~2009dc/ASASSN-15pz in the optical.}
\label{fig:SED_test}
\end{figure}

By comparing the SED (covering from $\sim$2000 to $\sim$8000~\AA) of \hy\ to those of well-studied \superc\ SNe, we further investigate whether the relative faintness of \hy\ in the optical is due to host-galaxy extinction. 
SN~2009dc \citep{Taubenberger11,Krisciunas17} and ASASSN-15pz \citep{Chen19} were chosen for this test due to their similarity in spectroscopic properties with \hy\ and the availability of photometry in the UV. They have host color excess $E(B-V)$ values of 0.1 \citep{Taubenberger11} and 0.0~mag \citep{Chen19}, determined by the EW of Na~{\sc i}~D, from the corresponding literature.
The SED of \hy\ is extinction and reddening ``corrected" to match those of SN~2009dc and ASASSN-15pz. 
The best matching extinction parameters, $R_V$ and $E(B-V)$, were selected by locating the minimum of the averaged Normalized Root Mean Square Error (NRMSE) between the comparison SEDs over the selected four epochs presented in Fig.~\ref{fig:SED_test}.
Note that the weight of the $uvm2$ was set to 0.05 when calculating the NRMSE to avoid bias due to the significant extinction in the UV.

Unusually high intrinsic UV flux would be required for \hy\ if it is to match the overall SED of the reference SNe, especially in the optical region, with the best-fit $R_V$ and $E(B-V)$.
The derived intrinsic SED of \hy\ has an average of 2 and 5 times of the flux in the $uvm2$ band compared to those of SN~2009dc and ASASSN-15pz respectively, which translates to $\sim$0.8 and $\sim$1.7~mag brighter in magnitudes.
Such high UV flux is not likely as SN~2009dc and ASASSN-15pz are already UV bright compared to normal SNe~Ia \citep{Taubenberger11,Brown14,Chen19}.
Given the above result from the SED extinction test and the minimal extinction derived from the nondetection of Na~{\sc i}~D, we suggest that the apparent differences in the SEDs are intrinsic to the SNe or their immediate surrounding environment.

\restartappendixnumbering 
\renewcommand{\thefigure}{D\arabic{figure}}
\renewcommand{\theHfigure}{D\arabic{figure}}

\section{Model Selection in the Core Degenerate Scenario}
\label{appx:models}

In the context of the CD scenario, we explain how the photometric and
spectral properties of \hy\ translate to our chosen model parameters in details in this section, with the main results summarized in Section~\ref{sec:model}.
Our goal is to show how we obtain our best fit model parameters and
not to present an overall model grid for the CD scenario.
In a star undergoing a common-envelope phase, we would expect that the hydrogen envelope is most stripped. We note however, that every AGB star removes its envelope in the final stages such that it forms a naked C/O core that later is seen as a white dwarf.
One of the goals of this investigation is to probe for the ignition mechanism in \hy.

\subsection{Model Construction and Free Parameters} \label{appx-model:parameters}
In this work, we use the parameterized framework of spherical envelope
models \citep{Hoeflich96} and make use of the spectral evolution of
elements and photometric properties to constrain model
parameters. Similar models were presented  for the \superc\ object LSQ14fmg \citep{Hsiao2020}. 
Here we list the main parameters and selection criteria as mentioned in Section~\ref{sec:model}.
The main parameters are:

\begin{enumerate}
    \item  the mass of the nondegenerate envelope $M_{\text{env}}$,
    \item  the radius of the nondegenerate envelope $R_{\text{env}}$,
    \item  the size of the He and C layers in mass and velocity space,
    \item  the initial metallicity $Z$,
    \item the mass of the hydrostatic (possibly rotating) core; referred to as the core mass, $M_{\text{core}}$,
    \item  the amount of mass burnt in the deflagration phase, $M_{\text{defl}}$
    or the transition density from the deflagration to detonation burning $\rho_{\text{tr}}$,
    \item possible interaction with the nearby environment or wind.
\end{enumerate}
 
The parameters of the presented models are determined by the observational constraints. 
The primary selection criteria allow us to determine the model parameter based on the observation properties are: 
\begin{enumerate}
    \item $M_{\text{env}}$ determines the final ``shell'' velocity indicated by the $Si$ line region formed in quasi statistical equilibrium (QSE; \citealt{Hoeflich96,Quimby07}),
    \item $R_{\text{env}}$ determines the width of the shell. With increasing radius, the shell becomes more confined in velocity space (Fig. \ref{fig:models}),
    \item $M_{\text{core}}$ determines the diffusion time scales, therefore, the rise time to maximum light \citep{Hoeflich96,Shen14,Dessart2014}. The detonation in a sub-$M_{\text{Ch}}$ ($\sim$1\Msun) model produces a rise time that is too short, and models with $1.8 M_\odot$ produces one that is too long by several days. The best agreement was found with a model of 1.47 $M_\odot$ that puts ASASSN-15hy near but over $M_{\text{Ch}}$.
\end{enumerate}

The secondary selection criteria in the case of DCD are: {\sl $\rho_{\text{tr}}$ or $M_{\text{defl}}$} which regulates the $^{56}$Ni production and therefore the luminosity, if all other parameters are kept the same, as in the classical delayed detonation models \citep[\eg][]{Hoeflich02}.
Note that in general within the CD scenario, Z and the relative amount of $M_{\text{He}}$ controls the color $B-V$, the UV and NIR flux, and the CO-formation.

\subsection{Structure of the Explosion Models} \label{appx-model:structure}

For the simulations, we used our HYDrodynamical RAdiation code (HYDRA) which utilizes a nuclear network of 218 isotopes during the early phases of the explosion and detailed, time-dependent non-LTE models for atomic level populations and diatomic molecules \citep{Sharp90,Hoeflich_KW1995}, including $\gamma$- and positron transport and radiation-hydrodynamics to calculate LCs and spectra \citep{hoeflich2003hydra,PH14,Hoeflich17}. 
 Here, 912 to 1812 depth points are used. For possible interaction, a
 recent module is used based on the work presented
 in \citet{gerardy07,dragulin2016}, and \citet{Hsiao2020}, covering a
 larger parameter space than existing published
 models \citep{Hoeflich96,Hsiao2020}. The ranges covered are
 $1.2~\Msun \le M_{\text{core}}\le  1.8~M_\odot$, where we assume that the
 rotationally supported degenerate core follows the solutions of \citet{YL05a};
$0.01~\Msun \le M_{\text{env}} \le 0.8$~\Msun; $5~R_{\text{core}}\le R_{\text{env}}\le 250~R_{\text{core}}$; $10^{-10}~\Zsun \le Z \le 1~Z_\odot$.  
The metallicity range is chosen to cover the range from extreme Population III to Population I stars. 
Only the baseline and  best matching model are presented here, but the
entire set of models will be discussed and applied to analyze a sample
of \superc\ SNe~Ia in future work.

\begin{figure}[tb!]
\centering
\includegraphics[width=\textwidth]{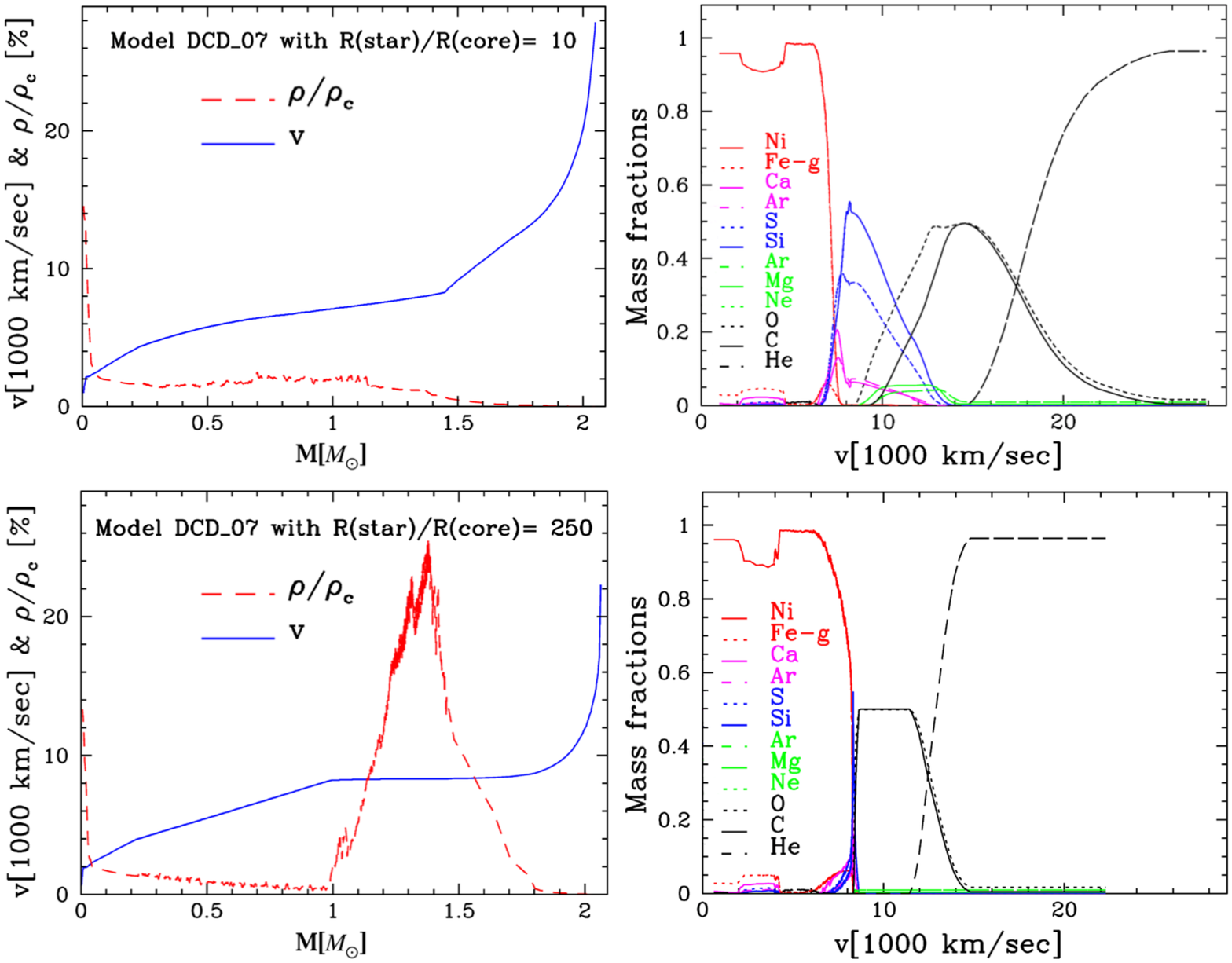}
\caption{Structure of the DCD07 envelope model for \hy\ (upper panels) and, for comparison, a model with the same parameters but an envelope 25 times its size (lower panels). 
The density and velocity structure is given as a function of mass (left panels), and the abundances of the $\alpha$ elements are given as a function of velocity (right panels). 
The model with the compact envelope shows no pronounced high-density shell and the IME, though confined, occupies a wider velocity range. 
In contrast, a large $R_{\text{env}}$ results in IME confined to a narrow velocity range (see text). 
Note that the bump in $^{56}$Ni corresponds to the refraction wave produced by the DDT transition as is an artifact of the spherical models.
}
\label{fig:models}
\end{figure}

Although the elemental contribution of a specific feature changes with time, we use the models to identify the velocity range in which specific elements are present.
Note that the model parameters have not been fine-tuned, and that Rayleigh-Taylor instabilities are suppressed in spherical models, but may smear-out the edges in velocity space by about $1{,}000$~\kms. 
The baseline model (DCD07) has the following parameters: $M_{\text{env}}=0.7~M_\odot$, $R_{\text{env}}=10~R_{\text{core}}$, $M_{\text{He,env}}=0.1~M_\odot$,  $M_{\text{core}}=1.47~M_\odot$, $Z=10^{-10}~Z_{\odot }$, $M_{\text{defl}}=0.42~M_\odot$, and the total $^{56}$Ni mass $=$ $0.87~M_\odot$.  
The resulting density and abundance structures are shown in Fig.~\ref{fig:models}. 

The overall structure consists of a central core of nuclear statistical equilibrium (NSE) burning products,  layers of IME produced by explosive O burning (Si/S \& Ca) and explosive C burning (Mg, Ne, O), and outer unburned C/O and He layers originating from the initial structure. 
As usual in this class of models, the presence of a nondegenerate envelope results in the compression in velocity space of all burning layers \citep{Hoeflich96,Quimby07}. 

Two peculiarities of the baseline model are: (1) the lack of a pronounced density shell when compared to a more typical envelope model or the large-amplitude pulsating delayed-detonation (DDT) models \citep{Hoeflich96,Quimby07,Hoeflich17,Hsiao2020}, and (2) that parts of the C/O envelope undergo burning.
Both are a direct consequence of the small $R_{\text{env}}$ of $10~R_{\text{core}}$ rather than an extended envelope with e.g. $100~R_{\text{core}}$ (see Fig.~\ref{fig:models}). 
Typically, an exploding WD reaches the initial stage of free expansion after about 10 to 20 seconds. 
However, in DCD07, the compact shell gets overrun by the ejecta within 1 to 2 seconds. 
Thus, a new pressure equilibrium between the shell and the inner material can be reached leading to a very shallow density shell-like feature between 0.7 and 1.4~\Msun\ (see the left panels of Fig. \ref{fig:models}). 
Moreover, the large envelope mass $M_{\text{env}}$, combined with the early interaction and lack of geometrical dilution of the ejecta, results in sufficiently high densities and peak temperature in the inner layers of the shell for explosive C burning (producing O, Ne, and Mg) to take place (see the right panels of Fig. \ref{fig:models}).   
In the model DCD07, the NSE, QSE and explosive C-burning products are below $7{,}500$~\kms, $6{,}500-14{,}000$~\kms, and $8{,}200-14{,}500$~\kms, respectively. The S velocity range  is shifted relative to Si by $\approx -500$ and $-1{,}000$~\kms\ at the inner and outer edges of explosive O burning, when the transitions to NSE and C-burning occur, respectively.

In this model, very limited He burning, below the abundance level of $10^{-3}$, takes place via helium  capture on carbon, $^{12}$C($\alpha,\gamma)^{16}$O, which also occurs during normal helium burning.
For models with similar parameters but smaller $M_{\text{env}}$, larger $M_{\text{He}}$, or smaller $R_{\text{env}}$, explosive He burning can lead to an increased outer abundance of NSE, QSE, and explosive C-burning products at high velocities, as is seen in some sub-$M_{\text{Ch}}$ or $M_{\text{Ch}}$-mass models with He-donors \citep{Nomoto82,Hoeflich96,Shen14,Hoeflich2019}.  
Corollary models with the same parameters as DCD07 but with a larger $R_{\text{env}}$ produce very few products of explosive C burning (lower right panel of Fig.~\ref{fig:models}) compared to our model that shows more O/Ne/Mg (upper-right panel of Fig.~\ref{fig:models}).\footnote{Note that $M_{\text{He}}$ can be slightly lower than $0.1~M_\odot$ without changing the discussion above. However, more C/O would result in an earlier onset of CO formation at $\approx 180$~d in this model.}

\subsection{Resulting Spectral Constraints} \label{appx-model:spectral_cons}
\label{model:spec_constrain}
Observed spectral measurements of \hy\ allow us to select the basic
parameters for the model. 
In the baseline model the Si abundance peaks at about $8{,}500$~\kms, which compares well with the observed \SiII\ $\lambda$6355 velocity plateau in Fig.~\ref{fig:Si_v} and allows us to set $M_{\text{env}}$. 
$R_{\text{env}}$ determines the velocity range of elements. Our choice of $R_{\text{env}}$ is determined such that silicon is produced between $7{,}000-14{,}000$~\kms. The absorption minimum indicates a velocity range for \SiII\ $\lambda$6355 of $5{,}000-9{,}500$~\kms. At later times, the feature around $\lambda6200$ becomes increasingly dominated by blends of Fe~II lines \citep{fisher91T99,Hoeflich95}. Therefore, the lower velocity limit is uncertain. 
The upper end of the Si velocity is better determined by the blue wing of the Si during early times \citep{Hoeflich95,Quimby06} because Si is needed to produce a wing. 
The observed value of the wing indicates a somewhat larger value ($\approx 14{,}000 -16{,}000$~\kms) than in our models.
This may indicate a slightly smaller $R_{\text{env}}$ than adopted. However, Shallow line wings are subject to blends and noise.
 
As discussed in Section~\ref{subsec:Si_velocity} and common in thermonuclear explosions \citep{Branch1983,Hoeflich95}, the spectral feature at roughly $\lambda6200$ transitions from being formed by \SiII\ to \FeII\ dominated when the photosphere recedes from S/Si to the NSE layers past maximum.
By +2~d from maximum light, the \SiII\ $\lambda$6355 feature is blended with \FeII\ $\lambda$6456, $\lambda$6516.
Note that the Fe is not primordial but produced by the radioactive
decay chain $^{56}$Ni $\rightarrow$ $^{56}$Co $\rightarrow$ $^{56}$Fe
with half-life times of 6.1 and 77.1~d, respectively. The decay
results in an Fe mass fraction $X_{\text{Fe}} \sim 0.1$ at $\sim40$~d past
the explosion. 
Therefore, the interpretation of this feature being blended by \FeII\ does not depend on the initial $Z$. 
The theoretical Mg velocities ($8{,}200-14{,}500$~\kms) are consistent with the observations discussed in the observational sections.
A potential challenge for the model comes from the observed velocity range of the optical \CII\ $\lambda$6580 from $9{,}300$~\kms\ at $-11$~d to about $5{,}000$~\kms\ at $\sim+10$~d past maximum light. 
If the weak feature in the NIR at 1.03~$\mu m$ is indeed \CI, it implies a higher velocity of $12{,}000$~\kms\ at $\sim -$8.5~d. The early \CI\ and \CII\ velocities are marginally consistent with the model which shows C down to $\sim9{,}000$~\kms, but the \CII\ velocity past maximum is not. 

The systematically lower velocity of \CII\ $\lambda$6580 may be understood as a limb effect. 
The optical \CII\ $\lambda$6580 feature is situated on the emission portion of the \SiII\ $\lambda$6355 P-Cygni profile, subjecting the \CII\ feature to a pronounced limb-brightening effect \citep{Hoeflich1990}. 
The \CII\ absorbs radiation from an underlying disk that brightens towards the edge. 
The result is that the \CII\ absorption profile is dominated by low absorption velocities, due to the fact that the observer sees the projected velocities \citep{Hoeflich02}. 

The possible early \CI\ velocity in the NIR agrees with the peak C abundance in the models.  
\CI\ has been previously identified in SN~1999by, a  subluminous SN~Ia with an extended C-rich layer \citep{Hoeflich02}.
SN~1999by and \hy\ are both red in their $B-V$ colors at maximum light. 
Rather than being dominated by ionized C, very strong \CI\ lines in the NIR are expected if the envelope of \hy\ is C rich. 
Since \hy\ does not show strong \CI\ lines but requires a large envelope mass for the overall abundance distribution, we conclude that the He layer provides the mass without additional spectral features.

By about maximum light, the photosphere is formed in the NSE and IME layers. 
As in SN~1999by, the ejecta are dominated by singly ionized iron-group elements, namely \FeII\ and \CoII.
Within the spectral constraints, we suggest a possible interpretation for the difference between the $H$- and $K$-band features in this object as an effect related to the velocity gradient ($\frac{dv}{dr}$). 
As discussed in section~\ref{subsec:NIR_spec},
\hy\ shows the individual components of Co/Fe multiplets in the $K$ but not in the $H$ band. 

\begin{figure}[tb!]
\begin{center}
\includegraphics[width=0.5\textwidth]{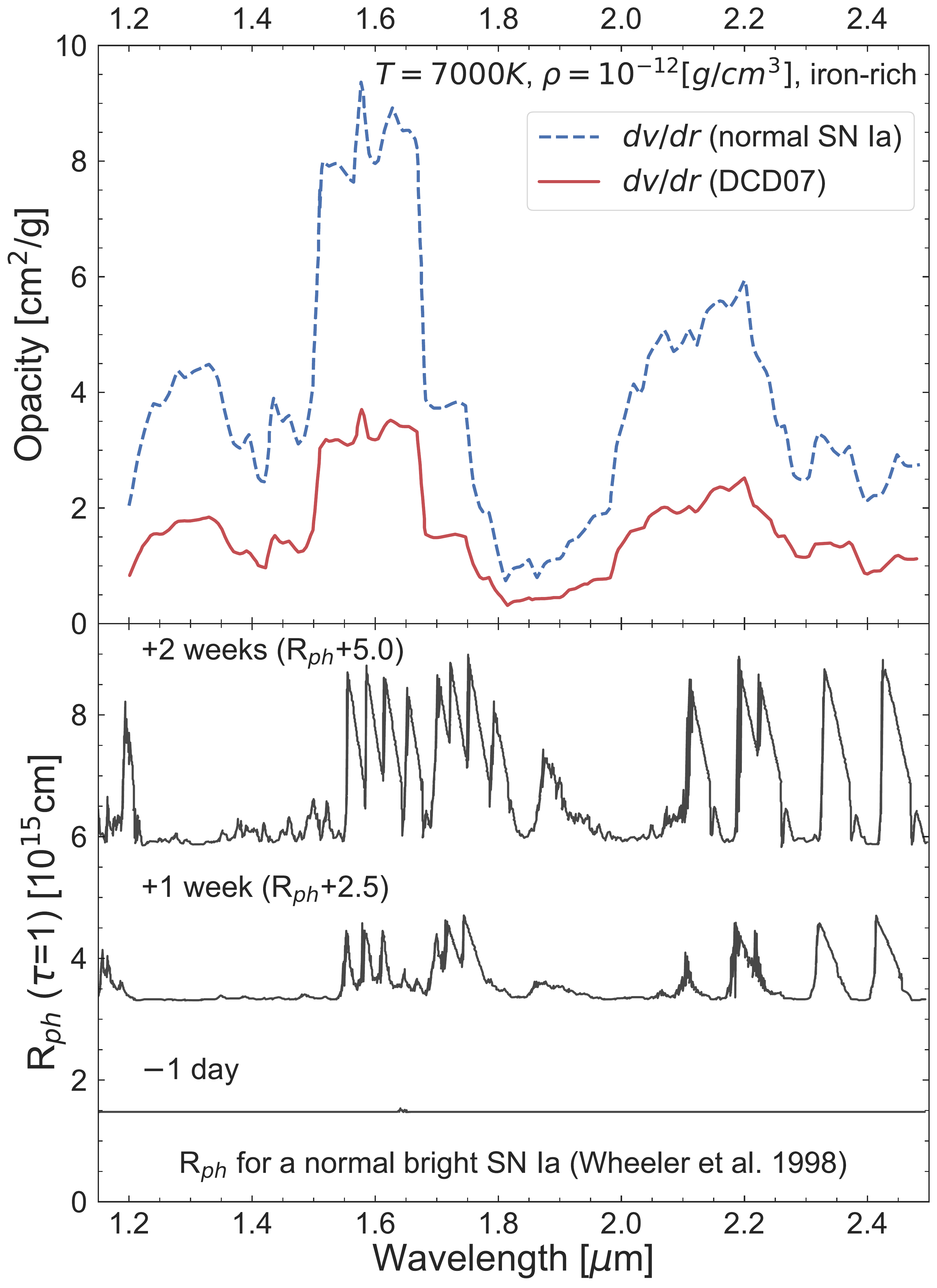}
\end{center}
\vskip -0.1cm
\caption{Formation of the $H$- and $K$-band features. \textit{Upper panel:} The effect of the velocity gradient on the quasi-continuum opacity in the NIR. For illustration, we show the broadband mean, quasi-continuum opacities (resolution R=10) as a function of wavelength for gradients corresponding to a normal SN~Ia for a $M_{\text{Ch}}$ mass explosion and our model (DCD07). 
\textit{Lower panel:}  We show the last scattering radius for a normal
bright SN~Ia calculated with HYDRA (adopted
from \citealt{Wheeler98}). Individual emission features are formed if
the individual components are  well above the continuum. 
Although the opacities of the individual components
of the multiplets are comparable, in the $K$ band the larger wavelength separation results in a smaller quasi-continuum opacity, by a factor of 2. As a result, in DCD07 and  subluminous SNe~Ia the $K$-band features are formed well above the continuum whereas the photosphere in $H$ remains close to the outer edge of the $^{56}$Ni region. Hence no emission features are formed early on.} 
\label{fig:kappa}
\end{figure}

The appearance of $H$-band emission has been attributed to the fact that the photosphere recedes within the NSE region and that the emission is powered by the incomplete Rosseland cycle \citep{mihalas78sa} well above the optical photosphere.  
Because the envelope is optically thick in the UV to the outer layers,
the Rosseland cycle  redistributes UV photons to the NIR.
The emission is expected to be formed at similar radii in both the $H$ ($1.6\pm 0.3~\mu$m) and $K$ ($2.2 \pm 0.4~\mu$m) bands because the lines originate from atomic levels with similar oscillator strengths and excitation energies (Fig.~\ref{fig:kappa}, lower panel).
As this process is dominated by atomic physics, one would assume it should also hold true for \superc\ SNe~Ia. 
However, observations show that this is not the case.
In \superc\ SNe, the $^{56}$Ni region is located in the deeper, high density layers where the photosphere recedes more slowly with time. 
Thus, less IGE are above the photosphere at comparable times, and the features are weaker and develop later in time.  

As a second effect, the quantitative difference between the $H$- and $K$-band properties in normal SNe~Ia versus \superc\ and  subluminous SNe~Ia can be understood in terms of differences in the velocity gradient.  
More precisely, the differences in the opacities in the $H$ and $K$ bands are manifested by the quasi-continuum produced by many overlapping lines.
The optical depth due to  expansion opacities is
$\propto \frac{dv}{dr}$ \citep{karp77,h93}, which is illustrated in  Fig.~\ref{fig:kappa}. 
The opacity is smaller by a factor of $2-3$ in \superc\ SNe compared to normal SNe~Ia, and the quasi-continuum over the $H$-band region is larger by a factor of two compared to the $K$-band region  (see Fig.~\ref{fig:kappa}, upper panel).
As a result, the quasi-continuum forms close to the outer edge of $^{56}$Ni in the $H$ band, whereas the photosphere recedes to well within the $^{56}$Ni region in the $K$ band.

This interpretation is supported by the similarity of \superc\ SNe and SN~1999by in their NIR evolution of the IGE spectral features \citep{Hoeflich02}. The larger emission in the $H$ band is mostly due to thermalization at the photosphere and results in a weak $H$-band break and weaker but present $K$-band features.
In both cases, the $^{56}$Ni is confined to low velocities layers (less than 6,000 to $7{,}000$~\kms),  whereas the $^{56}$Ni in normal SNe~Ia extends to about 12${,}$000 to $14{,}000$~\kms.  

Note that temperature in the outer layers of the envelope is sufficiently low and the density is sufficiently high for CO formation $\sim$1 month after the explosion.
However, with He substituting for the C/O mixture in the envelope, our
model forms CO only when the C/O is at low velocities and has
cooled down, that is, $5-6$ months after explosion.  
However, the exact details of CO formation are subject to numerical
uncertainties. 

\subsection{Resulting Light-curve Constraints} \label{appx-model:LC_cons}
The comparison between the DCD07 model and observed light curves is shown in Fig.~\ref{fig:LC}. 
For comparison to observations, we adopted a distance modulus of $\mu =34.33$~mag (see Section~\ref{subsec:host_z}), applied only the Galactic reddening correction of $E(B-V)_{\rm{MW}}=0.13$~mag (see Section~\ref{subsec:extinction}), and aided our choice of the best matching model with comparisons to the Doppler shifts of observed spectral features.
The model's post-maximum decline generally agrees with that of \hy. 
As usual for cool envelope models \citep{Hoeflich96,Hsiao2020}, they
do not show a strong secondary maximum, with the exception of
luminous SNe, small $M_{\text{env}}$, and high metallicity,  $Z\sim Z_{\sun}$. High metallicity is needed to produce a secondary maximum because iron group elements dominate the opacity.
The NIR flux is formed in the Wien tail of the energy distribution,
thus the NIR luminosity $L_{IR}$ scales as $\propto T\, R_{ph,IR}^2$.  
For bright SNe~Ia, the photospheric radius increases well beyond the
first maximum \citep{Hoeflich_KW1995}. However, for low temperatures,
the photosphere begins to recede near the time of maximum light and
the secondary maximum in the $i$-band merges with the primary
maximum \citep{Hoeflich_KW1995,Taubenberger17,Hsiao2020}.  
The challenge is to find a model within our framework that is red in $B-V$, dim in the $B$ band compared to other \superc\ objects, and luminous in both the UV and the NIR. while at the same time maintaining agreement in the spectroscopic properties that set $M_{\text{env}}$ and $R_{\text{env}}$.

As has been shown with previous envelope models 
with a detonation and a $1.2~M_{\odot }$ core (\citet{Hoeflich96}; see left panel of Fig.~\ref{fig:LC}), the rise times increase
with $M_{\text{env}}$, and the light curve becomes dimmer and broader
because geometrical dilution in the material of the degenerate WD
decreases.  However, for \hy, the core mass needed to be increased to
beyond $M_{\text{Ch}}$, $M_{\text{core}}=1.47~$\Msun\ to increase the
diffusion time scales in the core, leading to a slow pre-maximum rise
consistent with observations.

A large $M_{\text{env}}$ is inferred from the low \SiII\ velocity and
narrow \SiII\ distribution in the observed spectra.  However, we need
a fast receding photosphere and low opacities to avoid the photosphere
forming in the massive C/O layers.  Opacities are dominated by IGEs
with solar composition, or even typical Pop~II abundances with $Z=
Z_{\odot}/3$.  For low continuum opacities, we need He-rich material
and a C/O ratio larger than that in carbon stars, with a metallicity 
set to $Z=10^{-10}~Z_{\odot}$ for the baseline model.  The shell becomes
optically thin within days, even in the UV, resulting in a significant
amount of hard UV and X-rays ($\approx 40-50\%$ of the luminosity), 
during the phase dominated by $^{56}$Ni decay in the early part of the
light curve.  This avoids
the redistribution of the IR luminosity via the incomplete Rosseland
cycle mentioned above.  Virtually any metals would absorb the hard
radiation from the inner region, heating the shell and boosting the
optical and IR 
luminosity by volume driven bound-free and free-free emission.  The
need for a low-metallicity model is consistent with the low
metallicity of the host galaxy and the surrounding environment of \hy\
(see Section~\ref{subsec:host_env} and Appendix~\ref{appendix:host}).

The model $M_{\text{core}}$ places \hy\ in the regime of
super-$M_{\text{Ch}}$ degenerate cores but only slightly above $M_{\text{Ch}}$.  
Models with even larger
$M_{\text{core}}$, such as that produced by the merging of two WDs on
dynamical time scales, would start as detonations and produce more
$^{56}$Ni resulting in a brighter SN \citep{Pakmor12}.
Being close to $M_{\text{Ch}}$ makes the ignition of deflagration
burning likely \citep{khokhlov95,Niemeyer96}.  One of the key ingredients
of the CD model is an extended deflagration phase that results in 
pre-expansion of 
the core prior to the deflagration to detonation transition.  Here, we
use an extended deflagration burning of $M_{\text{defl}} \approx 0.42~
M_\odot$ to avoid a strong increase in the $^{56}$Ni production for a
small $R_{\text{env}}$.  Note that such a late transition from
deflagration to detonation has only been found in very  subluminous
SNe~Ia \citep{Hoeflich17}, which are rare in their own right.

The $Z=10^{-10}~Z_{\odot}$ model produced early light curves that are
approximately 1~mag brighter than the ASAS-SN nondetection limits.
Parts of the delay between the explosion and first light can be
understood in terms of the initial reheating phase.  At early times,
the temperature in the envelope cools rapidly, within minutes by
adiabatic expansion. Because of the small expansion velocity in the
nickel region, $\gamma$-rays are mostly trapped and don't contribute to
the heating of the outer material, resulting in delayed heating of the
outer layers compared to normal SNe~Ia.  The duration of this phase
mostly depends on the geometrical dilution of the $^{56}$Ni layers
and thus, on the expansion velocity of the $^{56}$Ni layers since
the $\gamma$-ray opacity does not depend on temperature.  This phase
lasts until $\approx 0.5 $ to $3.5 $ days after the explosion for
normal and  subluminous SNe~Ia,
respectively \citep{Hoeflich02,Hoeflich17}.  As a result of the low
temperature in these phases, the optical opacities are very low, and
this phase is characterized by an increasing photospheric velocity
with time, and rather blue $B-V$ colors compared to models with solar
metallicity and large envelope masses, which have $B-V \approx
0.3-0.4$~mag \citep{Hoeflich96}.  In \hy, the velocities are
similar to a  subluminous SNe~Ia, but the mass is larger and thus, the
density is higher, leading to high luminosity  from the degenerate core.
The $\gamma$-rays then start to heat the outer layers of the
degenerate core similar to  subluminous SNe~Ia after about $\sim 4-5$
days.  With $Z \sim 0$, the early radiation cannot excite C/O rich
material, due to the lack of strong line blanketing as commonly produced 
by lines of iron group elements.  To effectively block the early radiation, we need some
elements with lower ionization potential and strong lines such as
IGEs, for example, by mixing a small amount of IGEs into the inner
shell, boosting the opacity.  More extended mixing would shorten the
reheating.

The metallicity\footnote{For numerical reasons, a resolution
of only $\approx 10^{-4}$\Msun\ is used at the interface between the
core and the envelope.} was increased to
$Z=10^{-4}~Z_{\odot }$ at the interface of the degenerate core and the
C/O region of the shell. That is, the metallicity of the envelope was
increased. Since the degenerate core fully burns increasing the
metallicity there will only have small effects and here, we are
interested only in the opacity effects of the envelope on the light
curve, not in producing fully self-consistent models.  The result is an extended `dark(er)' phase,
lasting about one week after the explosion and an early red color
(see Fig.~\ref{fig:LC}).  The increase in $Z$ and the corresponding
dark phase bring this model within agreement of the ASAS-SN $V$-band
nondetections and reproduces the observed apparent rise time of $\sim24$ days.  
Thus, we choose this modified baseline model with
$Z=10^{-4}~Z_{\odot }$ as our best matching model. The $V$-band
last nondetection flux at 
16~d is slightly lower than the theoretical flux, but a change of 1 day
in phase from explosion between observations and the model is well
within the model uncertainties expected by mixing at the chemical
interface between QSE and IGE elements (see Fig.~\ref{fig:models}).
The increase in metallicity also brings the best matching model into closer
agreement with the UV observations (Section~\ref{sec:photometry}).
The early low luminosity excludes the presence of an ongoing interaction as
suggested for LSQ14fmg \citep{Hsiao2020}, which would add a constant early luminosity.
Using the same approach and
the nondetections being 2~mag below maximum light provides an upper limit of
the mass-loss rate of the AGB star to $\approx 2\times
10^{-7}$~\Msun~$yr^{-1}$.

The mass of the degenerate core suggests the ignition begins as a deflagration and subsequently turns into a detonation
as discussed above. To produce a subluminous SN we require a late transition density. 
A deflagration to detonation transition is a common characteristic in
reactive fluids in general, but the details in stellar explosions are
still under debate and may involve the Zel'dovich mechanism (i.e., the
mixing of burned and unburned material), crossing shock waves,
magnetohydrodynamic-instabilities, and shear flows in a highly
turbulent medium
\citep{Niemeyer96,khokhlov97,Livne1999,Poludnenko11,Hristov18}.
Recently, semi-analytical solutions have been developed that suggest
that a DDT is unavoidable under C/O-WD conditions and that the
transition density decreases with increasing specific nuclear energy
production \citep{polundenko19}.   
The model has low metallicity, resulting in a late DDT transition and a subluminous SN. 
Therefore, we expect these \hy-like objects are rare.

\subsection{Future Prospects} \label{appx-model:future_prospects}
We outline the limitations of this study, which
should be seen as a starting point towards understanding the nature of \superc\
SNe~Ia.  
Due to the sensitivity of the models to the rise time and the possible
`dark phase', early-time data are extremely important.  Such data may
be expected from ongoing and future campaigns such as and in
particular, early time UV spectra with HST to test for the chemical
signature of the outer few tenths of a solar mass. The simulations
are based on spherical explosions involving deflagrations and shell
interaction, both of which are Raleigh-Taylor (RT) unstable.  The
steep chemical gradients are likely to be smeared out on RT scales, but
full 3D hydro-simulations are needed to determine the details.  The picture
provided by the model is consistent with the observations of \hy,  but
forces the models into a particular corner of parameter 
space.  Two \superc\ objects, \hy\ and
LSQ14fmg are consistent with the scenario suggested here with clear
predictions for larger samples. The super-Chandrasekhar value of
$M_{\text{core}}$  lends credence to
the DCD scenario.
However, both \hy\ and LSQ14fmg are odd-balls in some respects. 
The lack of polarization in a few other \superc\ explosions, 
such as SN~2009dc \citep{Tanaka10} and SN~2007if \citep{Cikota19} 
is expected within the CD scenario, but not in merger scenarios
without massive envelopes. 
However, this does not exclude the possibility that a majority
of \superc\ SNe are the result of dynamical mergers. 
A larger sample of \superc\ events is needed for further analyses.
A key difference between scenarios should be expected in the
predictions of the nature of the nearby
environment.  
X-ray and radio observations and a better characterization of the
hosts are needed. 
Because of the similarity of the $BV$ bands with normal SNe~Ia, 03fg-like
SNe could hide in data sets where the light curves are followed in
only a few photometric bands.
Observations of redder bands such as the $i$ band are required in
order to distinguish these peculiar events as well as to exclude them for
cosmological use.


\end{CJK*}
\end{document}